\documentclass[%
 reprint,
 amsmath,amssymb,
 aps,
citeautoscript]{revtex4-2}

\setcitestyle{super}

\usepackage{amssymb}
\usepackage{amsthm}
\usepackage{graphicx}
\usepackage{dcolumn}
\usepackage{bm}
\usepackage{upgreek} 
\usepackage[colorlinks=true,bookmarks=false,citecolor=black,urlcolor=blue,linkcolor=black]{hyperref} 

\usepackage{color}
\usepackage{soul}
\soulregister\cite{7} 
\soulregister\ref{7} 
\soulregister\eqref{7} 

\begin{document}

\title{Dynamic gain and frequency comb formation in exceptional-point lasers}

\author{Xingwei Gao$^1$}\email{xingweig@usc.edu}
\author{Hao He$^1$}
\author{Scott Sobolewski$^1$}
\author{Alexander Cerjan$^2$}\email{awcerja@sandia.gov}
\author{Chia Wei Hsu$^1$}

\affiliation{
$^1$Ming Hsieh Department of Electrical and Computer Engineering, University of Southern California, Los Angeles, CA 90007, USA \\
$^2$Center for Integrated Nanotechnologies, Sandia National Laboratories, Albuquerque, NM 87185, USA
}

\begin{abstract}
\section*{Abstract}
Exceptional points (EPs)---singularities in the parameter space of non-Hermitian systems where two nearby eigenmodes coalesce---feature unique properties with applications such as sensitivity enhancement and chiral emission.
Existing realizations of EP lasers operate with static populations in the gain medium.
By analyzing the full-wave Maxwell--Bloch equations, here we show that in a laser operating sufficiently close to an EP, the nonlinear gain will spontaneously induce a multi-spectral multi-modal instability above a pump threshold, which initiates an oscillating population inversion and generates a frequency comb.
The efficiency of comb generation is enhanced by both the spectral degeneracy and the spatial coalescence of modes near an EP.
Such an ``EP comb'' has a widely tunable repetition rate, self-starts without external modulators or a continuous-wave pump, and can be realized with an ultra-compact footprint.
We develop an exact solution of the Maxwell--Bloch equations with an oscillating inversion, describing all spatiotemporal properties of the EP comb as a limit cycle.
We numerically illustrate this phenomenon in a 5-\textmu m-long gain-loss coupled AlGaAs cavity and adjust the EP comb repetition rate from 20 to 27~GHz. 
This work provides a rigorous spatiotemporal description of the rich laser behaviors that arise from the interplay between the non-Hermiticity, nonlinearity, and dynamics of a gain medium.
\end{abstract}

\maketitle

\section{Introduction}\label{Introduction}
\vspace{-4pt}

\begin{figure*}[t]
\centering
\includegraphics[trim={1.5cm 0 1.5cm 0}]{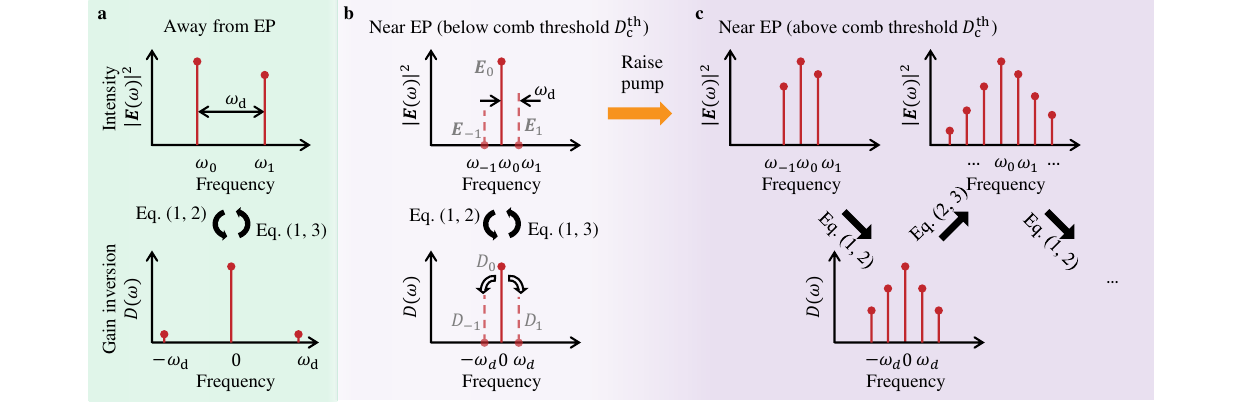}
\vspace{-4pt}
\caption{\textbf{Frequency comb formation in an exceptional-point (EP) laser.} \textbf{a} In an ordinary microcavity laser away from an EP, a second mode turns on when another resonance of the cavity receives enough gain to overcome its loss. 
Given the large frequency difference $\omega_{\rm d}$, the beating between the two modes is too fast to induce significant dynamics in the population inversion $D$.
\textbf{b} An EP boosts the dynamic inversion factor $\zeta \approx |\mathbf{E}_{-1}|^2/|\mathbf{E}_{1}|^2$ of Eq.~\eqref{eq:zeta}, creating a multi-spectral multi-modal perturbation (dashed lines) that induces a dynamic gain oscillating at the beat frequency $\omega_{\rm d}$.
\textbf{c} At pumping strengths above the comb threshold $D_{\textrm{c}}^{\text{th}}$, the multi-spectral perturbation grows into sustained oscillations (solid lines), which induce additional gain oscillations and cascade down to generate a comb at frequencies $\omega_m = \omega_0 + m \omega_\text{d}$.
The large-time behavior is described self-consistently by periodic-inversion \textit{ab~initio} laser theory (PALT) of Eqs.~\eqref{PALT:E}--\eqref{PALT:D}.
}
\vspace{-4pt}
\label{fig:1}
\end{figure*}

An exceptional point (EP) is a non-Hermitian degeneracy where not only do two eigenvalues coincide, but the spatial profiles of the two modes also become identical~\cite{2011_Moiseyev_book,Heiss_2012,feng2017non, el2018non, doi:10.1126/science.aar7709}. 
Realizing such non-Hermitian phenomena at steady state necessitates gain and loss, making microcavity lasers a fertile ground to explore EPs. 
The mode coalescence and corresponding topology of the eigenvalue landscape bestow EP lasers with unique properties such as reversed pump dependence~\cite{PhysRevLett.108.173901}, loss-induced lasing~\cite{doi:10.1126/science.1258004}, single-mode operation \cite{doi:10.1126/science.1258479, doi:10.1126/science.1258480}, chiral emission~\cite{doi:10.1073/pnas.1603318113, doi:10.1126/science.aaf8533, doi:10.1126/science.aba8996}, sensitivity enhancement \cite{doi:10.1038/nature23281, hodaei2017enhanced, hokmabadi2019non, lai2019observation, kononchuk2022exceptional,2023_Suntharalingam_ncomms}, spectral phase transitions~\cite{2021_Roy_ncomms}, and topological state transfer~\cite{schumer2022topological}.
In semiconductor microcavity lasers, the frequency separation between lasing modes is typically large enough that the cross beats between modes oscillate so fast that they average away before the gain medium can respond, leading to a static population inversion in the gain medium~\cite{PhysRevA.43.2446}.
Previous realizations of EP lasers operated in this regime, yielding stable single-mode or few-mode behavior (Fig.~\ref{fig:1}a); these static-inversion lasers can be modeled by the ``steady-state \textit{ab~initio} laser theory'' (SALT)~\cite{PhysRevA.74.043822, doi:10.1126/science.1155311, PhysRevA.82.063824, 2014_Esterhazy_PRA,Ge:08}. 

To enhance the performance of EP-related phenomena, such as the sensitivity of EP sensors~\cite{doi:10.1038/nature23281, hodaei2017enhanced, hokmabadi2019non, lai2019observation, kononchuk2022exceptional,2023_Suntharalingam_ncomms}, it is desirable to operate as close to an EP as possible. However, sufficiently close to an EP, the vanishingly small eigenvalue difference (namely, frequency difference) means that any two lasing modes of a multimode system necessarily produce beat notes slow enough to render the population inversion nonstationary.
In general, since the population inversion determines the laser's gain, any non-stationary inversion produced by beat notes acts as a periodic modulation over the effective complex refractive index of the laser system. If gain's periodic modulation frequency matches the cavity's free spectral range (FSR) such that its high-quality resonances can be excited, a frequency comb will form whose line spacing relies on the optical size of the cavities, examples of which include mode-locked laser combs~\cite{RevModPhys.75.325}, Kerr combs~\cite{2007_DelHaye_Nature,2018_Kippenberg_Science_review}, electro-optic combs~\cite{Parriaux:20}, and quantum cascade laser combs~\cite{hugi2012mid, silvestri2023frequency,2024_Opacak_Nature}.
Thus, a laser operating close to an EP has two competing frequency scales, one set by the eigenvalue splitting, and the other set by the cavity FSR; the former suggests that the system should become a comb due to population inversion dynamics, while the latter suggests that comb lines forming away from the cavity resonances will experience too much radiative loss to persist. Unfortunately, existing theories~\cite{PhysRevA.81.023822, PhysRevA.92.063829, 2016_Ge_srep, 2017_Teimourpour_srep, 2018_Kominis_APL, Horstman:20, 10.1093/nsr/nwac259, 2023_Drong_PRA, PhysRevLett.130.266901, ji2023tracking, 2022_Benzaouia_APLph} cannot describe both the spatial complexity and the temporal dynamics of the laser in this nonstationary-inversion regime. 




In this work, we develop a rigorous analysis of the full-wave Maxwell--Bloch equations and show that sufficiently close to an EP, a laser necessarily develops into a frequency comb when pumped above a comb threshold $D_{\textrm{c}}^{\text{th}}$.
In this operating regime, the nonlinear gain induces a multi-spectral multi-modal perturbation that destabilizes single-mode operation and initiates temporal oscillations in the population inversion (Fig.~\ref{fig:1}b).
The dynamic inversion then nonlinearly couples different frequencies to produce a frequency comb {above $D_{\textrm{c}}^{\text{th}}$} (Fig.~\ref{fig:1}c).  As such, our framework demonstrates that a comb must form even though the frequency of modulation driven by the dynamic inversion is typically orders of magnitude smaller than the FSR of the laser cavity.
Distinct from conventional combs, such an ``EP comb'' has a repetition rate independent of FSR, 
which enables a widely tunable repetition rate and a compact cavity size.
The EP comb oscillation self-starts, forming spontaneously above a pump threshold without an external modulator or an external continuous-wave laser.
Moreover, we find the efficiency of comb teeth generation, characterized by a $\zeta$ factor that we introduce, to be
enhanced by the spectral degeneracy 
and the spatial coalescence of the mode profile, conditions that are simultaneously met by operating near an EP.
Finally, as an example, we provide full-wave solutions of an EP comb in an AlGaAs gain-loss coupled 
cavity that is merely 5~\textmu m long, demonstrating a continuously tunable repetition rate from 20~GHz to 27~GHz, about 400 times smaller than the free spectral range of this small cavity. Overall, the EP comb phenomena we predict provides a rich and unexpected intersection between non-Hermitian photonics, laser physics, nonlinear dynamics, and frequency combs. 

\vspace{-8pt}
\section{Results}
\vspace{-4pt}

\vspace{-8pt}
\subsection{Dynamic inversion and comb formation near an exceptional point}\label{sec:SIA}
\vspace{-4pt}

To rigorously describe the wave physics and the spatiotemporal complexity of an EP laser, we adopt the Maxwell--Bloch (MB) equations~\cite{haken1985laser, PhysRevA.54.3347}
\begin{align}
&\frac{\partial}{\partial t}D = - \gamma_{\parallel}(D-D_\text{p})-\frac{i\gamma_{\parallel}}{2}(\mathbf{E}^*\cdot\mathbf{P}-\mathbf{E}\cdot\mathbf{P}^*), \label{MB:D}\\
&\frac{\partial}{\partial t}\mathbf{P} = -(i\omega_{ba} + \gamma_{\perp})\mathbf{P}-i\gamma_{\perp}D(\mathbf{E}\cdot \bm{\uptheta})\bm{\uptheta}^*, \label{MB:P} \\
-&\nabla\times\nabla\times \mathbf{E} - \frac{1}{c^2} \left(\varepsilon_c\frac{\partial^2}{\partial t^2}+\frac{\sigma}{\varepsilon_0} \frac{\partial}{\partial t}\right)\mathbf{E} = \frac{1}{c^2} \frac{\partial^2}{\partial t^2} \mathbf{P} \label{MB:E}.
\end{align}
The electrical field $\mathbf{E}(\mathbf{r},t)$ is described classically with Maxwell's equations.
The gain medium is described quantum mechanically as an ensemble of two-level atoms, leading to a population inversion $D(\mathbf{r},t)$ and inducing a polarization density $\mathbf{P}(\mathbf{r},t)$ that couple nonlinearly with $\mathbf{E}(\mathbf{r},t)$ through dipole interactions (Supplementary Sec.~1).
The $D$, $\mathbf{E}$, and $\mathbf{P}$ here are dimensionless as they have been normalized by $R^2/(\varepsilon_0\hbar\gamma_{\perp})$, $2R/(\hbar\sqrt{\gamma_{\perp}\gamma_\parallel})$, and $2R/(\varepsilon_0\hbar\sqrt{\gamma_{\perp}\gamma_\parallel})$, respectively,
with $R$ being the amplitude of the atomic dipole moment, $\varepsilon_0$ the vacuum permittivity, $\hbar$ the Planck constant, and $\gamma_\perp$ the dephasing rate of the gain-induced polarization ({\it i.e.}, the bandwidth of the gain).
Here, $D_\text{p}(\mathbf{r})$ is the normalized net pumping strength and profile, $\omega_{ba}$ is the frequency gap between the two atomic levels, 
${\bm{\uptheta}}$ is the unit vector of the atomic dipole moment with $\bm{\uptheta}\cdot\bm{\uptheta}^*=1$,
$\varepsilon_c(\mathbf{r})$ is the relative permittivity profile of the cold cavity, $\sigma(\mathbf{r})$ is a conductivity profile that produces linear absorption,
and $c$ is the vacuum speed of light. 
$\mathbf{E}$ and $\mathbf{P}$ satisfy an outgoing boundary condition outside the cavity.

When the pumping strength reaches the first lasing threshold $D_1^{\text{th}}$, the gain overcomes the radiation loss and absorption loss, and a single-mode lasing state
$\mathbf{E}(\mathbf{r},t)=\mathbf{E}_0(\mathbf{r})e^{-i\omega_0 t}$
emerges at a real-valued frequency $\omega_0$. 
Substituting this single-mode solution into the MB equations (Supplementary Sec.~2), we get
\begin{equation}
\hat{O}\left({\omega}_0\right)\mathbf{E}_{0}(\mathbf{r}) \equiv
\left[-\nabla\times\nabla\times + \frac{\omega_{0}^2}{c^2} 
\varepsilon_{\rm eff}(\mathbf{r},\omega_0)
\right] \mathbf{E}_{0}(\mathbf{r}) = 0.
\label{E0}
\end{equation}
Here, $\varepsilon_{\rm eff}(\mathbf{r},\omega) = \varepsilon_c(\mathbf{r})+i\sigma(\mathbf{r})/(\omega\varepsilon_0) + \Gamma(\omega)D_0(\mathbf{r})\bm{\uptheta}^*\bm{\uptheta}\cdot$
is an effective intensity-dependent and frequency-dependent permittivity profile of the active cavity, and
$\Gamma(\omega) \equiv {\gamma_\perp}/(\omega-\omega_{ba}+i\gamma_\perp)$.
The gain
$D(\mathbf{r},t)=D_0(\mathbf{r})
= D_\text{p}(\mathbf{r}) / [ 1+|\Gamma(\omega_0) \mathbf{E}_{0}(\mathbf{r})\cdot {\bm{\uptheta}}|^2 ]$ is nonlinearly saturated by the local intensity, referred to as spatial hole burning.
In this single-mode regime, Eq.~\eqref{E0} is an exact solution of the MB equations, 
the gain is static, and its relaxation rate $\gamma_\parallel$ plays no role at steady state.

One may freeze the nonlinearity by {considering a linear operator $\hat{O}\left({\omega}\right)$ in Eq.~\eqref{E0} that} uses a fixed saturated gain $D_0(\mathbf{r})= D_\text{p}(\mathbf{r}) / [ 1+|\Gamma(\omega_0) \mathbf{E}_{0}(\mathbf{r})\cdot {\bm{\uptheta}}|^2 ]$ for a fixed lasing intensity profile $|\mathbf{E}_0(\mathbf{r})|^2$.
{This linear $\hat{O}\left({\omega}\right)$ then admits eigenmodes $\{\mathbf{\psi}_n(\mathbf{r})\}_n$ with complex-valued eigen frequencies $\{\tilde{\omega}_n\}_n$, satisfying $\hat{O}(\tilde{\omega}_n) \mathbf{\psi}_n = 0$ with an outgoing boundary condition.}
We refer to them as the active-cavity resonances (also called quasinormal modes~\cite{2022_Sauvan_OE_review}).
{We also define operator $\hat{O}\left({\omega}\right)$ below the first lasing threshold $D_1^{\text{th}}$ simply using the linear unsaturated gain $D_0(\mathbf{r})= D_\text{p}(\mathbf{r})$.}
When we increase the pumping strength to $D_1^{\text{th}}$, the eigenvalue $\tilde{\omega}_0=\omega_0$ reaches the real-frequency axis, and that resonance becomes the first lasing mode
$\mathbf{E}_0(\mathbf{r}) \propto \mathbf{\psi}_0(\mathbf{r})$. 

In the following, we define an EP as where two eigenvalues $\{\tilde{\omega}_0, \tilde{\omega}_1\}$ of the linear operator $\hat{O}\left({\omega}\right)$ coalesce, at which point the corresponding mode profiles $\{\mathbf{\psi}_0, \mathbf{\psi}_1\}$ must also become the same given the non-Hermitian nature of $\hat{O}\left({\omega}\right)$.
An EP may exist at pumping strengths below the first lasing threshold $D_1^{\text{th}}$; such a below-threshold EP can indirectly affect laser properties~\cite{PhysRevLett.108.173901, doi:10.1126/science.1258004, 2014_El-Ganainy_PRA} but cannot be directly accessed since it does not correspond to a steady-state solution.
In this paper, we consider a laser close to an accessible EP at pumping strengths near or above $D_1^{\text{th}}$.

The SALT formalism assumes the population inversion to be static, $D(\mathbf{r},t) = D_0(\mathbf{r})$~\cite{PhysRevA.74.043822, doi:10.1126/science.1155311, PhysRevA.82.063824, 2014_Esterhazy_PRA}. Under SALT, the resonances $\{\mathbf{\psi}_n\}$ are the modes that turn on and lase when they receive enough gain. 
For a second mode $\mathbf{\psi}_1$ to turn on, it must have a spatial profile sufficiently different from the lasing mode $\mathbf{E}_0 \propto \mathbf{\psi}_0$ that it can amplify using the gain outside the spatial holes ({\it i.e.}, away from the peaks of $|\mathbf{\psi}_0(\mathbf{r})|^2$).
Near an EP, $\mathbf{\psi}_1$ necessarily has a similar spatial profile as $\mathbf{\psi}_0$ and so cannot turn on.
Therefore, SALT predicts an EP laser to stay single-mode. 
However, this single-mode prediction is based on the static-inversion assumption, which is questionable near an EP since
the slow beating between the two very-close-by frequencies may induce dynamics in the inversion $D(\mathbf{r},t)$. 
To find out what actually happens to a laser close to an EP, one must go beyond SALT and account for the inversion dynamics and its effects.

To do so, we start with a monochromatic perturbation $\mathbf{E}_1(\mathbf{r}) e^{-i\omega_1 t}$ (dashed line in Fig.~\ref{fig:1}b) to single-mode operation, so the total field is $\mathbf{E}(\mathbf{r},t)=\mathbf{E}_0(\mathbf{r})e^{-i\omega_0 t}+\mathbf{E}_1(\mathbf{r})e^{-i\omega_1 t}$.
The frequency difference $\omega_\text{d} = \omega_1 - \omega_0$ can be positive or negative.
With the inversion almost static, it follows from Eq.~\eqref{MB:P} that $\mathbf{P}(\mathbf{r},t)=\mathbf{P}_0(\mathbf{r})e^{-i\omega_0 t}+\mathbf{P}_1(\mathbf{r})e^{-i\omega_1 t}$ with $\mathbf{P}_m = \Gamma_m D_0E_{ m}{\bm{\uptheta}}^*$ for $m=0,1$, where $\Gamma_m = \Gamma(\omega_m)$ and $E_{ m} \equiv \mathbf{E}_{m}\cdot {\bm{\uptheta}}$. 
We then see from Eq.~\eqref{MB:D} that the inversion is no longer purely stationary;
as illustrated in Fig.~\ref{fig:1}b, we now have
$D(\mathbf{r},t) = D_{-1}(\mathbf{r})e^{i\omega_\text{d} t} + D_0(\mathbf{r}) + D_1(\mathbf{r})e^{-i\omega_\text{d} t}$ with a dynamic component induced by the perturbation, 
\begin{equation}
\vspace{-2pt}
D_{-1}(\mathbf{r}) = D_1^*(\mathbf{r}) = \frac{(\Gamma_0 -\Gamma_1^*)\gamma_{\parallel}}{2 (i\gamma_{\parallel} - \omega_\text{d})}E_{ 0}(\mathbf{r})E_{ 1}^*(\mathbf{r})D_0(\mathbf{r}). \label{CF}
\vspace{-2pt}
\end{equation}
This oscillating gain $D_{\pm1}(\mathbf{r})e^{\mp i\omega_\text{d} t}$ arises from cross beating in the nonlinear term $\mathbf{E}^*\cdot\mathbf{P}$ of Eq.~\eqref{MB:D}, so it is enhanced where $E_0(\mathbf{r})$ and $E_{1}(\mathbf{r})$ spatially overlap. 
Substituting $D(\mathbf{r},t)$ 
into Eq.~\eqref{MB:P} yields
a polarization $\mathbf{P}_{-1} = \Gamma_{-1}( D_{-1}E_{ 0} + D_0E_{ -1} ){\bm{\uptheta}}^*$ at new frequency $\omega_{-1}=\omega_0 -\omega_\text{d}$, 
which acts like a current source to produce a new field $\mathbf{E}_{-1}$ via Eq.~\eqref{MB:E}, 
\begin{equation}
\vspace{-1pt}
\hspace{-7pt}
{\small
\left[-\nabla\times\nabla\times + \frac{\omega_{-1}^2}{c^2} 
\varepsilon_{\rm eff}(\mathbf{r},\omega_{-1})
\right] \mathbf{E}_{-1} 
= -\frac{\omega_{-1}^2}{c^2}\Gamma_{-1} D_{-1} E_{ 0}\bm{\uptheta}^*. 
}
\label{E-1}
\end{equation}

This additional frequency component $\mathbf{E}_{-1}(\mathbf{r}) e^{-i\omega_{-1} t}$, generated in a four-wave-mixing~\cite{Boyd_book} fashion by the nonlinear gain (Fig.~\ref{fig:1}b), differentiates an EP laser from a conventional laser and marks the onset of dynamic inversion and comb formation.
To quantify the strength of this frequency generation, we solve Eq.~\eqref{E-1} 
to obtain (Supplementary Sec.~3)
\begin{equation}
\vspace{-2pt}
\hspace{-10pt}
\frac{\langle |\mathbf{E}_{-1}|^2 \rangle}{\langle |\mathbf{E}_{1}|^2 \rangle} \approx
\underbrace{\frac{\gamma_\parallel^2}{\omega_\text{d}^2+\gamma_\parallel^2}
\frac{\omega_{0}^2}{4\omega_\text{d}^2}}_{\text{I}} 
\underbrace{\frac{\langle |\mathbf{E}_0|^2 \rangle \, |\langle D_0 E_{ 0}^3E_{ 1}^* \rangle|^2}{\langle |\mathbf{E}_{1}|^2\rangle \, |\langle \varepsilon_c \mathbf{E}_0\cdot\mathbf{E}_0\rangle|^2}}_{\text{II}} 
\equiv \zeta, 
\vspace{-2pt}
\label{eq:zeta}
\end{equation}
which we denote as the dynamic inversion factor $\zeta$.
Here, $\langle \cdots \rangle$ = $\int(\cdots)dr^3 $ denotes integration over space.
We see $\zeta$ is proportional to the lasing intensity squared, $|\mathbf{E}_0|^4$,
but independent of the perturbation strength $|\mathbf{E}_{1}|$,
so $\zeta \neq 0$ even for an infinitesimal perturbation. 

The dynamic inversion factor $\zeta$ has two ingredients, Factor~I on the spectral dependence, and Factor~II on the spatial dependence and $|\mathbf{E}_0|^4$ laser intensity dependence.
When the perturbation $\mathbf{E}_1 e^{-i\omega_1 t}$ overlaps well with the long-lived resonances $\{\mathbf{\psi}_n e^{-i\tilde{\omega}_n t}\}$,
the response can sustain longer. 
So, the frequency difference $\omega_\text{d} = \omega_1 - \omega_0$ here correlates with the eigenvalue difference {$\tilde{\omega}_1 - \tilde{\omega}_0$}, which is minimized near an EP,
enhancing Factor~I through its 
$\omega_{\rm d}^{-4}$ scaling.
The resonances 
are biorthogonal with $\langle \varepsilon_c \mathbf{\psi}_0\cdot\mathbf{\psi}_1\rangle \approx 0$.
As the two resonances coalesce near an EP, $\mathbf{E}_0 \propto \mathbf{\psi}_0 \approx \mathbf{\psi}_1$, so $\langle \varepsilon_c \mathbf{E}_0\cdot\mathbf{E}_0\rangle \approx 0$,
which enhances Factor II of $\zeta$
in the same way as how an EP enhances
the Petermann factor $K \equiv |\langle \varepsilon_c |\mathbf{E}_0|^2\rangle/\langle \varepsilon_c \mathbf{E}_0\cdot\mathbf{E}_0\rangle|^2$
~\cite{1979_Petermann_JQE, 1989_Siegman_PRA_1,1996_Wenzel_JQE,2003_Berry_JMO,2008_Lee_PRA, 2017_Pick_OE, 2020_Wang_ncomms, 2022_Smith_PRA}.
Such a mode coalescence promotes coupling through the stronger field overlap.

In Supplementary Sec.~4, we perform a stability analysis~\cite{PhysRevA.92.013847,PhysRevA.95.023835} 
to determine the decay (or growth) rate of the multi-frequency perturbation
$\mathbf{E}_1(\mathbf{r}) e^{-i\omega_1 t} + \mathbf{E}_{-1}(\mathbf{r}) e^{-i\omega_{-1} t}$. 
As the pumping strength increases, the decay rate crosses over to become a growth rate, and the crossover marks the next threshold $D_2^{\text{th}}$. This is where the infinitesimal multi-frequency perturbation materializes into sustained oscillations at $\omega_{\pm 1}$.
As the pump increases further, the new frequencies induce higher-harmonic oscillations in the population inversion, which generates more lasing frequencies. The process cascades down to produce a frequency comb (Fig.~\ref{fig:1}c). 
Therefore, near an EP where the $\zeta$ factor is substantial, $D_2^{\text{th}} = D_{\textrm{c}}^{\text{th}}$ is also the threshold where the frequency comb (indicated by the subscript c) emerges,
corresponding to a supercritical Hopf bifurcation~\cite{strogatz2018nonlinear}. 

We note that while the $\zeta$ factor is resonantly enhanced, the beat frequency $\omega_{\rm d}$ and the coupled perturbation $\mathbf{E}_{\pm 1}(\mathbf{r})$ 
are determined by the linear stability eigenproblem (Supplementary Sec.~4), not by Eq.~\eqref{E0} as in SALT. 
Therefore, $\mathbf{E}_{\pm 1}(\mathbf{r})$ is generally {\it not} an active-cavity resonance  $\mathbf{\psi}_n$ but a superposition of multiple resonances,
and the comb spacing $\omega_{\rm d}$ is correlated with but not identical to the resonance spacing {$|\tilde{\omega}_1 - \tilde{\omega}_0|$}.

\vspace{-4pt}
\subsection{Exact dynamic solution: PALT} \label{PALT}
\vspace{-4pt}

The preceding analysis predicts comb formation near an EP and its threshold.
To additionally predict the laser behavior above $D_{\textrm{c}}^{\text{th}}$ such as the the evolution of the comb-line intensities, repetition rate, spatial profiles, and temporal dynamics,
one must address the coupling between all frequency components self-consistently.
Since the cascade process couples frequencies separated by $\omega_\text{d} = \omega_1 - \omega_0$, we postulate the following spatiotemporal dependence at large time~\cite{2023_Gao_CLEO,2023_Gao_NLO}
\begin{align}
\vspace{-2pt}
\mathbf{E}(\mathbf{r},t) &= e^{-i\omega_0 t}\sum_{m=-\infty}^{+\infty}\mathbf{E}_m(\mathbf{r})e^{-im\omega_\text{d} t},\label{CM:E} \\
\mathbf{P}(\mathbf{r},t) &= e^{-i\omega_0 t}\sum_{m=-\infty}^{+\infty}{\bm{\uptheta}}^*P_m(\mathbf{r})e^{-im\omega_\text{d} t},\label{CM:P} \\
D(\mathbf{r},t) &= \sum_{m=-\infty}^{+\infty}D_m(\mathbf{r})e^{-im\omega_\text{d} t}, \label{CM:D}
\vspace{-2pt}
\end{align} 
with $\omega_0$, $\omega_{\rm d}$, and $D(\mathbf{r},t)$ being real numbers.
This ansatz describes 
a limit cycle~\cite{strogatz2018nonlinear}, which is periodic in time and therefore represented rigorously by a Fourier series.
It also describes single-mode and two-mode operation as special cases but excludes operating with more than two cavity modes or a chaotic dynamics.
The temporal periodicity is $\tau = 2\pi/\omega_{\rm d}$.

We show in Sec.~\ref{sec:Method} that the ansatz of Eqs.~\eqref{CM:E}--\eqref{CM:D} forms an exact solution of the full-wave MB equations, Eqs.~\eqref{MB:D}--\eqref{MB:E}, with no approximation. Eliminating the gain-induced polarization yields a coupled nonlinear equation for $\{\mathbf{E}_{m}\}$ 
\begin{multline}
-\nabla\times\nabla\times \mathbf{E}_{m} + \frac{\omega_{m}^2}{c^2} \left(\varepsilon_c+\frac{i\sigma}{\omega_{m} \varepsilon_0}\right)\mathbf{E}_{m} \\
= -\frac{\omega_{m}^2}{c^2}\Gamma_m \sum_{n=-\infty}^{+\infty}{D_{m-n}(\mathbf{E}_n \cdot {\bm{\uptheta}}){\bm{\uptheta}}^*}, \label{PALT:E}
\end{multline}
and $\{D_m\}$
\begin{equation}
\bar{D}=D_\text{p}[\bar{\bar{I}}- 0.5 \bar{\bar{\Gamma}}_{\parallel}(\bar{\bar{E}}^\dag \bar{\bar{\Gamma}}_+ \bar{\bar{E}}-\bar{\bar{E}} \bar{\bar{\Gamma}}_-^\dag \bar{\bar{E}}^\dag)]^{-1}\bar{\delta}, \label{PALT:D}
\end{equation}
with $\omega_m = \omega_0 + m\omega_\text{d}$.
Different frequency components $\mathbf{E}_{m}$ are coherently coupled through a dynamic inversion $D_{m-n}$ oscillating at the frequency difference.
Here, $\bar{D}$ and $\bar{\delta}$ are column vectors with elements
$(\bar{D})_m=D_m$ and $(\bar{\delta})_m=\delta_{m}$, where $\delta_m$ is the Kronecker delta with $\delta_0 = 1$ and $\delta_{m\neq0} = 0$; 
$\bar{\bar{I}}$ is the identity matrix;
$\bar{\bar{E}}$ is a full matrix with elements $(\bar{\bar{E}})_{mn}=\mathbf{E}_{m-n}\cdot{\bm{\uptheta}}$;
$^\dagger$ denotes matrix conjugate transpose;
$\bar{\bar{{{\Gamma}}}}_{\parallel}$ and $\bar{\bar{{{\Gamma}}}}_{\pm}$ are diagonal matrices with $(\bar{\bar{\Gamma}}_{\parallel})_{mn}={\delta_{m-n}\gamma_\parallel}/{(m\omega_\text{d}+i\gamma_\parallel)}$ and 
$(\bar{\bar{\Gamma}}_{\pm})_{mn}= \delta_{m-n} \Gamma_{\pm m}$, where $\Gamma_m = \Gamma(\omega_m) = {\gamma_\perp}/(\omega_m-\omega_{ba}+i\gamma_\perp)$ was defined earlier.

Solving Eqs.~\eqref{PALT:E}--\eqref{PALT:D} for $\{\mathbf{E}_{m}(\mathbf{r})\}$, $\{D_m(\mathbf{r})\}$, $\omega_0$, and $\omega_\text{d}$ yields all properties of the laser comb, including the frequency spectrum, temporal dynamics, spatial profiles, and input-output curves.
To match the number of equations and the number of unknowns, we fix two gauge variables by recognizing that when $\mathbf{E}(\mathbf{r},t)$ is a solution, $e^{i\phi}\mathbf{E}(\mathbf{r},t-t_0)$ with any real-valued $\phi$ and $t_0$ is also a solution.
We name 
this formalism ``periodic-inversion \textit{ab~initio} laser theory'' (PALT), which overcomes the stationary-inversion limitation of SALT.

{Note there is no sharp transition between an ordinary two-mode laser and an EP comb.
An ordinary laser operating in the two-mode regime away from degeneracies is a trivial limit cycle with two dominant frequency components and is also rigorously described by Eqs.~\eqref{CM:E}}--{\eqref{PALT:D}. Such a laser features a small $\zeta$ factor, so the second threshold $D_2^{\text{th}}$ from the stability analysis reduces to the SALT threshold (Supplementary Sec.~4), and the intensities of the additional frequency components ($m \neq 0, 1$) are small enough to be neglected.
When $\zeta$ is raised, $D_2^{\text{th}}$ smoothly moves, and the additional frequency components above $D_2^{\text{th}}$ smoothly increase.}

Up to now, we have considered MB equations with an ensemble of two-level atoms.
In Supplementary Sec.~5, we generalize the MB equations to account for the band structure in semiconductor gain media and correspondingly generalize the PALT formalism, which does not change the conclusion on comb formation near an EP.

\vspace{-8pt}
\subsection{EP comb example} \label{sec:EP_comb}
\vspace{-4pt}

\begin{figure*}[t]
\centering
\includegraphics[trim={1.5cm 0 1.5cm 0}]{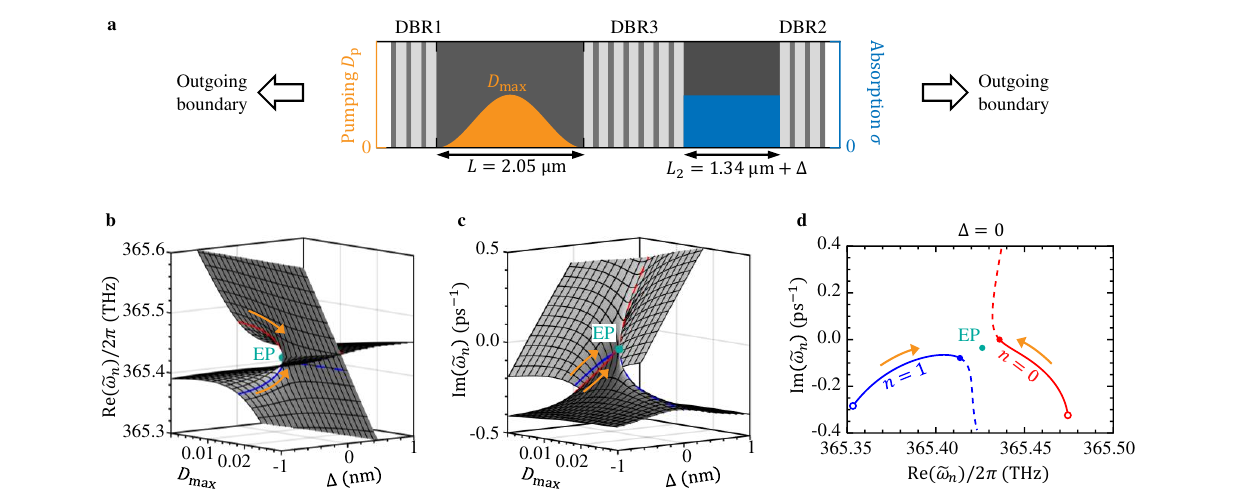}
\vspace{-18pt}
\caption{{\textbf{Exceptional point in a gain-loss coupled cavity}.
\textbf{a} A coupled 1D cavity separated by a distributed Bragg reflector (DBR), with gain in the left cavity and absorption in the right cavity.
Gray-scale colors indicate the cold-cavity permittivity profile $\varepsilon_c(x)$. Orange and blue shades show the gain and absorption profiles $D_\text{p}(x)$ and $\sigma(x)$, respectively. 
\textbf{b}--\textbf{c} The two relevant eigenvalues, $\tilde{\omega}_0$ and $\tilde{\omega}_1$, of the linear operator $\hat{O}\left({\omega}\right)$ in Eq.~\eqref{E0} with a linear gain $D_0(x)= D_\text{p}(x)$, as a function of the pumping strength $D_\text{max}$ and the length of the passive cavity, $L_2=1340~\rm nm + \Delta$. 
The two eigenvalues meet at an EP (green circle).
The absorption in the passive cavity is $\sigma/\varepsilon_0 = 4.9~\rm ps ^{-1}$. The red and blue curves indicate $\tilde{\omega}_0$ and $\tilde{\omega}_1$ with $\Delta=0$. 
\textbf{d} Eigenvalue trajectories on the complex-frequency plane for $\Delta = 0$, with the orange arrows indicating the directions of increasing $D_\text{max}$. Open circles indicate $D_\text{max}=0$, and filled red and blue circles indicate the first lasing threshold $D_\text{max}=D^{\text{th}}_1$ where $\tilde{\omega}_0$ reaches the real-frequency axis.
Dashed lines show the would-be above-threshold trajectories in the absence of gain saturation.}
}
\vspace{-2pt}
\label{fig:2}
\end{figure*}
We now use explicit full-wave examples for illustration.
We adopt a parity-time-symmetric-like configuration~\cite{feng2017non, el2018non, PhysRevLett.122.093901}, where a gain cavity is coupled to a passive cavity with material loss (Fig.~\ref{fig:2}\textbf{a}).
Supplementary Sec.~6 lists the system parameters.
The coupling and the gain-loss contrast are ingredients for an EP~\cite{2011_Moiseyev_book,Heiss_2012,feng2017non, el2018non, doi:10.1126/science.aar7709}.
Distributed Bragg reflectors (DBRs) are used to enclose the two cavities and to separate them. 
The gain cavity consists of AlGaAs 
(refractive index $\sqrt{\varepsilon_c}=3.4$~\cite{aspnes1986optical}, 
gain center $\tilde{\omega}_{ba} = 2\pi c/\omega_{ba} = 820$ nm,
gain width $\gamma_\perp=10^{13}$ s$^{-1}$,
and relaxation rate $\gamma_\parallel = 10^9$ s$^{-1}$)~\cite{YAO1995246}. 
The PALT formalism applies to any pumping profile $D_\text{p}(\mathbf{r})$; to improve the accuracy of the slow finite-difference time-domain (FDTD) simulations that we perform for validation, here we adopt a smooth profile $D_\text{p}(x)=0.5D_\text{max}[1-\text{cos}(2\pi x/L)]$.
The other cavity consists of passive GaAs ($\sqrt{\varepsilon_c}=3.67$)~\cite{aspnes1986optical} with a material absorption characterized by a conductivity $\sigma$.
The system is homogeneous in the transverse directions ($y$ and $z$), so it reduces to a 1D problem with $\mathbf{E}_m(\mathbf{r}) = E_m(x)\hat{z}$.

{In Fig.~\ref{fig:2}\textbf{b--c}, we show the two eigen frequencies $\{\tilde{\omega}_0, \tilde{\omega}_1\}$ of the linear operator $\hat{O}\left({\omega}\right)$ of Eq.~\eqref{E0} as a function of the pumping strength $D_\text{max}$ and the length of the passive cavity, $L_2=1340~\rm nm + \Delta$.
To illustrate the presence of an EP, in this figure (and this figure only) we adopt a linear gain $D_0(x)= D_\text{p}(x)$ with no saturation,
yielding two Riemann sheets that meet at an EP
at $D_\text{max}=0.0126, \Delta=0.01~\text{nm}$, 
$\tilde{\omega}_0 = \tilde{\omega}_1 = \tilde{\omega}_{\rm EP} = 2\pi \times 365.43~\text{THz} -i0.0356~\text{ps}^{-1}$ (green circle).}

{Next, we fix the length of the passive cavity at $L_2=1340$~nm ($\Delta=0$), for which the pump dependence of the two eigenvalues is shown by the red ($n=0$) and blue ($n=1$) curves in Fig.~\ref{fig:2}\textbf{b--d}.
At pumping strength $D_\text{max} = D_1^{\text{th}} = 0.0124$ (red and blue filled circles in Fig.~\ref{fig:2}\textbf{d}), $\tilde{\omega}_0$ reaches the real-frequency axis, and $E_0(x) \propto \psi_0(x)$ turns on as the first lasing mode;
the Petermann factor there is $K_0 \equiv |\langle \varepsilon_c |{\psi}_0|^2\rangle/\langle \varepsilon_c {\psi}_0^2\rangle|^2 = 28$.}

{Above the first threshold ($D_1^{\text{th}} < D_\text{max} < D_2^{\text{th}}$), the red and blue dashed lines in Fig.~\ref{fig:2}\textbf{b--d} show the would-be eigenvalue trajectories with a hypothetical linear gain, in which case the system enters a PT-broken phase where one mode is localized in the pumped cavity and the other mode is localized in the lossy cavity.
Gain saturation, however, clamps the saturated gain at the same level as the overall loss, which fixes the two nonlinearity-frozen eigenvalues $\{\tilde{\omega}_0, \tilde{\omega}_1\}$ near where they are at $D_1^{\text{th}}$ (red and blue filled circles in Fig.~\ref{fig:2}\textbf{d}), and this single-mode laser stays close to a nonlinear EP without entering the PT-broken phase.
}


\begin{figure*}[t]
\centering
\includegraphics[trim={3.1cm 0 3.1cm 0}]{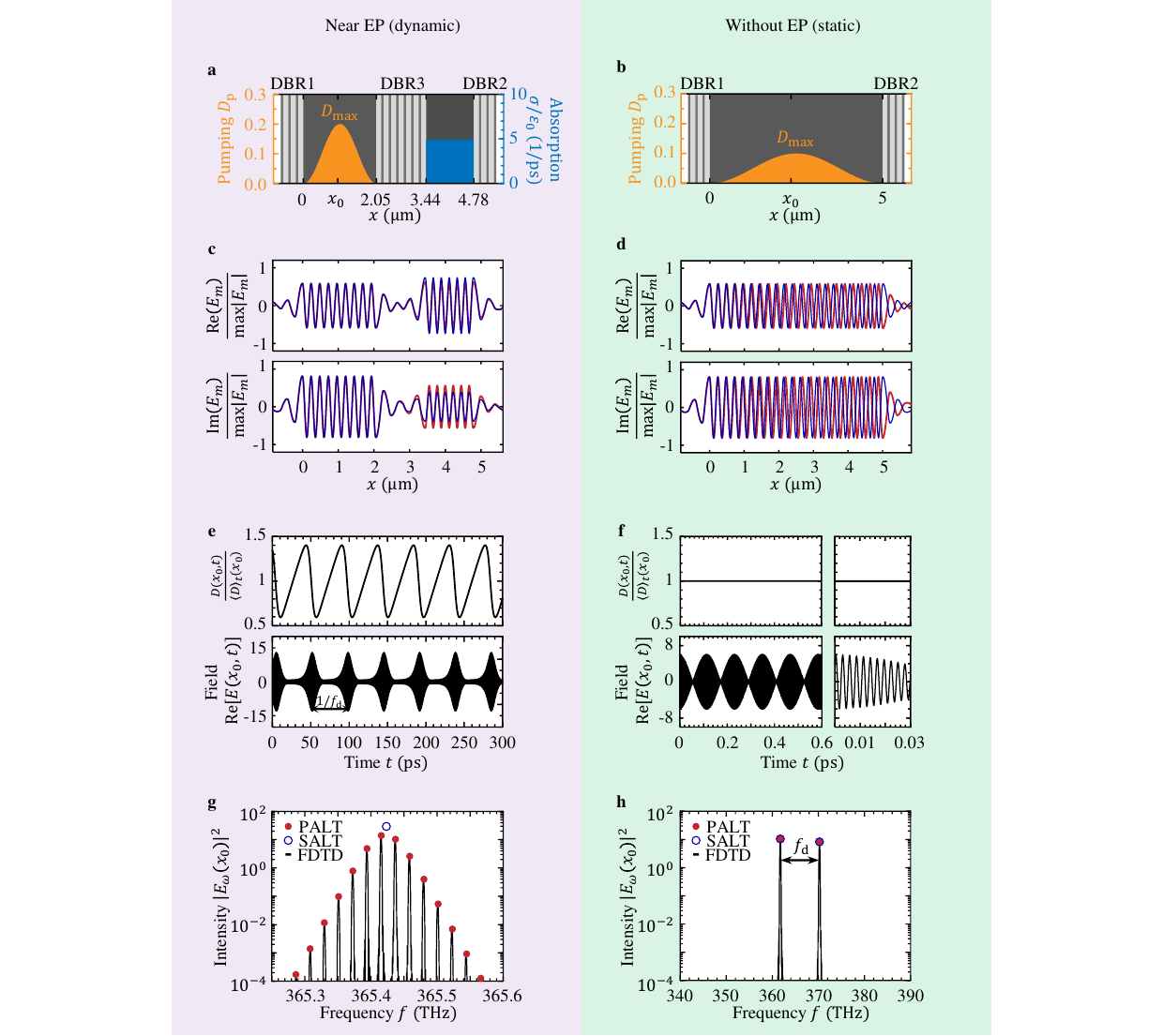}
\vspace{-6pt}
\caption{\textbf{Laser behavior above the second threshold $D_2^{\text{th}}$ near and away from an EP.} 
\textbf{a} The EP laser in Fig.~\ref{fig:2}\textbf{a} with $\Delta = 0$, $D_\text{max} = 0.2$, $\sigma/\varepsilon_0 = 8.1~\rm ps ^{-1}$.
\textbf{b} An ordinary AlGaAs laser cavity with DBR partial mirrors. 
\textbf{c}--\textbf{d} Spatial profiles $E_0$ (red) and $E_1$ (blue) at $\omega_0$ and $\omega_1$, from the full-wave PALT solution.
The two profiles are orthogonal in the ordinary laser but almost identical in the EP laser.
\textbf{e}--\textbf{f} Dynamics of the population inversion $D(x_0,t)$ and electrical field $E(x_0,t)$ at the location $x_0$ shown in \textbf{a}--\textbf{b}. $\langle D\rangle_t$ denotes the inversion averaged over time. \textbf{g}--\textbf{h} The intensity spectrum, comparing the PALT solution to the existing ``steady-state \textit{ab~initio} laser theory'' (SALT) 
and to FDTD simulations of the Maxwell--Bloch equations.
}
\vspace{-4pt}
\label{fig:3}
\end{figure*}  

\begin{figure*}[t]
\centering
\vspace{-4pt}
\includegraphics[trim={3cm 0 3cm 0}]{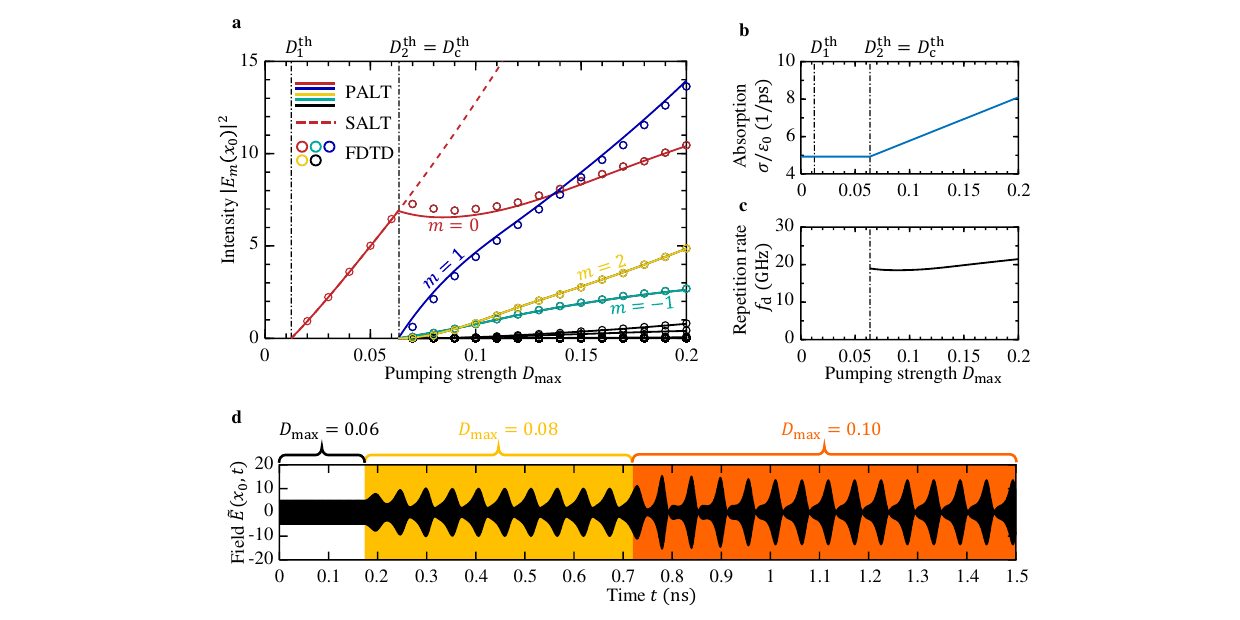}
\vspace{-8pt}
\caption{\textbf{Pump dependence of an EP laser}.
\textbf{a} Input-output curves: lasing intensity at frequencies $f_m = f_0 + m f_\text{d}$ as a function of the pumping strength, comparing the PALT solution (solid lines) to SALT (red dashed line) and FDTD simulations (circles).
\textbf{b}--\textbf{c} The level of absorption $\sigma$ is raised with the pump to keep the frequency difference $f_\text{d}$ small above the comb threshold $D_{\textrm{c}}^{\text{th}} = 0.064$.
\textbf{d} Field evolution in FDTD when the pump is raised across the comb threshold.}
\vspace{-4pt}
\label{fig:4}
\end{figure*}

As the pumping strength reaches above $D_\text{max} > D_2^{\text{th}} = D_{\textrm{c}}^{\text{th}} = 0.064$, 
the population inversion starts to oscillate (Fig.~\ref{fig:3}\textbf{e}), and a frequency comb emerges (Fig.~\ref{fig:3}\textbf{g}). 
Given the proximity to an EP, the repetition rate $|\omega_{\rm d}| \approx 1.35\times10^{11}$ rad/s at $D_{\textrm{c}}^{\text{th}}$ is around 400 times smaller than the FSR of the overall cavity, and the dynamic inversion factor $\zeta \approx 0.26$ is sizeable. 
For a complete characterization, we show in Fig.~\ref{fig:4}\textbf{a} the evolution of the intensity at different frequencies as a function of the pumping strength.
To keep the frequency difference $|\omega_{\rm d}|$ small, we raise the absorption level $\sigma$ when $D_\text{max} > D_{\textrm{c}}^{\text{th}}$ (Fig.~\ref{fig:4}\textbf{b--c}).
{The two center comb lines $\{{\omega}_0, {\omega}_1\}$ lie close to the two near-degenerate active-cavity resonances from SALT in Eq.~\eqref{E0}, $\{{\rm Re}(\tilde{\omega}_0), {\rm Re}(\tilde{\omega}_1)\}$; the remaining comb lines are generated by the nonlinear gain through four-wave mixing and are not lined up with any additional cavity modes (Supplementary Fig.~3).}
The spatial profiles at different frequencies are almost identical (Fig.~\ref{fig:3}\textbf{c}){; they remain comparable to the profiles near the EP in the single-mode regime $D_1^{\text{th}} < D_\text{max} < D_2^{\text{th}}$ where the two modes almost coalesce, without entering the PT-broken phase}.
Supplementary Sec.~6 shows the intensity and gain profiles at all frequencies and their relative phases.

It is commonly assumed~\cite{PhysRevA.43.2446,Ge:08,2014_Esterhazy_PRA,PhysRevA.92.013847,PhysRevA.95.023835} that the stationary-inversion approximation (SIA) of SALT is valid when $|\omega_\text{d}| > \gamma_{\parallel}$, namely when the beat notes oscillate faster than the gain relaxation rate.
However, such a reasoning does not account for the EP-enhanced frequency generation, as captured by the dynamic inversion factor $\zeta$ in Eq.~\eqref{eq:zeta}.
In the present example, $|\omega_{\rm d}| \approx 1.35\times10^{11}$~rad/s is two orders of magnitude greater than $\gamma_\parallel = 10^9$~s$^{-1}$, but SALT (blue circle in Fig.~\ref{fig:3}\textbf{g} and red dashed line in Fig.~\ref{fig:4}\textbf{a}) already fails above the comb threshold.
As described in Sec.~\ref{sec:SIA} and shown in Fig.~\ref{fig:2}\textbf{d},
SALT predicts the laser to stay single-mode because $\psi_1(x)$ has almost the same spatial profile as the lasing mode $E_0(x) \propto \psi_0(x)$ near an EP, so it experiences the same gain clamping as $E_0(x)$ and cannot turn on; this would indeed be the laser behavior when the system is near an EP but not close enough.
In the present example, given the very close proximity to an EP and the resulting large dynamic inversion factor $\zeta \approx 0.26$, 
what actually turns on at the comb threshold $D_2^{\text{th}} = D_{\textrm{c}}^{\text{th}}$ is not an isolated resonance ${\psi}_1$ of the operator $\hat{O}\left({\omega}\right)$ in Eq.~\eqref{E0} but the multi-spectral multi-modal perturbation ${E}_1(x) e^{-i\omega_1 t} + {E}_{-1}(x) e^{-i\omega_{-1} t}$ described in Sec.~\ref{sec:SIA},
which is a superposition of multiple resonances 
and
can amplify by additionally utilizing the dynamic gain $D_{\pm1}(x)e^{\mp i\omega_\text{d} t}$ of Eq.~\eqref{CF}. 

As a comparison to the near-EP laser above, we also consider an ordinary single-cavity laser (Fig.~\ref{fig:3}\textbf{b}) sandwiched between two DBR partial mirrors, operating in the two-mode regime. The active cavity has the same AlGaAs gain. At pumping strength $D_\text{max} > D_2^{\text{th}} = 0.033$, two modes that differ by one longitudinal order lase (Fig.~\ref{fig:3}\textbf{d}) and produce a sinusoidal beating pattern (Fig.~\ref{fig:3}\textbf{f}).
Here, the population inversion is static (Fig.~\ref{fig:3}\textbf{f}), and only two peaks appear in the spectrum (Fig.~\ref{fig:3}\textbf{h}).
There is no EP nearby in the parameter space.
The frequency separation $|\omega_{\rm d}| \approx \pi c/\sqrt{\varepsilon_c}L \approx 5.4 \times 10^{13}$ rad/s equals the free spectral range (FSR) of the cavity and is over four orders of magnitude greater than $\gamma_\parallel$, leading to a negligible dynamic inversion factor $\zeta \approx 3 \times 10^{-13}$.
The Petermann factor is $K=1.0$ here; the gain only balances the radiation loss and does not introduce mode non-orthogonality.
For such a two-mode laser away from degeneracies, PALT reduces to SALT (blue circles in Fig.~\ref{fig:3}\textbf{h}).

EPs feature a boosted sensitivity~\cite{doi:10.1038/nature23281, hodaei2017enhanced, hokmabadi2019non, lai2019observation, kononchuk2022exceptional, 2023_Suntharalingam_ncomms},
which also amplifies the numerical error, requiring an unusually high precision when solving Eqs.~\eqref{PALT:E}--\eqref{PALT:D}.
We find a finite-difference discretization~\cite{2014_Esterhazy_PRA} and the threshold constant-flux basis~\cite{PhysRevA.82.063824} to both require an impractically large basis to reach a satisfactory accuracy near an EP.
To improve the numerical efficiency, here we develop a volume-integral formalism that employs accurate semi-analytic Green's function of the passive system  to solve Eqs.~\eqref{PALT:E}--\eqref{PALT:D} (Supplementary Sec.~7).

To validate our prediction and to verify the stability of the single-mode and the comb solutions, we additionally carry out direct integration of the MB equations, Eqs.~\eqref{MB:D}--\eqref{MB:E}, using FDTD, where we evolve the system until all transient behaviors settle away (Supplementary Sec.~8).
The time-consuming FDTD simulations agree quantitatively with all of the PALT predictions (Figs.~\ref{fig:3}--\ref{fig:4}).
Figure~\ref{fig:4}\textbf{d} shows the field evolution in FDTD when the pump is raised across the comb threshold.

\begin{figure}[t]
\centering
\vspace{-6pt}
\includegraphics{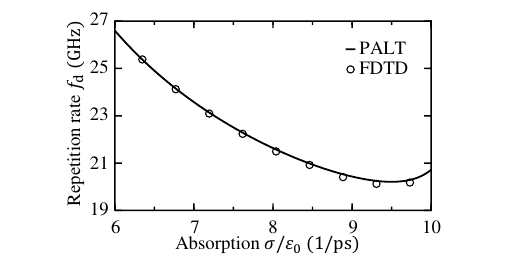}
\vspace{-20pt}
\caption{\textbf{Tunability of the EP comb}. 
Repetition rate of the preceding EP comb when the level of absorption is tuned while the pumping strength is fixed at $D_\text{max}=0.2$.}
\vspace{-4pt}
\label{fig:5}
\end{figure}

Since the EP comb repetition rate $f_\text{d} = |\omega_\text{d}|/(2\pi)$ is not tied to the cavity FSR, we can adjust it freely, for example by tuning the material absorption as shown in Fig.~\ref{fig:5}.
This is not possible with mode-locked combs, Kerr combs, and quantum cascade laser combs.

In the preceding example, we bring the laser close to an EP.
Supplementary Sec.~9 shows that the behavior is the same when we tune the system parameters with a higher precision such that the system has an almost exact EP above the first threshold, $D_{\textrm{EP}} > D_1^{\text{th}}$.
With increasing pump (while fixing the other system parameters), such a laser reaches the comb threshold $D_{\textrm{c}}^{\text{th}}$ and develops into a stable EP comb soon after $D_1^{\text{th}}$.
The exact-EP single-mode lasing state is unreachable as it lies at a higher pump (namely, $D_{\textrm{EP}} > D_{\textrm{c}}^{\text{th}} \gtrsim D_1^{\text{th}}$) and is unstable.

\vspace{-10pt}
\section{Discussion} \label{sec:discussion}
\vspace{-8pt}

In this work, we answer the question of what happens to a laser close to an EP.
Based on the full-wave MB equations, we show that the spectral degeneracy and the spatial coalescence of modes near an EP work with the nonlinearity of the gain medium to induce oscillations in the population inversion, resulting in an ``EP comb.'' The EP comb features a continuously tunable repetition rate, an ultra-compact cavity size, and self-starting operation with no need for an external modulator or continuous-wave laser.
The PALT formalism fully describes both the spatial complexity and the temporal dynamics of such a limit-cycle laser state, overcoming the stationary-inversion limitation of SALT.
This EP comb phenomenon uniquely bridges the subjects of non-Hermitian photonics, laser physics, nonlinear dynamics, and frequency combs.

{As EP sensors are more sensitive closer to an EP}~\cite{doi:10.1038/nature23281, hodaei2017enhanced, hokmabadi2019non, lai2019observation, kononchuk2022exceptional, 2023_Suntharalingam_ncomms},
{it may be desirable to operate such a sensor as close to an EP as possible. This work shows that when an EP laser is brought sufficiently close to an EP, it necessarily develops into a comb above a pump threshold. In such a comb regime, the optimal sensing scheme and the parametric dependence are nontrivial and can be the subject of a future study.}

Existing realizations of EP lasers had mode spacing above 100 GHz; given the $\omega_{\rm d}^{-4}$ scaling of the dynamic inversion factor $\zeta$ in Eq.~\eqref{eq:zeta}, the $\zeta$'s there were too small to induce the multi-spectral multi-modal instability responsible for comb formation, so those lasers exhibited static single-mode behavior.
Further reduction of the mode spacing requires finer tuning but is possible.
In fact, the self-pulsation 
observed in an InAs-quantum-dot Fano laser~\cite{2016_Yu_nphoton} 
may have been an EP comb {since that system has the features of an EP comb (self-starting comb formation in a compact microcavity) and all the EP ingredients: two modes with similar frequencies (one from a line-defect waveguide and one from a nanocavity), near-field coupling between the two modes, and differential gain (as only the waveguide is pumped).}

The repetition rate $|\omega_{\rm d}|$ of the EP comb 
{is determined by the stability eigenvalue problem (Supplementary Sec.~4) at the threshold $D_{\textrm{c}}^{\text{th}}$ and by solving the nonlinear Eqs.~\eqref{PALT:E}}--{\eqref{PALT:D} self-consistently above $D_{\textrm{c}}^{\text{th}}$.
While it is hard to extract insights from these complex equations, empirically we found the distance between the two linear SALT eigenvalues of Eq.~\eqref{E0} to provide a crude approximation, $|\omega_{\rm d}| \approx |\tilde{\omega}_1 - \tilde{\omega}_0|$, near $D_{\textrm{c}}^{\text{th}}$.
Close to an EP, the two linear eigenvalues are sensitive to all parameters of the system, so $\omega_{\rm d}$} can be tuned by changing the absorption, coupling strength, refractive index, {\it etc}. The repetition rate can also be reduced by considering larger cavities.
The minimal repetition rate is limited by the laser linewidth, which can be reduced with standard methods.

The limit of the linewidth itself would be an interesting subject for future investigations.
The divergent Petermann factor is known to broaden the linewidth~\cite{1979_Petermann_JQE, 1989_Siegman_PRA_1, 2020_Wang_ncomms, 2022_Smith_PRA, 2018_Zhang_nphoton}.
With an EP comb, we expect even richer noise properties since the dynamic population inversion can modify the spontaneous emission beyond noise models that
assume a linear gain~\cite{2008_Lee_PRA, PhysRevLett.123.180501} or a stationary inversion \cite{1986_Henry_JLT, 1989_Siegman_PRA_1, Pick:19, 2022_Smith_PRA, PhysRevLett.130.266901}. 
Additionally, the relation between noise and atomic populations is commonly derived at a local thermal equilibrium~\cite{1986_Henry_JLT, 1989_Siegman_PRA_1}, but such an equilibrium is no longer reached when the inversion fluctuates faster than the spontaneous emission rate. 

An EP comb provides a doorway to other nonlinear dynamics phenomena such as
bistability, period doubling, and chaos.
Future work can study the stability of the EP comb, its bifurcation properties, and the transition to other dynamic regimes.
The comb spectrum may be further analyzed and optimized.
The PALT formalism can also describe lasers near Hermitian degeneracies due to symmetry, going beyond perturbation theory~\cite{Ge:08} and stability analysis~\cite{PhysRevA.92.013847,PhysRevA.95.023835}.
We expect even richer behaviors near higher-order EPs and in spatially complex systems such as random lasers and chaotic-cavity lasers.
\vspace{-10pt}
\section{Methods} \label{sec:Method}
\vspace{-4pt}
\subsection{Derivation of PALT} 
\vspace{-8pt}
We show that the PALT ansatz in Eqs.~(\ref{CM:E}--\ref{CM:D}) 
forms an exact solution of the MB equations.
In doing so, we also derive Eqs.~(\ref{PALT:E}--\ref{PALT:D}).

We substitute Eqs.~\eqref{CM:E}--\eqref{CM:D} into the MB equations, Eqs.~\eqref{MB:D}--\eqref{MB:E}, and match terms with the same time dependence. 
Solving Eq.~\eqref{MB:D}, we get
\begin{multline}
D_m = D_\text{p}\delta_{m} + \\
\frac{1}{2}\frac{\gamma_\parallel}{m\omega_d+i\gamma_\parallel}\sum_{n=-\infty}^{+\infty}\bigg(\mathbf{E}_{-m+n}^*\cdot \mathbf{\uptheta}^*P_n -\mathbf{E}_{m-n}\cdot \mathbf{\uptheta}P^*_{-n}\bigg), \label{Dm}
\end{multline}
where $\delta_m$ is the Kronecker delta with $\delta_0 = 1$ and $\delta_{m\neq0} = 0$.
From Eq.~\eqref{MB:P}, we get
\begin{align}
P_m= \Gamma_m\sum_{n=-\infty}^{+\infty}D_n\mathbf{E}_{m-n}\cdot\mathbf{\uptheta}, \label{Pm}
\end{align}
where $\Gamma_m= \Gamma(\omega_m) = \gamma_\perp/(\omega_m-\omega_{ba}+i\gamma_\perp)$, and $\omega_m = m\omega_\text{d}+\omega_0$. 
From Eq.~\eqref{MB:E}, we get
\begin{equation}
-\nabla\times\nabla\times \mathbf{E}_{m} + \frac{\omega_{m}^2}{c^2} \left(\varepsilon_c+\frac{i\sigma}{\omega_{m}\varepsilon_0}\right)\mathbf{E}_{m} = -\frac{\omega_{m}^2}{c^2} P_m\hat{\mathbf{\uptheta}}^* \label{PALT:E-P}.
\end{equation}
This confirms that all of the MB equations, Eqs.~\eqref{MB:D}--\eqref{MB:E}, are satisfied with no approximation.
Substituting Eq.~\eqref{Pm} into Eq.~(\eqref{PALT:E-P}), we get Eq.~(\ref{PALT:E}), where we have applied the commutativity of convolution, 
\begin{equation}
\sum_{n=-\infty}^{+\infty}D_n\mathbf{E}_{m-n}\cdot\mathbf{\uptheta} = \sum_{n=-\infty}^{+\infty}D_{m-n}\mathbf{E}_{n}\cdot\mathbf{\uptheta}.
\end{equation}

To eliminate the gain-induced polarization, we first recognize that since $D(\mathbf{r},t)$ is real-valued, its Fourier components have to be symmetric, $D_n^* =D_{-n}$. With this fact, we take the complex conjugate of Eq.~\eqref{Pm} and then replace the dummy variable $n$ by $-n$,
\begin{equation}
\begin{split}
P_{-m}^*&= \Gamma_{-m}^*\sum_{n=-\infty}^{+\infty}D_{n}^*\mathbf{E}_{-m-n}^*\cdot\mathbf{\uptheta} \\
&= \Gamma_{-m}^*\sum_{n=-\infty}^{+\infty}D_{-n}\mathbf{E}_{-m-n}^*\cdot\mathbf{\uptheta} \\
&=  \Gamma_{-m}^*\sum_{n=-\infty}^{+\infty}D_{n}\mathbf{E}_{-m+n}^*\cdot\mathbf{\uptheta}. \label{P-m}
\end{split}
\end{equation} 

Eqs.~\eqref{Dm}, \eqref{Pm}, \eqref{P-m} can be summarized in matrix form as
\begin{align}
\bar{D}&=D_\text{p}\bar{\delta}+\frac{1}{2}\bar{\bar{\Gamma}}_\parallel(\bar{\bar{E}}^\dag\bar{P}-\bar{\bar{E}}\bar{P}^*_-), \label{vec:D} \\
\bar{P}&= \bar{\bar{\Gamma}}_{+} \bar{\bar{E}}\bar{D}, \label{vec:P} \\
\bar{P}^*_-&=\bar{\bar{\Gamma}}_{-}^\dag\bar{\bar{E}}^\dag\bar{D} \label{vec:P*}, 
\end{align}
where $\dag$ denotes matrix conjugate transpose, with
\begin{itemize}
\item Column vectors: $(\bar{P})_m=P_m$, $(\bar{P}^*_-)_m=P_{-m}^*$, $(\bar{D})_m=D_m$, and $(\bar{\delta})_m=\delta_{m}$.
\item Matrices: $(\bar{\bar{E}})_{mn}=\mathbf{E}_{m-n}\cdot\hat{\mathbf{\uptheta}}$, $(\bar{\bar{\Gamma}}_{\parallel})_{mn}=\delta_{m-n}\gamma_\parallel/(m\omega_\text{d}+i\gamma_\parallel)$, $(\bar{\bar{\Gamma}}_{\pm})_{mn}=\delta_{m-n}\Gamma_{\pm m} $.
\end{itemize}
Substituting Eqs.~\eqref{vec:P}--\eqref{vec:P*} into Eq.~\eqref{vec:D}, we can solve for $\bar{D}$ to obtain Eq.~\eqref{PALT:D}. 

\subsection{Slow-gain limit}
In the slow-gain limit of $|\omega_d| \gg \gamma_\parallel$, all entries of the diagonal matrix $\bar{\bar{\Gamma}}_\parallel$ are approximately $0$ except $(\bar{\bar{\Gamma}}_\parallel)_{00}= -i$. In this limit, Eq.~\eqref{PALT:D} simplifies to $\bar{D}\approx D_0\bar{\delta} \label{SAI:D}$
with
\begin{equation}
D_0(\mathbf{r})\approx\frac{D_\text{p}(\mathbf{r})}{1+\sum_{m}
|\Gamma_m \mathbf{E}_m(\mathbf{r})\cdot \mathbf{\uptheta}|^2}. \label{SIA:D0}
\end{equation} 
Then, Eq.~\eqref{vec:P} yields
$P_m\approx \Gamma_m D_0 (\mathbf{E}_m\cdot \mathbf{\uptheta})$,
so Eq.~\eqref{PALT:E} becomes
\begin{equation}
\left[ -\nabla\times\nabla\times + \frac{\omega_m^2}{c^2} \left(\varepsilon_c+ \frac{i\sigma}{\omega_{m}\varepsilon_0}+\Gamma_mD_0 \mathbf{\uptheta}^*\mathbf{\uptheta}\cdot\right) \right]\mathbf{E}_{m} \approx0 \label{SIA:E}. 
\end{equation}
These Eqs.~\eqref{SIA:D0}--\eqref{SIA:E} reduce to SALT~\cite{PhysRevA.82.063824,2014_Esterhazy_PRA} in the single-mode or two-mode regime (with two indices, $m=0,1$; $\mathbf{E}_m$ with $m\ne 0$ or $1$ has to be zero unless $\omega_m$ happens to be the resonant frequency of a third lasing mode). 

\subsection{Fast-gain limit}
In the fast-gain limit where the lasing bandwidth is much smaller than both $\gamma_\parallel$ and $\gamma_\perp$, we can show that $\bar{\bar{\Gamma}}_\parallel \approx -i\bar{\bar{I}}$, $\bar{\bar{\Gamma}}_\pm \approx \Gamma_0\bar{\bar{I}}$, and $\bar{\bar{E}}^\dag\bar{\bar{E}} = \bar{\bar{E}}\bar{\bar{E}}^\dag$. In this limit, Eq.~\eqref{PALT:D} simplifies to
\begin{equation}
\bar{D}\approx D_\text{p}(\bar{\bar{I}}+|\Gamma_0|^2\bar{\bar{E}}^\dag \bar{\bar{E}})^{-1}\bar{\delta}, \label{GL:D}
\end{equation} 
where we have applied $\Gamma_0 - \Gamma_0^* = -2i|\Gamma_0|^2$. Eq.~\eqref{GL:D} yields
\begin{equation}
\vspace{-2pt}
\bar{D}+|\Gamma_0|^2\bar{\bar{E}}^\dag \bar{\bar{E}}\bar{D}\approx D_\text{p}\bar{\delta}. \vspace{-2pt}
\label{GL:D1}
\end{equation} 
The entries of the column vector $\bar{\bar{E}}^\dag \bar{\bar{E}}\bar{D}$ are the Fourier components of $|\mathbf{E}(\mathbf{r},t)\cdot \hat{\mathbf{\uptheta}}|^2D(\mathbf{r},t)$. Therefore, if we multiply Eq.~\eqref{GL:D1} to the left with the row vector $[...,e^{2i\omega_\text{d}t}, e^{i\omega_\text{d}t},1,e^{-i\omega_\text{d}t},e^{-2i\omega_\text{d}t},...]$, we obtain the time evolution
$D(\mathbf{r},t)+|\Gamma_0\mathbf{E}(\mathbf{r},t)\cdot \hat{\mathbf{\uptheta}}|^2D(\mathbf{r},t) \approx D_\text{p}(\mathbf{r})$, 
namely
\begin{equation}
\vspace{-2pt}
D(\mathbf{r},t)\approx \frac{D_\text{p}(\mathbf{r})}{1+|\Gamma_0\mathbf{E}(\mathbf{r},t)\cdot \hat{\mathbf{\uptheta}}|^2}. \vspace{-2pt}
\label{GL:Dt}
\end{equation} 
In this fast-gain limit, the instantaneous population inversion is given by the instantaneous intensity at that time.



\vspace{8pt}
\noindent {\bf Acknowledgments}
We thank M.~Khajavikhan, A.~D.~Stone, L.~Ge, S.~G.~Johnson, and M.~Yu for helpful discussions.
This work was supported by the National Science Foundation CAREER award (ECCS-2146021) and the University of Southern California.
A.C.\ acknowledges support from the U.S.\ Department of Energy, Office of Basic Energy Sciences, Division of Materials Sciences and Engineering.
This work was performed, in part, at the Center for Integrated Nanotechnologies, an Office of Science User Facility operated for the U.S. Department of Energy (DOE) Office of Science. Sandia National Laboratories is a multimission laboratory managed and operated by National Technology \& Engineering Solutions of Sandia, LLC, a wholly owned subsidiary of Honeywell International, Inc., for the U.S.\ DOE's National Nuclear Security Administration under contract DE-NA-0003525. The views expressed in the article do not necessarily represent the views of the U.S.\ DOE or the United States Government.

\vspace{3pt}
\noindent {\bf Author contributions}
X.G.~developed the theory on the $\zeta$ factor, stability analysis, PALT, integral equation solver, and performed the SALT and PALT calculations. H.H., assisted by A.C., developed the FDTD code. X.G., H.H., and S.S.~performed the FDTD simulations. H.H.~developed the spectral analysis of the FDTD results. C.W.H.~conceived of the project. C.W.H.~and A.C.~supervised the project. C.W.H.~and X.G.~wrote the paper with inputs from the other coauthors. All authors discussed the results.

\vspace{3pt}
\noindent {\bf Competing interests}
The authors declare no competing interests.

\vspace{3pt}
\noindent {\bf Data availability}
The data of PALT calculation and FDTD simulation results presented in the paper are available on OSF database [\url{https://osf.io/jptza/}].

\vspace{3pt}
\noindent {\bf Code availability}
Codes that reproduce the results in this study, including the PALT integral equation solver, stability eigenvalue solver, and Maxwell--Bloch FDTD simulations, are available on GitHub [\url{https://github.com/complexphoton/PALT}]

\def\bibsection{\section*{\refname}} 
\bibliographystyle{naturemag}

\begin{thebibliography}{}
\expandafter\ifx\csname url\endcsname\relax
  \def\url#1{\texttt{#1}}\fi
\expandafter\ifx\csname urlprefix\endcsname\relax\def\urlprefix{URL }\fi
\providecommand{\bibinfo}[2]{#2}
\providecommand{\eprint}[2][]{\url{#2}}

\end{thebibliography}


\begin{thebibliography}{10}
\expandafter\ifx\csname url\endcsname\relax
  \def\url#1{\texttt{#1}}\fi
\expandafter\ifx\csname urlprefix\endcsname\relax\def\urlprefix{URL }\fi
\providecommand{\bibinfo}[2]{#2}
\providecommand{\eprint}[2][]{\url{#2}}

\bibitem{PhysRevA.82.063824}
\bibinfo{author}{Ge, L.}, \bibinfo{author}{Chong, Y.~D.} \&
  \bibinfo{author}{Stone, A.~D.}
\newblock \bibinfo{title}{Steady-state \textit{ab initio} laser theory:
  Generalizations and analytic results}.
\newblock \emph{\bibinfo{journal}{Phys. Rev. A}} \textbf{\bibinfo{volume}{82}},
  \bibinfo{pages}{063824} (\bibinfo{year}{2010}).

\bibitem{2014_Esterhazy_PRA}
\bibinfo{author}{Esterhazy, S.} \emph{et~al.}
\newblock \bibinfo{title}{Scalable numerical approach for the steady-state
  \textit{ab initio} laser theory}.
\newblock \emph{\bibinfo{journal}{Phys. Rev. A}} \textbf{\bibinfo{volume}{90}},
  \bibinfo{pages}{023816} (\bibinfo{year}{2014}).

\bibitem{2022_Sauvan_OE_review}
\bibinfo{author}{Sauvan, C.}, \bibinfo{author}{Wu, T.},
  \bibinfo{author}{Zarouf, R.}, \bibinfo{author}{Muljarov, E.~A.} \&
  \bibinfo{author}{Lalanne, P.}
\newblock \bibinfo{title}{Normalization, orthogonality, and completeness of
  quasinormal modes of open systems: the case of electromagnetism [invited]}.
\newblock \emph{\bibinfo{journal}{Opt. Express}} \textbf{\bibinfo{volume}{30}},
  \bibinfo{pages}{6846--6885} (\bibinfo{year}{2022}).

\bibitem{balanis2012advanced}
\bibinfo{author}{Balanis, C.~A.}
\newblock \emph{\bibinfo{title}{Advanced engineering electromagnetics,
  {A}ppendix {II}}} (\bibinfo{publisher}{John Wiley \& Sons},
  \bibinfo{year}{2012}).

\bibitem{2017_Pick_OE}
\bibinfo{author}{Pick, A.} \emph{et~al.}
\newblock \bibinfo{title}{General theory of spontaneous emission near
  exceptional points}.
\newblock \emph{\bibinfo{journal}{Opt. Express}} \textbf{\bibinfo{volume}{25}},
  \bibinfo{pages}{12325--12348} (\bibinfo{year}{2017}).

\bibitem{PhysRevA.92.013847}
\bibinfo{author}{Burkhardt, S.}, \bibinfo{author}{Liertzer, M.},
  \bibinfo{author}{Krimer, D.~O.} \& \bibinfo{author}{Rotter, S.}
\newblock \bibinfo{title}{Steady-state \textit{ab initio} laser theory for
  fully or nearly degenerate cavity modes}.
\newblock \emph{\bibinfo{journal}{Phys. Rev. A}} \textbf{\bibinfo{volume}{92}},
  \bibinfo{pages}{013847} (\bibinfo{year}{2015}).

\bibitem{PhysRevA.95.023835}
\bibinfo{author}{Liu, D.} \emph{et~al.}
\newblock \bibinfo{title}{Symmetry, stability, and computation of degenerate
  lasing modes}.
\newblock \emph{\bibinfo{journal}{Phys. Rev. A}} \textbf{\bibinfo{volume}{95}},
  \bibinfo{pages}{023835} (\bibinfo{year}{2017}).

\bibitem{2020_Benzaouia_APL}
\bibinfo{author}{Benzaouia, M.}, \bibinfo{author}{Cerjan, A.} \&
  \bibinfo{author}{Johnson, S.~G.}
\newblock \bibinfo{title}{{Is single-mode lasing possible in an infinite
  periodic system?}}
\newblock \emph{\bibinfo{journal}{Appl. Phys. Lett.}}
  \textbf{\bibinfo{volume}{117}}, \bibinfo{pages}{051102}
  (\bibinfo{year}{2020}).

\bibitem{2022_Benzaouia_APLph}
\bibinfo{author}{Benzaouia, M.}, \bibinfo{author}{Stone, A.~D.} \&
  \bibinfo{author}{Johnson, S.~G.}
\newblock \bibinfo{title}{{Nonlinear exceptional-point lasing with \textit{ab
  initio} Maxwell–Bloch theory}}.
\newblock \emph{\bibinfo{journal}{APL Photonics}} \textbf{\bibinfo{volume}{7}},
  \bibinfo{pages}{121303} (\bibinfo{year}{2022}).

\bibitem{Cerjan:12}
\bibinfo{author}{Cerjan, A.}, \bibinfo{author}{Chong, Y.}, \bibinfo{author}{Ge,
  L.} \& \bibinfo{author}{Stone, A.~D.}
\newblock \bibinfo{title}{Steady-state \textit{ab initio} laser theory for
  {N}-level lasers}.
\newblock \emph{\bibinfo{journal}{Opt. Express}} \textbf{\bibinfo{volume}{20}},
  \bibinfo{pages}{474--488} (\bibinfo{year}{2012}).

\bibitem{chow2013semiconductor}
\bibinfo{author}{Chow, W.~W.} \& \bibinfo{author}{Koch, S.~W.}
\newblock \emph{\bibinfo{title}{Semiconductor-laser fundamentals: physics of
  the gain materials, {C}hapter 2}} (\bibinfo{publisher}{Springer Science \&
  Business Media}, \bibinfo{year}{2013}).

\bibitem{Cerjan:CSALT}
\bibinfo{author}{Cerjan, A.}, \bibinfo{author}{Chong, Y.~D.} \&
  \bibinfo{author}{Stone, A.~D.}
\newblock \bibinfo{title}{Steady-state \textit{ab initio} laser theory for
  complex gain media}.
\newblock \emph{\bibinfo{journal}{Opt. Express}} \textbf{\bibinfo{volume}{23}},
  \bibinfo{pages}{6455--6477} (\bibinfo{year}{2015}).

\bibitem{Cerjan:15}
\bibinfo{author}{Cerjan, A.}, \bibinfo{author}{Pick, A.},
  \bibinfo{author}{Chong, Y.~D.}, \bibinfo{author}{Johnson, S.~G.} \&
  \bibinfo{author}{Stone, A.~D.}
\newblock \bibinfo{title}{Quantitative test of general theories of the
  intrinsic laser linewidth}.
\newblock \emph{\bibinfo{journal}{Opt. Express}} \textbf{\bibinfo{volume}{23}},
  \bibinfo{pages}{28316--28340} (\bibinfo{year}{2015}).

\bibitem{mur1981absorbing}
\bibinfo{author}{Mur, G.}
\newblock \bibinfo{title}{Absorbing boundary conditions for the
  finite-difference approximation of the time-domain electromagnetic-field
  equations}.
\newblock \emph{\bibinfo{journal}{IEEE Trans. Electromagn. Compat.}}
  \textbf{\bibinfo{volume}{EMC-23}}, \bibinfo{pages}{377--382}
  (\bibinfo{year}{1981}).

\bibitem{2005_Frigo_FFTW3}
\bibinfo{author}{Frigo, M.} \& \bibinfo{author}{Johnson, S.}
\newblock \bibinfo{title}{The design and implementation of {FFTW3}}.
\newblock \emph{\bibinfo{journal}{Proc. IEEE}} \textbf{\bibinfo{volume}{93}},
  \bibinfo{pages}{216--231} (\bibinfo{year}{2005}).

\end{thebibliography}


\begin{thebibliography}{10}
\expandafter\ifx\csname url\endcsname\relax
  \def\url#1{\texttt{#1}}\fi
\expandafter\ifx\csname urlprefix\endcsname\relax\def\urlprefix{URL }\fi
\providecommand{\bibinfo}[2]{#2}
\providecommand{\eprint}[2][]{\url{#2}}

\bibitem{2011_Moiseyev_book}
\bibinfo{author}{Moiseyev, N.}
\newblock \emph{\bibinfo{title}{Non-Hermitian Quantum Mechanics}}
  (\bibinfo{publisher}{Cambridge University Press}, \bibinfo{year}{2011}).

\bibitem{Heiss_2012}
\bibinfo{author}{Heiss, W.~D.}
\newblock \bibinfo{title}{The physics of exceptional points}.
\newblock \emph{\bibinfo{journal}{J. Phys. A: Math. Theor.}}
  \textbf{\bibinfo{volume}{45}}, \bibinfo{pages}{444016}
  (\bibinfo{year}{2012}).

\bibitem{feng2017non}
\bibinfo{author}{Feng, L.}, \bibinfo{author}{El-Ganainy, R.} \&
  \bibinfo{author}{Ge, L.}
\newblock \bibinfo{title}{Non-{H}ermitian photonics based on parity--time
  symmetry}.
\newblock \emph{\bibinfo{journal}{Nat. Photon.}} \textbf{\bibinfo{volume}{11}},
  \bibinfo{pages}{752--762} (\bibinfo{year}{2017}).

\bibitem{el2018non}
\bibinfo{author}{El-Ganainy, R.} \emph{et~al.}
\newblock \bibinfo{title}{Non-{H}ermitian physics and {PT} symmetry}.
\newblock \emph{\bibinfo{journal}{Nat. Phys.}} \textbf{\bibinfo{volume}{14}},
  \bibinfo{pages}{11--19} (\bibinfo{year}{2018}).

\bibitem{doi:10.1126/science.aar7709}
\bibinfo{author}{Miri, M.-A.} \& \bibinfo{author}{Alù, A.}
\newblock \bibinfo{title}{Exceptional points in optics and photonics}.
\newblock \emph{\bibinfo{journal}{Science}} \textbf{\bibinfo{volume}{363}},
  \bibinfo{pages}{eaar7709} (\bibinfo{year}{2019}).

\bibitem{PhysRevLett.108.173901}
\bibinfo{author}{Liertzer, M.} \emph{et~al.}
\newblock \bibinfo{title}{Pump-induced exceptional points in lasers}.
\newblock \emph{\bibinfo{journal}{Phys. Rev. Lett.}}
  \textbf{\bibinfo{volume}{108}}, \bibinfo{pages}{173901}
  (\bibinfo{year}{2012}).

\bibitem{doi:10.1126/science.1258004}
\bibinfo{author}{Peng, B.} \emph{et~al.}
\newblock \bibinfo{title}{Loss-induced suppression and revival of lasing}.
\newblock \emph{\bibinfo{journal}{Science}} \textbf{\bibinfo{volume}{346}},
  \bibinfo{pages}{328--332} (\bibinfo{year}{2014}).

\bibitem{doi:10.1126/science.1258479}
\bibinfo{author}{Feng, L.}, \bibinfo{author}{Wong, Z.~J.}, \bibinfo{author}{Ma,
  R.-M.}, \bibinfo{author}{Wang, Y.} \& \bibinfo{author}{Zhang, X.}
\newblock \bibinfo{title}{Single-mode laser by parity-time symmetry breaking}.
\newblock \emph{\bibinfo{journal}{Science}} \textbf{\bibinfo{volume}{346}},
  \bibinfo{pages}{972--975} (\bibinfo{year}{2014}).

\bibitem{doi:10.1126/science.1258480}
\bibinfo{author}{Hodaei, H.}, \bibinfo{author}{Miri, M.-A.},
  \bibinfo{author}{Heinrich, M.}, \bibinfo{author}{Christodoulides, D.~N.} \&
  \bibinfo{author}{Khajavikhan, M.}
\newblock \bibinfo{title}{Parity-time–symmetric microring lasers}.
\newblock \emph{\bibinfo{journal}{Science}} \textbf{\bibinfo{volume}{346}},
  \bibinfo{pages}{975--978} (\bibinfo{year}{2014}).

\bibitem{doi:10.1073/pnas.1603318113}
\bibinfo{author}{Peng, B.} \emph{et~al.}
\newblock \bibinfo{title}{Chiral modes and directional lasing at exceptional
  points}.
\newblock \emph{\bibinfo{journal}{Proc. Natl. Acad. Sci. U.S.A.}}
  \textbf{\bibinfo{volume}{113}}, \bibinfo{pages}{6845--6850}
  (\bibinfo{year}{2016}).

\bibitem{doi:10.1126/science.aaf8533}
\bibinfo{author}{Miao, P.} \emph{et~al.}
\newblock \bibinfo{title}{Orbital angular momentum microlaser}.
\newblock \emph{\bibinfo{journal}{Science}} \textbf{\bibinfo{volume}{353}},
  \bibinfo{pages}{464--467} (\bibinfo{year}{2016}).

\bibitem{doi:10.1126/science.aba8996}
\bibinfo{author}{Zhang, Z.} \emph{et~al.}
\newblock \bibinfo{title}{Tunable topological charge vortex microlaser}.
\newblock \emph{\bibinfo{journal}{Science}} \textbf{\bibinfo{volume}{368}},
  \bibinfo{pages}{760--763} (\bibinfo{year}{2020}).

\bibitem{doi:10.1038/nature23281}
\bibinfo{author}{Chen, W.}, \bibinfo{author}{Kaya~{\"O}zdemir, {\c{S}}.},
  \bibinfo{author}{Zhao, G.}, \bibinfo{author}{Wiersig, J.} \&
  \bibinfo{author}{Yang, L.}
\newblock \bibinfo{title}{Exceptional points enhance sensing in an optical
  microcavity}.
\newblock \emph{\bibinfo{journal}{Nature}} \textbf{\bibinfo{volume}{548}},
  \bibinfo{pages}{192--196} (\bibinfo{year}{2017}).

\bibitem{hodaei2017enhanced}
\bibinfo{author}{Hodaei, H.} \emph{et~al.}
\newblock \bibinfo{title}{Enhanced sensitivity at higher-order exceptional
  points}.
\newblock \emph{\bibinfo{journal}{Nature}} \textbf{\bibinfo{volume}{548}},
  \bibinfo{pages}{187--191} (\bibinfo{year}{2017}).

\bibitem{hokmabadi2019non}
\bibinfo{author}{Hokmabadi, M.~P.}, \bibinfo{author}{Schumer, A.},
  \bibinfo{author}{Christodoulides, D.~N.} \& \bibinfo{author}{Khajavikhan, M.}
\newblock \bibinfo{title}{Non-{H}ermitian ring laser gyroscopes with enhanced
  {S}agnac sensitivity}.
\newblock \emph{\bibinfo{journal}{Nature}} \textbf{\bibinfo{volume}{576}},
  \bibinfo{pages}{70--74} (\bibinfo{year}{2019}).

\bibitem{lai2019observation}
\bibinfo{author}{Lai, Y.-H.}, \bibinfo{author}{Lu, Y.-K.},
  \bibinfo{author}{Suh, M.-G.}, \bibinfo{author}{Yuan, Z.} \&
  \bibinfo{author}{Vahala, K.}
\newblock \bibinfo{title}{Observation of the exceptional-point-enhanced
  {S}agnac effect}.
\newblock \emph{\bibinfo{journal}{Nature}} \textbf{\bibinfo{volume}{576}},
  \bibinfo{pages}{65--69} (\bibinfo{year}{2019}).

\bibitem{kononchuk2022exceptional}
\bibinfo{author}{Kononchuk, R.}, \bibinfo{author}{Cai, J.},
  \bibinfo{author}{Ellis, F.}, \bibinfo{author}{Thevamaran, R.} \&
  \bibinfo{author}{Kottos, T.}
\newblock \bibinfo{title}{Exceptional-point-based accelerometers with enhanced
  signal-to-noise ratio}.
\newblock \emph{\bibinfo{journal}{Nature}} \textbf{\bibinfo{volume}{607}},
  \bibinfo{pages}{697--702} (\bibinfo{year}{2022}).

\bibitem{2023_Suntharalingam_ncomms}
\bibinfo{author}{Suntharalingam, A.},
  \bibinfo{author}{Fern{\'a}ndez-Alc{\'a}zar, L.}, \bibinfo{author}{Kononchuk,
  R.} \& \bibinfo{author}{Kottos, T.}
\newblock \bibinfo{title}{Noise resilient exceptional-point voltmeters enabled
  by oscillation quenching phenomena}.
\newblock \emph{\bibinfo{journal}{Nat. Commun.}} \textbf{\bibinfo{volume}{14}},
  \bibinfo{pages}{5515} (\bibinfo{year}{2023}).

\bibitem{2021_Roy_ncomms}
\bibinfo{author}{Roy, A.}, \bibinfo{author}{Jahani, S.},
  \bibinfo{author}{Langrock, C.}, \bibinfo{author}{Fejer, M.} \&
  \bibinfo{author}{Marandi, A.}
\newblock \bibinfo{title}{Spectral phase transitions in optical parametric
  oscillators}.
\newblock \emph{\bibinfo{journal}{Nat. Commun.}} \textbf{\bibinfo{volume}{12}},
  \bibinfo{pages}{835} (\bibinfo{year}{2021}).

\bibitem{schumer2022topological}
\bibinfo{author}{Schumer, A.} \emph{et~al.}
\newblock \bibinfo{title}{Topological modes in a laser cavity through
  exceptional state transfer}.
\newblock \emph{\bibinfo{journal}{Science}} \textbf{\bibinfo{volume}{375}},
  \bibinfo{pages}{884--888} (\bibinfo{year}{2022}).

\bibitem{PhysRevA.43.2446}
\bibinfo{author}{Fu, H.} \& \bibinfo{author}{Haken, H.}
\newblock \bibinfo{title}{Multifrequency operations in a short-cavity
  standing-wave laser}.
\newblock \emph{\bibinfo{journal}{Phys. Rev. A}} \textbf{\bibinfo{volume}{43}},
  \bibinfo{pages}{2446--2454} (\bibinfo{year}{1991}).

\bibitem{PhysRevA.74.043822}
\bibinfo{author}{T\"ureci, H.~E.}, \bibinfo{author}{Stone, A.~D.} \&
  \bibinfo{author}{Collier, B.}
\newblock \bibinfo{title}{Self-consistent multimode lasing theory for complex
  or random lasing media}.
\newblock \emph{\bibinfo{journal}{Phys. Rev. A}} \textbf{\bibinfo{volume}{74}},
  \bibinfo{pages}{043822} (\bibinfo{year}{2006}).

\bibitem{doi:10.1126/science.1155311}
\bibinfo{author}{Türeci, H.~E.}, \bibinfo{author}{Ge, L.},
  \bibinfo{author}{Rotter, S.} \& \bibinfo{author}{Stone, A.~D.}
\newblock \bibinfo{title}{Strong interactions in multimode random lasers}.
\newblock \emph{\bibinfo{journal}{Science}} \textbf{\bibinfo{volume}{320}},
  \bibinfo{pages}{643--646} (\bibinfo{year}{2008}).

\bibitem{PhysRevA.82.063824}
\bibinfo{author}{Ge, L.}, \bibinfo{author}{Chong, Y.~D.} \&
  \bibinfo{author}{Stone, A.~D.}
\newblock \bibinfo{title}{Steady-state \textit{ab initio} laser theory:
  Generalizations and analytic results}.
\newblock \emph{\bibinfo{journal}{Phys. Rev. A}} \textbf{\bibinfo{volume}{82}},
  \bibinfo{pages}{063824} (\bibinfo{year}{2010}).

\bibitem{2014_Esterhazy_PRA}
\bibinfo{author}{Esterhazy, S.} \emph{et~al.}
\newblock \bibinfo{title}{Scalable numerical approach for the steady-state
  \textit{ab initio} laser theory}.
\newblock \emph{\bibinfo{journal}{Phys. Rev. A}} \textbf{\bibinfo{volume}{90}},
  \bibinfo{pages}{023816} (\bibinfo{year}{2014}).

\bibitem{Ge:08}
\bibinfo{author}{Ge, L.}, \bibinfo{author}{Tandy, R.~J.},
  \bibinfo{author}{Stone, A.~D.} \& \bibinfo{author}{T\"{u}reci, H.~E.}
\newblock \bibinfo{title}{Quantitative verification of \textit{ab initio}
  self-consistent laser theory}.
\newblock \emph{\bibinfo{journal}{Opt. Express}} \textbf{\bibinfo{volume}{16}},
  \bibinfo{pages}{16895--16902} (\bibinfo{year}{2008}).

\bibitem{RevModPhys.75.325}
\bibinfo{author}{Cundiff, S.~T.} \& \bibinfo{author}{Ye, J.}
\newblock \bibinfo{title}{Colloquium: Femtosecond optical frequency combs}.
\newblock \emph{\bibinfo{journal}{Rev. Mod. Phys.}}
  \textbf{\bibinfo{volume}{75}}, \bibinfo{pages}{325--342}
  (\bibinfo{year}{2003}).

\bibitem{2007_DelHaye_Nature}
\bibinfo{author}{Del'Haye, P.} \emph{et~al.}
\newblock \bibinfo{title}{Optical frequency comb generation from a monolithic
  microresonator}.
\newblock \emph{\bibinfo{journal}{Nature}} \textbf{\bibinfo{volume}{450}},
  \bibinfo{pages}{1214--1217} (\bibinfo{year}{2007}).

\bibitem{2018_Kippenberg_Science_review}
\bibinfo{author}{Kippenberg, T.~J.}, \bibinfo{author}{Gaeta, A.~L.},
  \bibinfo{author}{Lipson, M.} \& \bibinfo{author}{Gorodetsky, M.~L.}
\newblock \bibinfo{title}{Dissipative {K}err solitons in optical
  microresonators}.
\newblock \emph{\bibinfo{journal}{Science}} \textbf{\bibinfo{volume}{361}},
  \bibinfo{pages}{eaan8083} (\bibinfo{year}{2018}).

\bibitem{Parriaux:20}
\bibinfo{author}{Parriaux, A.}, \bibinfo{author}{Hammani, K.} \&
  \bibinfo{author}{Millot, G.}
\newblock \bibinfo{title}{Electro-optic frequency combs}.
\newblock \emph{\bibinfo{journal}{Adv. Opt. Photon.}}
  \textbf{\bibinfo{volume}{12}}, \bibinfo{pages}{223--287}
  (\bibinfo{year}{2020}).

\bibitem{hugi2012mid}
\bibinfo{author}{Hugi, A.}, \bibinfo{author}{Villares, G.},
  \bibinfo{author}{Blaser, S.}, \bibinfo{author}{Liu, H.~C.} \&
  \bibinfo{author}{Faist, J.}
\newblock \bibinfo{title}{Mid-infrared frequency comb based on a quantum
  cascade laser}.
\newblock \emph{\bibinfo{journal}{Nature}} \textbf{\bibinfo{volume}{492}},
  \bibinfo{pages}{229--233} (\bibinfo{year}{2012}).

\bibitem{silvestri2023frequency}
\bibinfo{author}{Silvestri, C.}, \bibinfo{author}{Qi, X.},
  \bibinfo{author}{Taimre, T.}, \bibinfo{author}{Bertling, K.} \&
  \bibinfo{author}{Rakić, A.~D.}
\newblock \bibinfo{title}{{Frequency combs in quantum cascade lasers: An
  overview of modeling and experiments}}.
\newblock \emph{\bibinfo{journal}{APL Photonics}} \textbf{\bibinfo{volume}{8}},
  \bibinfo{pages}{020902} (\bibinfo{year}{2023}).

\bibitem{2024_Opacak_Nature}
\bibinfo{author}{Opa{\v c}ak, N.} \emph{et~al.}
\newblock \bibinfo{title}{{N}ozaki--{B}ekki solitons in semiconductor lasers}.
\newblock \emph{\bibinfo{journal}{Nature}} \textbf{\bibinfo{volume}{625}},
  \bibinfo{pages}{685--690} (\bibinfo{year}{2024}).

\bibitem{PhysRevA.81.023822}
\bibinfo{author}{Zaitsev, O.} \& \bibinfo{author}{Deych, L.}
\newblock \bibinfo{title}{Diagrammatic semiclassical laser theory}.
\newblock \emph{\bibinfo{journal}{Phys. Rev. A}} \textbf{\bibinfo{volume}{81}},
  \bibinfo{pages}{023822} (\bibinfo{year}{2010}).

\bibitem{PhysRevA.92.063829}
\bibinfo{author}{Malik, O.}, \bibinfo{author}{Makris, K.~G.} \&
  \bibinfo{author}{T\"ureci, H.~E.}
\newblock \bibinfo{title}{Spectral method for efficient computation of
  time-dependent phenomena in complex lasers}.
\newblock \emph{\bibinfo{journal}{Phys. Rev. A}} \textbf{\bibinfo{volume}{92}},
  \bibinfo{pages}{063829} (\bibinfo{year}{2015}).

\bibitem{2016_Ge_srep}
\bibinfo{author}{Ge, L.} \& \bibinfo{author}{El-Ganainy, R.}
\newblock \bibinfo{title}{Nonlinear modal interactions in parity-time ({PT})
  symmetric lasers}.
\newblock \emph{\bibinfo{journal}{Sci. Rep.}} \textbf{\bibinfo{volume}{6}},
  \bibinfo{pages}{24889} (\bibinfo{year}{2016}).

\bibitem{2017_Teimourpour_srep}
\bibinfo{author}{Teimourpour, M.~H.}, \bibinfo{author}{Khajavikhan, M.},
  \bibinfo{author}{Christodoulides, D.~N.} \& \bibinfo{author}{El-Ganainy, R.}
\newblock \bibinfo{title}{Robustness and mode selectivity in parity-time ({PT})
  symmetric lasers}.
\newblock \emph{\bibinfo{journal}{Sci. Rep.}} \textbf{\bibinfo{volume}{7}},
  \bibinfo{pages}{10756} (\bibinfo{year}{2017}).

\bibitem{2018_Kominis_APL}
\bibinfo{author}{Kominis, Y.}, \bibinfo{author}{Choquette, K.~D.},
  \bibinfo{author}{Bountis, A.} \& \bibinfo{author}{Kovanis, V.}
\newblock \bibinfo{title}{Exceptional points in two dissimilar coupled diode
  lasers}.
\newblock \emph{\bibinfo{journal}{Appl. Phys. Lett.}}
  \textbf{\bibinfo{volume}{113}}, \bibinfo{pages}{081103}
  (\bibinfo{year}{2018}).

\bibitem{Horstman:20}
\bibinfo{author}{Horstman, L.}, \bibinfo{author}{Hsu, N.},
  \bibinfo{author}{Hendrie, J.}, \bibinfo{author}{Smith, D.} \&
  \bibinfo{author}{Diels, J.-C.}
\newblock \bibinfo{title}{Exceptional points and the ring laser gyroscope}.
\newblock \emph{\bibinfo{journal}{Photon. Res.}} \textbf{\bibinfo{volume}{8}},
  \bibinfo{pages}{252--256} (\bibinfo{year}{2020}).

\bibitem{10.1093/nsr/nwac259}
\bibinfo{author}{Bai, K.} \emph{et~al.}
\newblock \bibinfo{title}{{Nonlinearity-enabled higher-order exceptional
  singularities with ultra-enhanced signal-to-noise ratio}}.
\newblock \emph{\bibinfo{journal}{Natl. Sci. Rev.}}
  \textbf{\bibinfo{volume}{10}}, \bibinfo{pages}{nwac259}
  (\bibinfo{year}{2022}).

\bibitem{2023_Drong_PRA}
\bibinfo{author}{Drong, M.} \emph{et~al.}
\newblock \bibinfo{title}{Spin vertical-cavity surface-emitting lasers with
  linear gain anisotropy: Prediction of exceptional points and nontrivial
  dynamical regimes}.
\newblock \emph{\bibinfo{journal}{Phys. Rev. A}}
  \textbf{\bibinfo{volume}{107}}, \bibinfo{pages}{033509}
  (\bibinfo{year}{2023}).

\bibitem{PhysRevLett.130.266901}
\bibinfo{author}{Bai, K.} \emph{et~al.}
\newblock \bibinfo{title}{Nonlinear exceptional points with a complete basis in
  dynamics}.
\newblock \emph{\bibinfo{journal}{Phys. Rev. Lett.}}
  \textbf{\bibinfo{volume}{130}}, \bibinfo{pages}{266901}
  (\bibinfo{year}{2023}).


\bibitem{ji2023tracking}
\bibinfo{author}{Ji, K.} \emph{et~al.}
\newblock \bibinfo{title}{Tracking exceptional points above the lasing threshold}.
\newblock \emph{\bibinfo{journal}{Nat. Commun.}} \textbf{\bibinfo{volume}{14}},
  \bibinfo{pages}{8304} (\bibinfo{year}{2023}).

\bibitem{2022_Benzaouia_APLph}
\bibinfo{author}{Benzaouia, M.}, \bibinfo{author}{Stone, A.~D.} \&
  \bibinfo{author}{Johnson, S.~G.}
\newblock \bibinfo{title}{{Nonlinear exceptional-point lasing with \textit{ab
  initio} Maxwell–Bloch theory}}.
\newblock \emph{\bibinfo{journal}{APL Photonics}} \textbf{\bibinfo{volume}{7}},
  \bibinfo{pages}{121303} (\bibinfo{year}{2022}).

\bibitem{haken1985laser}
\bibinfo{author}{Haken, H.}
\newblock \emph{\bibinfo{title}{Laser light dynamics, Vol. 2}}
  (\bibinfo{publisher}{North-Holland Amsterdam}, \bibinfo{year}{1985}).

\bibitem{PhysRevA.54.3347}
\bibinfo{author}{Hess, O.} \& \bibinfo{author}{Kuhn, T.}
\newblock \bibinfo{title}{{M}axwell-{B}loch equations for spatially
  inhomogeneous semiconductor lasers. {I}. {T}heoretical formulation}.
\newblock \emph{\bibinfo{journal}{Phys. Rev. A}} \textbf{\bibinfo{volume}{54}},
  \bibinfo{pages}{3347--3359} (\bibinfo{year}{1996}).

\bibitem{2022_Sauvan_OE_review}
\bibinfo{author}{Sauvan, C.}, \bibinfo{author}{Wu, T.},
  \bibinfo{author}{Zarouf, R.}, \bibinfo{author}{Muljarov, E.~A.} \&
  \bibinfo{author}{Lalanne, P.}
\newblock \bibinfo{title}{Normalization, orthogonality, and completeness of
  quasinormal modes of open systems: the case of electromagnetism [invited]}.
\newblock \emph{\bibinfo{journal}{Opt. Express}} \textbf{\bibinfo{volume}{30}},
  \bibinfo{pages}{6846--6885} (\bibinfo{year}{2022}).

\bibitem{2014_El-Ganainy_PRA}
\bibinfo{author}{El-Ganainy, R.}, \bibinfo{author}{Khajavikhan, M.} \&
  \bibinfo{author}{Ge, L.}
\newblock \bibinfo{title}{Exceptional points and lasing self-termination in
  photonic molecules}.
\newblock \emph{\bibinfo{journal}{Phys. Rev. A}} \textbf{\bibinfo{volume}{90}},
  \bibinfo{pages}{013802} (\bibinfo{year}{2014}).

\bibitem{Boyd_book}
\bibinfo{author}{Boyd, R.~W.}
\newblock \emph{\bibinfo{title}{Nonlinear Optics}}
  (\bibinfo{publisher}{Academic Press}, \bibinfo{year}{2020}),
  \bibinfo{edition}{4} edn.

\bibitem{1979_Petermann_JQE}
\bibinfo{author}{Petermann, K.}
\newblock \bibinfo{title}{Calculated spontaneous emission factor for
  double-heterostructure injection lasers with gain-induced waveguiding}.
\newblock \emph{\bibinfo{journal}{IEEE J. Quantum Electron.}}
  \textbf{\bibinfo{volume}{15}}, \bibinfo{pages}{566--570}
  (\bibinfo{year}{1979}).

\bibitem{1989_Siegman_PRA_1}
\bibinfo{author}{Siegman, A.~E.}
\newblock \bibinfo{title}{Excess spontaneous emission in non-{H}ermitian
  optical systems. {I}. {L}aser amplifiers}.
\newblock \emph{\bibinfo{journal}{Phys. Rev. A}} \textbf{\bibinfo{volume}{39}},
  \bibinfo{pages}{1253--1263} (\bibinfo{year}{1989}).

\bibitem{1996_Wenzel_JQE}
\bibinfo{author}{Wenzel, H.}, \bibinfo{author}{Bandelow, U.},
  \bibinfo{author}{Wunsche, H.-J.} \& \bibinfo{author}{Rehberg, J.}
\newblock \bibinfo{title}{Mechanisms of fast self pulsations in two-section
  {DFB} lasers}.
\newblock \emph{\bibinfo{journal}{IEEE J. Quantum Electron.}}
  \textbf{\bibinfo{volume}{32}}, \bibinfo{pages}{69--78}
  (\bibinfo{year}{1996}).

\bibitem{2003_Berry_JMO}
\bibinfo{author}{Berry, M.~V.}
\newblock \bibinfo{title}{Mode degeneracies and the {P}etermann excess-noise
  factor for unstable lasers}.
\newblock \emph{\bibinfo{journal}{J. Mod. Opt.}} \textbf{\bibinfo{volume}{50}},
  \bibinfo{pages}{63--81} (\bibinfo{year}{2003}).

\bibitem{2008_Lee_PRA}
\bibinfo{author}{Lee, S.-Y.} \emph{et~al.}
\newblock \bibinfo{title}{Divergent {P}etermann factor of interacting
  resonances in a stadium-shaped microcavity}.
\newblock \emph{\bibinfo{journal}{Phys. Rev. A}} \textbf{\bibinfo{volume}{78}},
  \bibinfo{pages}{015805} (\bibinfo{year}{2008}).

\bibitem{2017_Pick_OE}
\bibinfo{author}{Pick, A.} \emph{et~al.}
\newblock \bibinfo{title}{General theory of spontaneous emission near
  exceptional points}.
\newblock \emph{\bibinfo{journal}{Opt. Express}} \textbf{\bibinfo{volume}{25}},
  \bibinfo{pages}{12325--12348} (\bibinfo{year}{2017}).

\bibitem{2020_Wang_ncomms}
\bibinfo{author}{Wang, H.}, \bibinfo{author}{Lai, Y.-H.},
  \bibinfo{author}{Yuan, Z.}, \bibinfo{author}{Suh, M.-G.} \&
  \bibinfo{author}{Vahala, K.}
\newblock \bibinfo{title}{Petermann-factor sensitivity limit near an
  exceptional point in a {B}rillouin ring laser gyroscope}.
\newblock \emph{\bibinfo{journal}{Nat. Commun.}} \textbf{\bibinfo{volume}{11}},
  \bibinfo{pages}{1610} (\bibinfo{year}{2020}).

\bibitem{2022_Smith_PRA}
\bibinfo{author}{Smith, D.~D.}, \bibinfo{author}{Chang, H.},
  \bibinfo{author}{Mikhailov, E.} \& \bibinfo{author}{Shahriar, S.~M.}
\newblock \bibinfo{title}{Beyond the {P}etermann limit: {P}rospect of
  increasing sensor precision near exceptional points}.
\newblock \emph{\bibinfo{journal}{Phys. Rev. A}}
  \textbf{\bibinfo{volume}{106}}, \bibinfo{pages}{013520}
  (\bibinfo{year}{2022}).

\bibitem{PhysRevA.92.013847}
\bibinfo{author}{Burkhardt, S.}, \bibinfo{author}{Liertzer, M.},
  \bibinfo{author}{Krimer, D.~O.} \& \bibinfo{author}{Rotter, S.}
\newblock \bibinfo{title}{Steady-state \textit{ab initio} laser theory for
  fully or nearly degenerate cavity modes}.
\newblock \emph{\bibinfo{journal}{Phys. Rev. A}} \textbf{\bibinfo{volume}{92}},
  \bibinfo{pages}{013847} (\bibinfo{year}{2015}).

\bibitem{PhysRevA.95.023835}
\bibinfo{author}{Liu, D.} \emph{et~al.}
\newblock \bibinfo{title}{Symmetry, stability, and computation of degenerate
  lasing modes}.
\newblock \emph{\bibinfo{journal}{Phys. Rev. A}} \textbf{\bibinfo{volume}{95}},
  \bibinfo{pages}{023835} (\bibinfo{year}{2017}).

\bibitem{strogatz2018nonlinear}
\bibinfo{author}{Strogatz, S.~H.}
\newblock \emph{\bibinfo{title}{Nonlinear dynamics and chaos: With applications
  to physics, biology, chemistry, and engineering}} (\bibinfo{publisher}{CRC
  press}, \bibinfo{year}{2015}), \bibinfo{edition}{2} edn.

\bibitem{2023_Gao_CLEO}
\bibinfo{author}{Gao, X.}, \bibinfo{author}{He, H.},
  \bibinfo{author}{Sobolewski, S.} \& \bibinfo{author}{Hsu, C.~W.}
\newblock \bibinfo{title}{\textit{Ab initio} theory for exceptional-point
  lasers and periodic-inversion lasers}.
\newblock In \emph{\bibinfo{booktitle}{2023 Conference on Lasers and
  Electro-Optics (CLEO)}}, \bibinfo{pages}{paper SF2N.8}
  (\bibinfo{year}{2023}).

\bibitem{2023_Gao_NLO}
\bibinfo{author}{Gao, X.}, \bibinfo{author}{He, H.},
  \bibinfo{author}{Sobolewski, S.} \& \bibinfo{author}{Hsu, C.~W.}
\newblock \bibinfo{title}{Frequency comb generation with lasers near
  exceptional points}.
\newblock In \emph{\bibinfo{booktitle}{Optica Nonlinear Optics Topical Meeting
  2023}}, \bibinfo{pages}{paper Tu1A.3} (\bibinfo{year}{2023}).

\bibitem{PhysRevLett.122.093901}
\bibinfo{author}{Sweeney, W.~R.}, \bibinfo{author}{Hsu, C.~W.},
  \bibinfo{author}{Rotter, S.} \& \bibinfo{author}{Stone, A.~D.}
\newblock \bibinfo{title}{Perfectly absorbing exceptional points and chiral
  absorbers}.
\newblock \emph{\bibinfo{journal}{Phys. Rev. Lett.}}
  \textbf{\bibinfo{volume}{122}}, \bibinfo{pages}{093901}
  (\bibinfo{year}{2019}).

\bibitem{aspnes1986optical}
\bibinfo{author}{Aspnes, D.~E.}, \bibinfo{author}{Kelso, S.~M.},
  \bibinfo{author}{Logan, R.~A.} \& \bibinfo{author}{Bhat, R.}
\newblock \bibinfo{title}{{Optical properties of AlGaAs}}.
\newblock \emph{\bibinfo{journal}{J. Appl. Phys.}}
  \textbf{\bibinfo{volume}{60}}, \bibinfo{pages}{754--767}
  (\bibinfo{year}{1986}).

\bibitem{YAO1995246}
\bibinfo{author}{Yao, J.}, \bibinfo{author}{Agrawal, G.~P.},
  \bibinfo{author}{Gallion, P.} \& \bibinfo{author}{Bowden, C.~M.}
\newblock \bibinfo{title}{Semiconductor laser dynamics beyond the rate-equation
  approximation}.
\newblock \emph{\bibinfo{journal}{Opt. Commun.}}
  \textbf{\bibinfo{volume}{119}}, \bibinfo{pages}{246--255}
  (\bibinfo{year}{1995}).

\bibitem{2016_Yu_nphoton}
\bibinfo{author}{Yu, Y.}, \bibinfo{author}{Xue, W.}, \bibinfo{author}{Semenova,
  E.}, \bibinfo{author}{Yvind, K.} \& \bibinfo{author}{Mork, J.}
\newblock \bibinfo{title}{Demonstration of a self-pulsing photonic crystal
  {F}ano laser}.
\newblock \emph{\bibinfo{journal}{Nat. Photon.}} \textbf{\bibinfo{volume}{11}},
  \bibinfo{pages}{81--84} (\bibinfo{year}{2017}).

\bibitem{2018_Zhang_nphoton}
\bibinfo{author}{Zhang, J.} \emph{et~al.}
\newblock \bibinfo{title}{A phonon laser operating at an exceptional point}.
\newblock \emph{\bibinfo{journal}{Nat. Photon.}} \textbf{\bibinfo{volume}{12}},
  \bibinfo{pages}{479--484} (\bibinfo{year}{2018}).

\bibitem{PhysRevLett.123.180501}
\bibinfo{author}{Zhang, M.} \emph{et~al.}
\newblock \bibinfo{title}{Quantum noise theory of exceptional point amplifying
  sensors}.
\newblock \emph{\bibinfo{journal}{Phys. Rev. Lett.}}
  \textbf{\bibinfo{volume}{123}}, \bibinfo{pages}{180501}
  (\bibinfo{year}{2019}).

\bibitem{1986_Henry_JLT}
\bibinfo{author}{Henry, C.}
\newblock \bibinfo{title}{Theory of spontaneous emission noise in open
  resonators and its application to lasers and optical amplifiers}.
\newblock \emph{\bibinfo{journal}{J. Lightwave Technol.}}
  \textbf{\bibinfo{volume}{4}}, \bibinfo{pages}{288--297}
  (\bibinfo{year}{1986}).

\bibitem{Pick:19}
\bibinfo{author}{Pick, A.}, \bibinfo{author}{Cerjan, A.} \&
  \bibinfo{author}{Johnson, S.~G.}
\newblock \bibinfo{title}{\textit{ab initio} theory of quantum fluctuations and
  relaxation oscillations in multimode lasers}.
\newblock \emph{\bibinfo{journal}{J. Opt. Soc. Am. B}}
  \textbf{\bibinfo{volume}{36}}, \bibinfo{pages}{C22--C40}
  (\bibinfo{year}{2019}).




\end{thebibliography}

\end{document}


\title{Supplementary Information for Dynamic gain and frequency comb formation in exceptional-point lasers}

\maketitle

\onecolumngrid

\vspace{-26pt}
\tableofcontents

\thispagestyle{plain} 

\vspace{-2pt}
\section{Maxwell--Bloch equations in 3D} \label{sec:MB}
\vspace{-2pt}
In this section, we derive the 3D Maxwell--Bloch (MB) equations, namely Eqs.~(1)--(3) of the main text, from the evolution equation of the atomic density matrix. 
Here, the active part of the gain medium is modeled by an ensemble of two-level atoms. For each atom, let $\hat{H}_0$ be the atomic Hamiltonian in the absence of external electric fields, with $|a \rangle$ and $|b \rangle$ being the ground state and excited state respectively, at energies $\hbar \omega_a$ and $\hbar \omega_b$. In the presence of an external electrical field $\tilde{\mathbf{E}}$, the total Hamiltonian $\hat{H}$ includes the electric dipole energy $e \hat{\mathbf{r}}  \cdot \tilde{\mathbf{E}}$, 
\begin{CEquation}
\hat{H}=\hat{H}_0 + e \tilde{\mathbf{E}}\cdot \hat{\mathbf{r}}, \label{Hamiltonian}
\end{CEquation} 
where $\hat{\mathbf{r}}$ is the position operator of the two-level atom, with $\langle a|\hat{\mathbf{r}}|a\rangle = \langle b|\hat{\mathbf{r}}|b\rangle = \mathbf{0}$ given its spatial symmetry. 

The ensemble of atoms can be described by a density matrix $\hat{\rho}$, which satisfies
\begin{CEquation}
 \frac{\partial \hat{\rho}}{\partial t} = \frac{1}{i\hbar}[\hat{H},\hat{\rho}] + \text{incoherent terms}, \label{rho}
\end{CEquation}
with the brackets denoting a commutator.
The ``incoherent terms'' here include pumping from the ground state to the excited state at rate $\gamma_{ab}$, spontaneous emission from the excited state to the ground state at rate $\gamma_{ba}$, and dephasing of the off-diagonal elements of the density matrix at rate $\gamma_\perp$.
In the basis of $|a\rangle$ and $|b\rangle$, 
\begin{CEquation}
\begin{bmatrix}
\rho_{aa} & \rho_{ab} \\
\rho_{ba} & \rho_{bb} 
\end{bmatrix}
=
\begin{bmatrix}
\langle a|\hat{\rho}|a\rangle & \langle a|\hat{\rho}|b\rangle \\
\langle b|\hat{\rho}|a\rangle & \langle b|\hat{\rho}|b\rangle \label{two-level expansion}
\end{bmatrix},
\end{CEquation}
the elements of the density matrix then evolve as
\begin{CAlign}
\frac{\partial}{\partial t} \rho_{aa}&=\gamma_{ba}\rho_{bb}-\gamma_{ab}\rho_{aa}-\frac{1}{i\hbar}\tilde{\mathbf{E}}\cdot(\rho_{ba}\mathbf{R}^*-\text{c.c.}), \label{aa}\\
\frac{\partial}{\partial t} \rho_{bb}&=\gamma_{ab}\rho_{aa}-\gamma_{ba}\rho_{bb}+\frac{1}{i\hbar}\tilde{\mathbf{E}}\cdot(\rho_{ba}\mathbf{R}^*-\text{c.c.}), \label{bb}\\
\frac{\partial}{\partial t} \rho_{ba}&=-(i\omega_{ba}+\gamma_\perp)\rho_{ba}+\frac{1}{i\hbar}(\rho_{bb}-\rho_{aa})\tilde{\mathbf{E}}\cdot \mathbf{R} \label{ba},
\end{CAlign}
where 
$\mathbf{R}=-e\langle b|\hat{\mathbf{r}}|a\rangle$ is the atomic dipole moment, $\omega_{ba}=\omega_b-\omega_a$ is the frequency gap, and \text{c.c.} denotes complex conjugate. Let $N$ be the number density of the two-level atoms, so that $N_a = N \rho_{aa}$ and $N_b = N \rho_{bb}$ are the population densities at the ground level and the excited level, respectively.
The expectation of the gain-induced polarization is then $\tilde{\mathbf{P}} = -eN\text{tr}(\hat{\rho}\hat{\mathbf{r}})=\mathbf{P}+\mathbf{P}^*$, where $\mathbf{P}=N\rho_{ba}\mathbf{R}^*$. 
Multiplying Eqs.~\eqref{aa}--\eqref{bb} with $N$ and Eq.~\eqref{ba} with $N\mathbf{R}^*$, we get
\begin{CAlign}
\frac{\partial}{\partial t} N_{a} &=\gamma_{ba}N_{b}-\gamma_{ab}N_{a}-\frac{1}{i\hbar}\tilde{\mathbf{E}}\cdot(\mathbf{P}-\mathbf{P}^*), \label{Na}\\
\frac{\partial}{\partial t} N_{b} &=\gamma_{ab}N_{a}-\gamma_{ba}N_{b}+\frac{1}{i\hbar}\tilde{\mathbf{E}}\cdot(\mathbf{P}-\mathbf{P}^*), \label{Nb}\\
\frac{\partial}{\partial t} {\mathbf{P}} &=-(i\omega_{ba}+\gamma_\perp)\mathbf{P}+\frac{1}{i\hbar}D(\tilde{\mathbf{E}}\cdot \mathbf{R})\mathbf{R}^*, \label{P}
\end{CAlign}
where $D=N_b-N_a$ is the population inversion. Subtracting Eq.~\eqref{Na} from Eq.~\eqref{Nb} yields
\begin{CAlign}
\frac{\partial}{\partial t} D &=-\gamma_\parallel(D-D_\text{p})+\frac{2}{i\hbar}\tilde{\mathbf{E}}\cdot(\mathbf{P} - \mathbf{P}^*), \label{D}
\end{CAlign}
where $\gamma_\parallel = \gamma_{ab}+\gamma_{ba}$ is the decay rate of the inversion, and $D_\text{p} = N(\gamma_{ab}-\gamma_{ba})/(\gamma_{ab}+\gamma_{ba})$ is the net pumping strength. 
Eqs.~\eqref{P}--\eqref{D} together with Maxwell's equations constitute the MB equations in 3D.

For the analytical and numerical analysis, it is more convenient to work with MB equations with dimensionless units.
Let $R=\sqrt{\mathbf{R}\cdot\mathbf{R}^*}$ be the amplitude and $\mathbf{\uptheta}=\mathbf{R}/R$ be the unit vector of the atomic dipole moment. We normalize $D$ and $D_\text{p}$ in the units of $R^2/(\varepsilon_0\hbar\gamma_{\perp})$, $\tilde{\mathbf{E}}$ in the units of $2R/(\hbar\sqrt{\gamma_{\perp}\gamma_\parallel})$, and $\mathbf{P}$ in the units of $2R/(\varepsilon_0\hbar\sqrt{\gamma_{\perp}\gamma_\parallel})$. Then the dimensionless MB equations read
\begin{CAlign}
\frac{\partial}{\partial t} D &=-\gamma_\parallel(D-D_\text{p})-\frac{i\gamma_\parallel}{2}\tilde{\mathbf{E}}\cdot(\mathbf{P} - \mathbf{P}^*), \label{FDTD:D}\\
\frac{\partial}{\partial t} {\mathbf{P}} &=-(i\omega_{ba}+\gamma_\perp)\mathbf{P}-i\gamma_\perp D(\tilde{\mathbf{E}}\cdot \mathbf{\uptheta})\mathbf{\uptheta}^*, \label{FDTD:P}\\
\nabla &\times \tilde{\mathbf{B}} =\frac{1}{c^2} \left(\varepsilon_c\frac{\partial \tilde{\mathbf{E}}}{\partial t}+\frac{\sigma}{\varepsilon_0} \tilde{\mathbf{E}} + \frac{\partial \tilde{\mathbf{P}}}{\partial t}\right),\label{FDTD:B}\\
\nabla &\times \tilde{\mathbf{E}} = -\frac{\partial \tilde{\mathbf{B}}}{\partial t}. \label{FDTD:E}
\end{CAlign}
Our implementation of FDTD simulations (Sec.~\ref{sec:FDTD} below) is based directly on Eqs.~\eqref{FDTD:D}--\eqref{FDTD:E}.

We can write the real-valued field $\tilde{\mathbf{E}}$ as $\tilde{\mathbf{E}}(\mathbf{r},t)=\mathbf{E}(\mathbf{r},t)+\mathbf{E}^*(\mathbf{r},t)$.
The spectrum of the lasing field $\mathbf{E}(\mathbf{r},t)$ is centered near the peak-gain frequency of $\omega_{ba}$ with a bandwidth no greater than the gain bandwidth $\gamma_\perp \ll \omega_{ba}$. 
Meanwhile, the spectrum of $\mathbf{E}^*(\mathbf{r},t)$ is centered near $-\omega_{ba}$.
Similarly for $\mathbf{P}$ and $\mathbf{P}^*$.
Therefore, the $\mathbf{E} \cdot \mathbf{P}$ term and the $\mathbf{E}^* \cdot \mathbf{P}^*$ term in Eq.~\eqref{FDTD:D} will oscillate approximately as $e^{-2i\omega_{ba}t}$ and $e^{2i\omega_{ba}t}$ respectively.
Since $\omega_{ba} \sim 10^{15}$ rad/s is six orders of magnitude greater than $\gamma_\parallel \sim 10^9$ s$^{-1}$ for semiconductor lasers operating at optical frequencies, these oscillations average away before the gain medium can respond, and we can drop the $\mathbf{E} \cdot \mathbf{P}$ and $\mathbf{E}^* \cdot \mathbf{P}^*$ terms in Eq.~\eqref{FDTD:D}.
Similarly the $\mathbf{E}^*$ term in the right-hand side of Eq.~\eqref{FDTD:P} will oscillate near $-\omega_{ba}$ while the gain-induced polarization $\mathbf{P}$ oscillates near $\omega_{ba}$, so we can also drop the $\mathbf{E}^*$ term in the right-hand side of Eq.~\eqref{FDTD:P}.
With this rotating-wave approximation, Eqs.~\eqref{FDTD:D}--\eqref{FDTD:E} become
\begin{CAlign}
\frac{\partial}{\partial t}D &= - \gamma_{\parallel}(D-D_\text{p})-\frac{i\gamma_{\parallel}}{2}(\mathbf{E}^*\cdot\mathbf{P}-\mathbf{E}\cdot\mathbf{P}^*), \label{MB:D}\\
\frac{\partial}{\partial t}\mathbf{P} &= -(i\omega_{ba} + \gamma_{\perp})\mathbf{P}-i\gamma_{\perp}D(\mathbf{E}\cdot\mathbf{\uptheta})\mathbf{\uptheta}^*, \label{MB:P}\\
-\nabla&\times\nabla\times \mathbf{E} - \frac{1}{c^2} \left(\varepsilon_c\frac{\partial^2}{\partial t^2}+\frac{\sigma}{\varepsilon_0} \frac{\partial}{\partial t}\right)\mathbf{E} = \frac{1}{c^2} \frac{\partial^2}{\partial t^2} \mathbf{P} \label{MB:E},
\end{CAlign}
These Eq.~\eqref{MB:D}--\eqref{MB:E} are the MB equations, Eqs.~(1)--(3) of the main text, that we analyze below.

\vspace{-2pt}
\section{Single-mode lasing solution and active-cavity resonance operator}
\vspace{-2pt}
In the subsequent sections, we will build our analysis upon the steady-state solution in the single-mode lasing regime, which features a real-valued lasing frequency $\omega_0$ with~\cite{PhysRevA.82.063824,2014_Esterhazy_PRA}
\begin{CAlign}
D(\mathbf{r},t) &= D_0(\mathbf{r}), \label{eq:D0} \\
\mathbf{P}(\mathbf{r},t) &= P_0(\mathbf{r})\mathbf{\uptheta}^* e^{-i\omega_0 t}, \\
\mathbf{E}(\mathbf{r},t) &= \mathbf{E}_0(\mathbf{r}) e^{-i\omega_0 t}. \label{eq:E0}
\end{CAlign}
We can verify that Eqs.~\eqref{eq:D0}--\eqref{eq:E0} constitute an exact solution of the MB equations, Eqs.~\eqref{MB:D}--\eqref{MB:E}, when
\begin{CAlign}
& \qquad \qquad \qquad D_0(\mathbf{r}) = \frac{D_{\rm p}(\mathbf{r})}{1+|\Gamma_0E_0(\mathbf{r})|^2}, \label{M1:D0}\\
& \qquad \qquad \qquad P_0(\mathbf{r}) = \Gamma_0 D_0(\mathbf{r}) E_0(\mathbf{r}), \label{M1:P0} \\
&\left[-\nabla\times\nabla\times + \frac{\omega_{0}^2}{c^2} \left(\varepsilon_c(\mathbf{r})+\frac{i\sigma(\mathbf{r})}{\omega_{0}\varepsilon_0} + \Gamma_{0}D_0(\mathbf{r}) \mathbf{\uptheta}^*\mathbf{\uptheta}\cdot \right)\right] \mathbf{E}_{0}(\mathbf{r}) = 0, \label{M1:E0}
\end{CAlign}
where $\Gamma_0 =\Gamma(\omega_0)$,
$E_0 \equiv \mathbf{E}_0\cdot \mathbf{\uptheta}$, and
\begin{CEquation}
\Gamma(\omega) \equiv \frac{\gamma_\perp}{\omega-\omega_{ba}+i\gamma_\perp}. \label{eq:Gamma}
\end{CEquation}
Eq.~\eqref{M1:E0} is repeated as Eq.~(4) in the main text.

To facilitate discussions in the next two sections, we introduce the active-cavity wave operator in this single-mode regime
\begin{CEquation}
\hat{O}({\omega})
=-\nabla\times\nabla\times + \frac{\omega^2}{c^2}\varepsilon_{\rm eff}(\mathbf{r},\omega), \label{H}
\end{CEquation}
with an outgoing boundary condition, where the effective permittivity profile is
\begin{CEquation}
\varepsilon_{\rm eff}(\mathbf{r},\omega)
= \varepsilon_c(\mathbf{r})+\tilde{\mathbf{\varepsilon}}(\mathbf{r},\omega),
\qquad
\tilde{\mathbf{\varepsilon}}(\mathbf{r},\omega)=
\frac{i\sigma(\mathbf{r})}{\omega\varepsilon_0} + 
\Gamma(\omega)
D_0(\mathbf{r}) \mathbf{\uptheta}^*\mathbf{\uptheta}\cdot. \label{gain-loss}
\end{CEquation}
We separate out the gain and loss contributions in $\tilde{\mathbf{\varepsilon}}(\mathbf{r},\omega)$, which is generally much smaller than the passive contribution $\varepsilon_c(\mathbf{r})$.
Then, Eq.~\eqref{M1:E0} can be written as
\begin{CEquation}
\hat{O}(\omega_0)\mathbf{E}_0 = 0. \label{M1:E0_O}
\end{CEquation}

Eq.~\eqref{M1:E0_O} is nonlinear in $\mathbf{E}_0(\mathbf{r})$ because the saturated gain $D_0(\mathbf{r})$ in $\tilde{\mathbf{\varepsilon}}(\mathbf{r},\omega)$ of the operator $\hat{O}({\omega})$ depends on $\mathbf{E}_0(\mathbf{r})$.
For a given pump strength, after obtaining $\omega_0$ and $\mathbf{E}_0(\mathbf{r})$ by solving Eq.~\eqref{M1:E0_O} as a nonlinear equation, it is convenient to consider operator $\hat{O}({\omega})$ with a $D_0(\mathbf{r})$ fixed (frozen) by that pre-computed $\mathbf{E}_0(\mathbf{r})$ but with a variable $\omega$.
This ``active cavity wave operator'' $\hat{O}({\omega})$ defines the resonances~\cite{2022_Sauvan_OE_review} $\{\mathbf{\psi}_n(\mathbf{r}) \}$ (namely, eigenmodes with an outgoing boundary condition) of the active cavity at that pump strength,
\begin{CEquation}
\hat{O}(\tilde{\omega}_n) \mathbf{\psi}_n = 0, \label{psi}
\end{CEquation}
where $\tilde{\omega}_n$ is the complex-valued eigen frequency of resonance $\mathbf{\psi}_n$. 
Since the gain in operator $\hat{O}$ is fixed, Eq.~\eqref{psi} is linear in $\mathbf{\psi}_n$.
By construction, the lasing mode $\mathbf{E}_0$ is a resonance of the active cavity with a real-valued eigenvalue of $\tilde{\omega}_n = \omega_0$.

When the pumping strength is below the first lasing threshold $D_1^{\text{th}}$, the steady-state solution is $\mathbf{E}(\mathbf{r},t) = 0$, which trivially satisfies the MB equations, Eqs.~\eqref{MB:D}--\eqref{MB:E}.
Below the first threshold $D_1^{\text{th}}$, we can consider operator $\hat{O}$ with an unsaturated gain (since $\mathbf{E} = 0$), and track its set of eigenvalues $\{\tilde{\omega}_n\}$.
All of the eigenvalues should have a negative imaginary part, corresponding to an exponential decay in time.
The first threshold is reached when one of the eigenvalues $\tilde{\omega}_n$ reaches the real-frequency axis, corresponding to a cavity resonance having enough gain to overcome its radiation loss and absorption loss.
With pumping at and above $D_1^{\text{th}}$, that eigenvalue stays on the real-frequency axis as $\omega_0$, and the corresponding resonance becomes the lasing mode $\mathbf{E}_0$. The amplitude of the lasing mode is determined self-consistently by solving the nonlinear Eq.~\eqref{M1:E0_O}.

In the ``steady-state \textit{ab~initio} laser theory'' (SALT)~\cite{PhysRevA.82.063824,2014_Esterhazy_PRA}, the same recipe is continued above the first lasing threshold $D_1^{\text{th}}$. One would track the eigenvalues of the resonances of the active cavity given $\mathbf{E}_0$ at each pump strength. The next threshold $D_2^{\text{th}}$ is reached when the eigenvalue of another resonance (excluding $\mathbf{E}_0$) of the active cavity reaches the real-frequency axis.
We will show in Sec.~\ref{sec:stability} that such a recipe is rigorous in the limit of $\gamma_\parallel \to 0$, where the gain medium is static.
When $\gamma_\parallel$ is not negligible, the dynamics of the gain medium can modify the stability of the single-mode solution such that the next threshold is no longer simply a resonance of the active cavity reaching the real-frequency axis.

\vspace{-4pt}
\section{Dynamic inversion factor $\zeta$}
\vspace{-4pt}
In Sec.~II of the main text, we consider a perturbation to the single-mode lasing solution, Eqs.~\eqref{eq:D0}--\eqref{M1:E0} above.
Here, we derive the dynamic inversion factor $\zeta$ that quantifies the degree of frequency coupling, producing $\mathbf{E}_{-1}e^{-i\omega_{-1} t}$ from $\mathbf{E}_{1}e^{-i\omega_1 t}$ in the perturbation.
Our starting point is Eq.~(6) of the main text, rewritten below as 
\begin{CEquation}
\begin{split}
\hat{O}({\omega_{-1}})\mathbf{E}_{-1} 
&= -\frac{\omega_{-1}^2}{c^2} \frac{\Gamma_{-1}(\Gamma_0-\Gamma_1^*)}{2}\frac{\gamma_{\parallel}}{i\gamma_{\parallel}-\omega_d}D_0E_{ 0}^2E_{ 1}^*\mathbf{\uptheta}^* \\
&\approx \frac{\omega_{-1}^2}{c^2} \frac{\gamma_{\parallel}}{i\gamma_{\parallel}-\omega_d}D_0E_{ 0}^2E_{ 1}^*\mathbf{\uptheta}^*, \label{E-1}
\end{split}
\end{CEquation}
where $E_m \equiv \mathbf{E}_m\cdot \mathbf{\uptheta}$, $\Gamma_m =\Gamma(\omega_m)$, $\omega_m = \omega_0 + m \omega_d $.
We consider $|\omega_m-\omega_{ba}| \ll \gamma_\perp$, so that $\Gamma_m \approx -i$.

To solve Eq.~\eqref{E-1}, we expand $\mathbf{E}_{-1}$ in the non-orthogonal basis of the active-cavity resonances $\{\mathbf{\psi}_n\}$ in Eq.~\eqref{psi},
 \begin{CEquation}
\mathbf{E}_{-1} = \sum_n \alpha_n \mathbf{\psi}_n. \label{E-1_ep}
\end{CEquation}
Substituting Eq.~\eqref{E-1_ep} into Eq.~\eqref{E-1}, subtracting Eq.~\eqref{psi}, and approximating $\varepsilon_{\rm eff}(\mathbf{r},\omega) \approx \varepsilon_c(\mathbf{r})$, we obtain
\begin{CEquation}
\sum_n\alpha_n(\omega_{-1}^2-\tilde{\omega}_n^2)\varepsilon_c\mathbf{\psi}_n 
\approx \omega_{-1}^2 \frac{\gamma_{\parallel}}{i\gamma_{\parallel}-\omega_d}D_0E_{ 0}^2E_{ 1}^*\mathbf{\uptheta}^*. \label{eq:abc}
\end{CEquation}

To proceed, we want to solve Eq.~\eqref{eq:abc} for the expansion coefficients $\{\alpha_n\}$.
The set of resonances $\{\mathbf{\psi}_n\}$ are not orthogonal but are biorthogonal~\cite{2022_Sauvan_OE_review}, which we can utilize to project out $\alpha_n$ in Eq.~\eqref{eq:abc}.
Consider the Green's vector identity~\cite{balanis2012advanced},
\begin{CEquation}
\mathbf{\psi}_m \cdot (\nabla \times \nabla \times \mathbf{\psi}_n) - \mathbf{\psi}_n \cdot (\nabla \times \nabla \times \mathbf{\psi}_m) = \nabla \cdot [\mathbf\psi_n \times (\nabla \times \mathbf\psi_m) - \mathbf\psi_m \times (\nabla \times \mathbf\psi_n)]. \label{curl}
\end{CEquation}
Using the approximation $\varepsilon_{\rm eff}(\mathbf{r},\omega) \approx \varepsilon_c(\mathbf{r})$ again, from Eq.~\eqref{psi}, we get
$\nabla \times \nabla \times \mathbf{\psi}_n \approx \frac{\tilde{\omega}^2_n}{c^2} \varepsilon_c \mathbf{\psi}_n$,
so
\begin{CEquation}
\mathbf{\psi}_m \cdot (\nabla \times \nabla \times \mathbf{\psi}_n) - \mathbf{\psi}_n \cdot (\nabla \times \nabla \times \mathbf{\psi}_m) \approx \frac{\tilde{\omega}^2_n - \tilde{\omega}^2_m}{c^2} \varepsilon_c \mathbf{\psi}_m \cdot \mathbf{\psi}_n \label{eq:psimn}
\end{CEquation}
Substituting Eq.~\eqref{eq:psimn} into Eq.~\eqref{curl}, taking the volume integration over the laser cavity, and applying the divergence theorem, we get
\begin{CEquation}
\frac{\tilde{\omega}^2_n - \tilde{\omega}^2_m}{c^2} \int_V \varepsilon_c \mathbf{\psi}_m \cdot \mathbf{\psi}_n dr^3 \approx \oiint_{\mathbf{S}} [\mathbf{\psi_n} \times (\nabla \times \mathbf{\psi_m}) - \mathbf{\psi_m} \times (\nabla \times \mathbf{\psi_n})] \cdot d \mathbf{s}, \label{sf}
\end{CEquation}
where $V$ is the volume of the laser cavity and $\mathbf{S}$ is its surface. 
To proceed, we assume that the resonances $\{\mathbf{\psi}_n\}$ of interest have high-enough quality factors that the radiation field on the cavity surface is much weaker than the interior field, and the right-hand side of Eq.~\eqref{sf} is negligible compared to the left-hand-side.
Under such approximation, we obtain the biorthogonality relationship,
\begin{CEquation}
\int_V\varepsilon_c \mathbf{\psi}_m \cdot \mathbf{\psi}_n dr^3 \approx 0 \ \text{when}\ \ \tilde{\omega}_m\ne \tilde{\omega}_n. \label{orth}
\end{CEquation}

Multiplying Eq.~\ref{eq:abc} with $\mathbf{\psi}_n$, integrating over $V$, and using the biorthogonality relationship Eq.~\eqref{orth}, we obtain the coefficient $\alpha_n$ of interest,
\begin{CEquation}
\alpha_n \approx \frac{\omega_{-1}^2}{\omega_{-1}^2-\tilde{\omega}_n^2} \frac{\gamma_{\parallel}}{i\gamma_{\parallel}-\omega_d}\frac{\langle D_0 E_{ 0}^2E_{ 1}^*\mathbf{\psi}_n\cdot\mathbf{\uptheta}^* \rangle}{\langle \varepsilon_c \mathbf{\psi}_n\cdot\mathbf{\psi}_n\rangle}, \label{alpha_n}
\end{CEquation}
where $\langle \cdots \rangle = \int_V \cdots dr^3$ denotes integration over $V$. 
We can see that the coefficient $\alpha_n$ is enhanced when (1) the resonance eigen frequency $\tilde{\omega}_n$ is close to $\omega_{-1}$, (2) the frequency detuning $\omega_d = \omega_1 - \omega_0$ between the perturbation and the lasing frequency is small, (3) the spatial overlap $|\langle D_0 E_{ 0}^2E_{ 1}^*\mathbf{\psi}_n\cdot\mathbf{\uptheta}^* \rangle|^2$ between the excitation source profile and the resonance profile is large, and (4) the Petermann factor $K_n \equiv |\langle \varepsilon_c |\mathbf{\psi_n}|^2\rangle/\langle \varepsilon_c \mathbf{\psi_n}\cdot\mathbf{\psi_n}\rangle|^2$ of the resonance is large.

Among the set $\{\alpha_n\}$, typically one coefficient will be much larger than the other coefficients.
We denote the largest coefficient as $\alpha_\psi$, with $\mathbf{\psi}$ denoting the corresponding resonance and $\omega_\psi$ its complex-valued eigen frequency.
Such a $\omega_\psi$ is typically close to $\omega_{-1}$, so
  \begin{CEquation}
\alpha_\psi \approx \frac{\omega_{-1}}{2(\omega_{-1}-\omega_\psi)} \frac{\gamma_\parallel}{i\gamma_\parallel-\omega_d}\frac{\langle D_0 E_{ 0}^2E_{ 1}^*\mathbf{\psi}\cdot\mathbf{\uptheta}^* \rangle}{\langle \varepsilon_c \mathbf{\psi}\cdot\mathbf{\psi}\rangle}. \label{alpha_psi}
\end{CEquation}

Keeping only the dominant contribution, $\mathbf{E}_{-1} \approx \alpha_\psi \mathbf{\psi}$, we obtain the stationary-inversion factor $\zeta$,
  \begin{CEquation}
\frac{\langle |\mathbf{E}_{-1}|^2 \rangle}{\langle |\mathbf{E}_{1}|^2\rangle}
\approx |\alpha_\psi|^2\frac{\langle |\mathbf{\psi}|^2 \rangle}{\langle |\mathbf{E}_{1}|^2\rangle} 
\approx \frac{\gamma_\parallel^2}{\omega_d^2+\gamma_\parallel^2} 
\frac{\omega_{-1}^2/4}{|\omega_{-1}-\omega_\psi|^2} 
\frac{\langle |\mathbf{\psi}|^2 \rangle}{\langle |\mathbf{E}_{1}|^2\rangle}
\bigg|\frac{\langle D_0 E_{ 0}^2E_{ 1}^*\mathbf{\psi}\cdot\mathbf{\uptheta}^* \rangle}{\langle \varepsilon_c \mathbf{\psi}\cdot\mathbf{\psi}\rangle}\bigg|^2 \equiv \zeta. \label{zeta}
\end{CEquation} 
Taking $\mathbf{\psi}=\mathbf{E}_{0}$ to be the resonance closest to $\omega_{-1}$, approximating $\omega_{-1}^2 \approx \omega_{0}^2$, and taking $\mathbf{\uptheta}$ to be real-valued, we obtain Eq.~(7) of the main text.

When $\omega_{-1}$ is near a degeneracy, there may be two resonances that both contribute significant. 
If it is a Hermitian degeneracy due to symmetry, we can take $\mathbf{\psi}$ to be the superposition of the two near-degenerate modes that maximizes the spatial overlap $|\langle D_0 E_{ 0}^2E_{ 1}^*\mathbf{\psi}_n\cdot\mathbf{\uptheta}^* \rangle|^2$ with the excitation source profile.
If it is a non-Hermitian degeneracy, the two resonances will have similar spatial profiles, so the $\zeta$ factor will be comparable for the two resonances; using either one can give an order-of-magnitude estimate of ${\langle |\mathbf{E}_{-1}|^2 \rangle}/{\langle |\mathbf{E}_{1}|^2\rangle}$.
Because of the non-orthogonality of the basis $\{\mathbf{\psi}_n\}$, projecting onto only one resonance $\mathbf{\psi}$ is not quantitatively accurate.
For example, approaching an EP, the Petermann factor (and the associated projection onto one resonance of the EP pair) can diverge, even though the total projection onto both resonances remains finite due to a cancellation of divergences~\cite{2017_Pick_OE}.
However, the single-resonance projection can already provide an order-of-magnitude estimate. For the system considered in Figs.~2--3 of the main text, we obtain $\zeta \approx 0.26$ from Eq.~\eqref{zeta}, reasonably close to the actual ${\langle |\mathbf{E}_{-1}|^2 \rangle}/{\langle |\mathbf{E}_{1}|^2\rangle} \approx 0.10$ computed from PALT just above the comb threshold $D_2^{\text{th}} = D_\textrm{c}^{\text{th}}$, even though the Petermann factor $K_n = 28$ is already very large.

\vspace{-4pt}
\section{Single-mode stability analysis and the comb threshold} \label{sec:stability}
\vspace{-4pt}

Here, we perform a general stability analysis on the single-mode steady-state lasing solution, Eqs.~\eqref{eq:D0}--\eqref{M1:E0} above.
As the pumping strength is increased above the first threshold $D_1^{\text{th}}$, the next threshold $D_2^{\text{th}}$ will be reached when the single-mode solution becomes unstable.
This second threshold $D_2^{\text{th}}$ is typically when a second mode turns on.
For a laser sufficiently close to an EP that the $\zeta$ factor is not negligible, this threshold $D_2^{\text{th}} = D_\textrm{c}^{\text{th}}$ is where a frequency comb emerges.
In any of these scenarios, the stability can be determined by perturbing Eqs.~\eqref{eq:D0}--\eqref{M1:E0} with a small perturbation~\cite{PhysRevA.92.013847},
\begin{CAlign}
D(\mathbf{r},t) &=D_0(\mathbf{r})+\delta D(\mathbf{r},t), \label{Perp:D} \\
\mathbf{P}(\mathbf{r},t) &= [{P}_0(\mathbf{r})+\delta {P}(\mathbf{r},t)]\mathbf{\uptheta}^* e^{-i\omega_0 t} , \label{Perp:P}\\
\mathbf{E}(\mathbf{r},t) &= [\mathbf{E}_0(\mathbf{r})+\delta\mathbf{E}(\mathbf{r},t)] e^{-i\omega_0 t}, \label{Perp:E}
\end{CAlign}
and analyzing whether the perturbation decays or grows in time.
Plugging Eqs.~\eqref{Perp:D}--\eqref{Perp:E} into Eqs.~\eqref{MB:D}--\eqref{MB:E} and keeping only terms linear to the perturbation (since the perturbation is small), we get
\begin{CAlign}
\frac{\partial}{\partial t} \delta D &= - \gamma_{\parallel}\delta D-\frac{i\gamma_{\parallel}}{2}(\delta E^*{P}_0+E_{0}^*\delta P-\delta E{P}_0^*-E_{0}\delta P^*), \label{Pt:D} \\
\frac{\partial}{\partial t}{\delta P} &= i(\omega_{0}-\omega_{ba} + i \gamma_{\perp})\delta{P}-i\gamma_{\perp}(\delta D E_{ 0} +D_0\delta E_{} ),  \label{Pt:P}\\
-\nabla\times\nabla\times \delta\mathbf{E} &- \frac{1}{c^2} \left[\varepsilon_c\left(\frac{\partial}{\partial t}-i\omega_0\right)^2+\frac{\sigma}{\varepsilon_0} \left(\frac{\partial}{\partial t}-i\omega_0\right) \right]\delta\mathbf{E} = \frac{1}{c^2} \left(\frac{\partial}{\partial t}-i\omega_0\right)^2 \delta{P} \mathbf{\uptheta}^*,  \label{Pt:E}
\end{CAlign}
where $\delta E = \delta\mathbf{E}\cdot\mathbf{\uptheta}$. 
The zeroth-order terms reduce to Eq.~\eqref{M1:D0}--\eqref{M1:E0} and are always satisfied.
Eqs.~\eqref{Pt:D}--\eqref{Pt:E} apply to any perturbation. Since they define a linear dynamical system for $\delta\mathbf{E}(\mathbf{r},t)$, $\delta\mathbf{E}^*(\mathbf{r},t)$, $\delta{P}(\mathbf{r},t)$, $\delta{P}^*(\mathbf{r},t)$, and $\delta{D}(\mathbf{r},t)$, the time evolution of any perturbation can be written as a superposition of the eigenvectors of the linear system. Therefore, for the purpose of a stability analysis, it suffices to characterize the dynamics of the eigenvectors (namely, the eigenvalues).

In previous work~\cite{PhysRevA.92.013847,PhysRevA.95.023835,2020_Benzaouia_APL,2022_Benzaouia_APLph}, such an eigen analysis was carried out by separating the real and imaginary parts of Eqs.~\eqref{Pt:D}--\eqref{Pt:E} and considering an aggregated real-valued vector $u = ({\rm Re}\,\delta\mathbf{E},{\rm Im}\,\delta\mathbf{E},{\rm Re}\,\delta\mathbf{P},{\rm Im}\,\delta\mathbf{P},\delta D)$.
The resulting equations are suitable for numerical solution but have changed so much from the original wave equations that it is difficult to analyze them, interpret the spectrum and other physical quantities, or relate them to the single-resonance-turn-on picture of SALT.

In the preceding derivation of the dynamic inversion factor $\zeta \approx |\mathbf{E}_{-1}|^2/|\mathbf{E}_{1}|^2$, we showed that a monochromatic perturbation $\mathbf{E}_1 e^{-i(\omega_0 + \omega_{\rm d}) t}$ necessarily results in another frequency component $\mathbf{E}_{-1} e^{-i (\omega_0 - \omega_{\rm d}) t}$ when $\zeta$ is not negligible.
We also showed that $\mathbf{E}_{\pm 1}$, $\mathbf{P}_{\pm 1}$, and $D_{\pm 1}$ are all proportional to the perturbation.
Also, from the $\delta E_{}^*{P}_0$ term in Eq.~\eqref{Pt:D}, we see that a perturbation with $e^{-i\omega_d t}$ time dependence [note that $e^{-i\omega_0 t}$ is already factored out from the perturbation in Eqs.~\eqref{Perp:P}--\eqref{Perp:E}] will generate $e^{i\omega_d^* t}$ time dependence.
Therefore, here we consider a multi-spectral perturbation with both $\omega_d$ and $-\omega_d^*$ frequency components as the trial eigenvector,
\begin{CAlign}
\delta D(\mathbf{r},t) &=D_1(\mathbf{r})e^{-i\omega_d t}+ D_1^*(\mathbf{r})e^{i\omega_d^* t}, \label{wd:D}\\
\delta P(\mathbf{r},t) &=P_1(\mathbf{r})e^{-i\omega_d t}+ P_{-1}(\mathbf{r})e^{i\omega_d^* t}, \label{wd:P}\\
\delta \mathbf{E}(\mathbf{r},t) &=\mathbf{E}_1(\mathbf{r})e^{-i\omega_d t}+ \mathbf{E}_{-1}(\mathbf{r})e^{i\omega_d^* t}. \label{wd:E}
\end{CAlign}
The ``frequency difference'' $\omega_d$ here is the eigenvalue of the linear perturbation problem and is complex-valued in general.
Substituting into Eqs.~\eqref{Pt:D}--\eqref{Pt:E}, we can verify that the ansatz of Eqs.~\eqref{wd:D}--\eqref{wd:E} has the correct time dependencies and constitutes a valid eigenvector for the linear dynamical system.

From the $e^{-i\omega_d t}$ and $e^{i\omega_d^* t}$ terms of Eq.~\eqref{Pt:D}--\eqref{Pt:E}, we obtain
\begin{CAlign}
D_{1} &= 0.5\Gamma_\parallel(E_{-1}^* P_0 + E_0^* P_1 - E_1 P_0^* - E_0 P_{-1}^*), \label{M1:Dep} \\
P_{\pm 1} &= \Gamma_{\pm}(D_{\pm 1}E_0+D_0E_{\pm 1}), \label{M1:P} \\
-\nabla\times\nabla&\times \mathbf{E}_{\pm 1} + \frac{\omega_{\pm 1}}{c^2} \left(\varepsilon_c+\frac{i\sigma}{\omega_{\pm 1}\varepsilon_0}\right)\mathbf{E}_{\pm 1} = -\frac{\omega_{\pm 1}}{c^2}{P}_{\pm 1} \mathbf{\uptheta}^*, \label{M1:E}
\end{CAlign}
where 
$\omega_1 = \omega_0 + \omega_d$, $\omega_{-1} = \omega_0 - \omega_d^*$,
$\Gamma_\parallel \equiv \gamma_\parallel / (\omega_d +i\gamma_\parallel)$,
$\Gamma_\pm = \Gamma(\omega_{\pm 1})$,
and $D_{-1} = D_1^*$.
Substituting Eqs.~\eqref{M1:P0} and \eqref{M1:P} into Eq.~\eqref{M1:Dep}, we can eliminate the gain-induced polarization to yield
\begin{CEquation}
D_{1} = 0.5\Gamma_\parallel\left[
(\Gamma_{+}-\Gamma_{-}^*)|E_0|^2D_1
+(\Gamma_{+}-\Gamma_{0}^*)D_0E_0^*E_1
+(\Gamma_{0}-\Gamma_{-}^*)D_0E_0E_{-1}^*
\right],\label{M1:Dep2}
\end{CEquation}
from which $D_1$ can be solved as
\begin{CEquation}
D_{1} = D_0(\chi_+E_1 + \chi_-E_{-1}^*), \label{M1:D1}
\end{CEquation}
with
\begin{CAlign}
\chi_+ = \frac{0.5\Gamma_\parallel(\Gamma_{+}-\Gamma_{0}^*)E_0^*}{1-0.5\Gamma_\parallel(\Gamma_{+}-\Gamma_{-}^*)|E_0|^2}, \label{chi+} \\
\chi_- = \frac{0.5\Gamma_\parallel(\Gamma_{0}-\Gamma_{-}^*)E_0}{1-0.5\Gamma_\parallel(\Gamma_{+}-\Gamma_{-}^*)|E_0|^2}. \label{chi-}
\end{CAlign}
Substituting Eqs.~\eqref{M1:P} and \eqref{M1:D1} into Eq.~\eqref{M1:E}, we can further eliminate $D_1$ and obtain an eigenvalue problem for $\mathbf{E}_{\pm 1}(\mathbf{r})$ along:
\begin{CAlign}
\hat{O}(\omega_1) \mathbf{E}_1 &+\frac{\omega_{1}^2}{c^2}\Gamma_{+}D_0E_0(\chi_+E_1 + \chi_-E_{-1}^*)\mathbf{\uptheta}^* = 0, \label{M1:E1}\\
\hat{O}(\omega_{-1}) \mathbf{E}_{-1} &+\frac{\omega_{-1}^2}{c^2}\Gamma_{-}D_0E_0(\chi_+^*E_1^* + \chi_-^*E_{-1})\mathbf{\uptheta}^* = 0, \label{M1:E-1}
\end{CAlign}
where $\hat{O}(\omega)$ is the active-cavity wave operator defined in Eq.~\eqref{H}.

Eqs.~\eqref{M1:E1}--\eqref{M1:E-1} show that the two spectral components $\mathbf{E}_{\pm 1}(\mathbf{r}) e^{-i \omega_{\pm 1} t}$ of the perturbation are explicitly coupled through $E_0 \chi_\pm$, which arises from the dynamic inversion $D_1 e^{-i\omega_d t}$ and is proportional to its relaxation rate $\gamma_\parallel$. 
We can take the complex conjugate of Eq.~\eqref{M1:E-1}, so that Eq.~\eqref{M1:E1} and [Eq.~\eqref{M1:E-1}$]^*$ together forms a coupled eigenproblem for $\mathbf{E}_{1}$ (with an outgoing boundary condition) and $\mathbf{E}_{-1}^*$ (with an incoming boundary condition).
Solving this coupled eigenproblem yields the complex eigenvalue $\omega_d$ and the associated perturbation in Eqs.~\eqref{wd:D}--\eqref{wd:E}.
As a reminder, $\omega_{\pm 1}$, $\Gamma_\pm$, $\chi_\pm$, and the $\Gamma_\parallel$ in $\chi_\pm$ all depend on $\omega_d$, so this is a nonlinear eigenvalue problem.

Eqs.~\eqref{M1:E1}--\eqref{M1:E-1} always have a trivial solution with $\omega_d=0$ and $(\mathbf{E}_1,\mathbf{E}_{-1}) = (A,-A^*)\mathbf{E}_0$, where $A = A_r + i A_i$ is any constant, such that the perturbation $\delta \mathbf{E}(\mathbf{r},t) e^{-i\omega_0 t} = 2i A_i \mathbf{E}_0(\mathbf{r})e^{-i\omega_0 t}$ is marginally stable and is $\pm 90$-degree out-of-phase with the lasing mode $\mathbf{E}_0(\mathbf{r}) e^{-i\omega_0 t}$.
Here, the coupling terms are cancelled, $\chi_+E_1 + \chi_-E_{-1}^* = 0$, and Eqs.~\eqref{M1:E1}--\eqref{M1:E-1} are satisfied since $\hat{O}(\omega_0)\mathbf{E}_0 = 0$.
This trivial solution must exist because there is a gauge degree of freedom in the single-mode solution Eq.~\eqref{eq:E0}:
$\mathbf{E}(\mathbf{r},t) = e^{i\phi} \mathbf{E}_0(\mathbf{r}) e^{-i\omega_0 t}$
with any real-valued global phase $\phi$ is also a solution of the MB equations.

At the first threshold $D_1^{\text{th}}$, $\mathbf{E}_0 = 0$, so the coupling strength $E_0 \chi_\pm$ is zero, and Eqs.~\eqref{M1:E1}--\eqref{M1:E-1} decouple to $\hat{O}(\omega_{\pm1}) \mathbf{E}_{\pm1} = 0$, same as Eq.~\eqref{psi}.
Therefore, the set of stability eigenvalues $\{\omega_d\}$ above $D_1^{\text{th}}$ is continuously connected to $\{\tilde{\omega}_n - \omega_0\}$ below $D_1^{\text{th}}$, where $\{\tilde{\omega}_n\}$ is the set of eigen frequencies of the active-cavity resonances with an unsaturated gain, defined in Eq.~\eqref{psi}.
Above the first threshold $D_1^{\text{th}}$, $\mathbf{E}_0 \neq 0$, so Eqs.~\eqref{M1:E1}--\eqref{M1:E-1} are no longer equivalent to Eq.~\eqref{psi}, and each stability eigenmode evolves into a superposition of the cavity resonances.

Above $D_1^{\text{th}}$, we can track how the set of complex eigenvalues $\{\omega_d\}$ evolves with the pumping strength.
Near $D_1^{\text{th}}$, all eigenvalues (except the trivial one, $\omega_{\rm d} = 0$) have a negative imaginary part, so any perturbation $\delta\mathbf{E}(\mathbf{r},t)$, $\delta{P}(\mathbf{r},t)$, and $\delta{D}(\mathbf{r},t)$ has to either decay exponentially in time or merely change the global phase of the lasing mode; here, the single-mode lasing solution Eqs.~\eqref{eq:D0}--\eqref{M1:E0} is stable.
When one of the non-trivial $\omega_d$ reaches the real-frequency axis, single-mode lasing is no longer stable since any perturbation that overlaps with the eigenvector with that real-valued $\omega_d$ will start to grow exponentially.
Therefore, the crossing of $\omega_d$ with the real-frequency axis marks the second threshold $D^\text{th}_2$. Just above that threshold, the perturbations $\mathbf{E}_1$ and $\mathbf{E}_{-1}$ continuously evolve into two comb lines in Eq.~(11) of the main text while that $\omega_d$ stays on the real axis and becomes the comb spacing. 

A stationary-inversion system corresponds to the limit of $\gamma_\parallel \to 0$, where $\chi_\pm \to 0$, the dynamic inversion $D_1$ in the perturbation eigenvector vanishes, and $\mathbf{E}_{\pm 1}$ in Eqs.~\eqref{M1:E1}--\eqref{M1:E-1} decouple.
Each of the decoupled $\mathbf{E}_{\pm 1}$ follows
$\hat{O}(\omega_{\pm1}) \mathbf{E}_{\pm1} = 0$, same as Eq.~\eqref{psi}.
In this limit, the stability eigenmodes are the resonances $\{\mathbf{\psi}_n\}$ of the active cavity $\varepsilon_{\rm eff}$.
This reduces to the formalism of SALT~\cite{PhysRevA.82.063824,2014_Esterhazy_PRA}, where the second threshold $D_2^{\text{th}}$ corresponds to a second individual resonance of the active cavity receiving enough gain to overcome its loss and reach the real-frequency axis.

For a dynamic system where $\gamma_\parallel$ is not negligible---more specifically, when the dynamic inversion factor $\zeta \approx |\mathbf{E}_{-1}|^2/|\mathbf{E}_{1}|^2$ is not negligible---the eigenmode $(\mathbf{E}_{1}e^{-i \omega_{1} t}, \mathbf{E}_{-1}e^{-i \omega_{-1} t})$ in Eqs.~\eqref{M1:E1}--\eqref{M1:E-1} is a multi-spectral perturbation that comes with a dynamic inversion $D_1$.
In this case, $\mathbf{E}_{1}$ (and similarly for $\mathbf{E}_{-1}$) satisfies Eqs.~\eqref{M1:E1}--\eqref{M1:E-1} but not Eq.~\eqref{psi}, so it is no longer an isolated resonance $\{\mathbf{\psi}_n\}$ of the active cavity; it is a mixture of resonances.
One can quantify the amount of mixing by projecting $\mathbf{E}_1$ onto the two dominant resonances,
\begin{CEquation}
\frac{\mathbf{E}_1}{\sqrt{\langle \varepsilon_c \mathbf{E}_1^2 \rangle}} \approx \sum_{n=1}^2{\frac{\langle \varepsilon_c \mathbf{E}_1 \cdot \mathbf{\psi}_n \rangle}{\sqrt{\langle \varepsilon_c \mathbf{E}_1^2 \rangle}\sqrt{\langle \varepsilon_c \mathbf{\psi}_n^2 \rangle}} \frac{\mathbf{\psi}_n}{\sqrt{\langle \varepsilon_c \mathbf{\psi}_n^2 \rangle}}},
\label{projection:E1}
\end{CEquation}
where $\mathbf{E}_1^2 = \mathbf{E}_1 \cdot \mathbf{E}_1$ and $\mathbf{\psi}_n^2 = \mathbf{\psi}_n \cdot \mathbf{\psi}_n$.
For the near-EP laser in the main text at $D_2^{\text{th}}$, the normalized projection factors are
\begin{CEquation}
\bigg| {\frac{\langle \varepsilon_c \mathbf{E}_1 \cdot \mathbf{\psi}_1 \rangle}{\sqrt{\langle \varepsilon_c \mathbf{E}_1^2 \rangle}\sqrt{\langle \varepsilon_c \mathbf{\psi}_1^2 \rangle}}} \bigg|^2=0.26, \    \
\bigg| {\frac{\langle \varepsilon_c \mathbf{E}_1 \cdot \mathbf{\psi}_2 \rangle}{\sqrt{\langle \varepsilon_c \mathbf{E}_1^2 \rangle}\sqrt{\langle \varepsilon_c \mathbf{\psi}_2^2 \rangle}}} \bigg|^2=0.80, 
\label{projections}
\end{CEquation}
which confirms that $\mathbf{E}_1$ has significant contributions from both resonances.

An intermediate case is possible for Hermitian degeneracies arising from symmetry~\cite{PhysRevA.92.013847,PhysRevA.95.023835}, where $\zeta$ can be small because there is no Petermann factor enhancement, but the ratio $\gamma_\parallel/\omega_{\rm d}$ can be large.
Since $\zeta$ is small, $\mathbf{E}_{-1}$ is negligible, and Eqs.~\eqref{M1:E1}--\eqref{M1:E-1} approximately decouple.
The decoupled eigenproblem Eq.~\eqref{M1:E1} for $\mathbf{E}_{1}$ differs from Eq.~\eqref{psi} because of the nonzero $\chi_+$ (which is proportional to $\gamma_\parallel$) induced by a dynamic inversion in the perturbation.
Here, the stability eigenmode $\mathbf{E}_{1}$ can also be a mixture of the near-degenerate active-cavity resonances~\cite{PhysRevA.92.013847,PhysRevA.95.023835} but is monochromatic.

\vspace{-6pt}
\section{PALT for direct-bandgap semiconductor lasers}\label{sec:multi-level}
\begin{figure}[b]
\vspace{-20pt}
\includegraphics[width=0.42\textwidth]{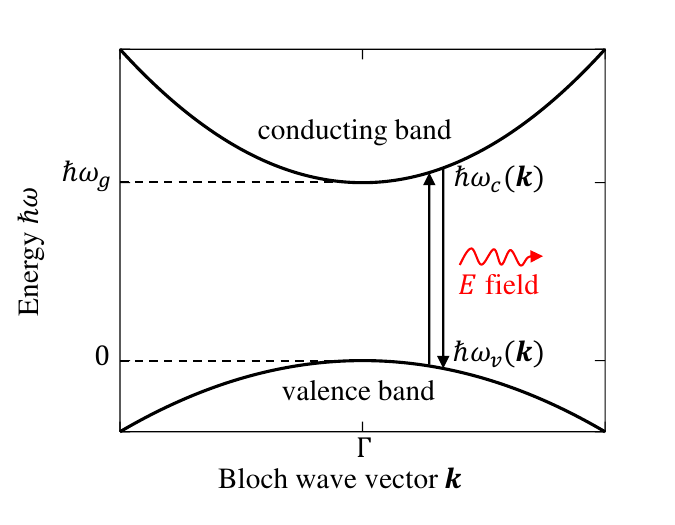}
\centering
\vspace{-12pt}
\caption{The band structure of a direct-bandgap semiconductor.}
\label{fig:s3}
\end{figure}

\vspace{-6pt}
In the preceding sections, we considered the Maxwell--Bloch equations for an ensemble of two-level atoms, which captures the essential properties of the nonlinear gain in a minimal form. 
Here, we consider a more quantitative model for a semiconductor gain medium~\cite{Cerjan:12, chow2013semiconductor, Cerjan:CSALT}.
\textbf{Supplementary Fig.}~\ref{fig:s3} shows the band structure of a direct-bandgap semiconductor near the bandgap. The electrons comply with the density matrix equation, Eq.~\eqref{rho}, with the same Hamiltonian expressed by Eq.~\eqref{Hamiltonian}. But unlike the two-level system, here the electron's Hilbert space has more than two dimensions. Let $\{|B,\mathbf{k}\rangle\}$ be the eigenstates of $\hat{H}_0$, where $B$ represents valence ($B=v$) or conduction ($B=c$) band and $\mathbf{k}$ the Bloch wave vector. Let $N$ be the total density of electrons on both bands. $N_{Bk} = N\langle B,\mathbf{k}|\hat{\rho}|B,\mathbf{k}\rangle$ is the electron population at band $B$ at $\mathbf{k}$. The light emission/absorption occurs when electrons transit between the valence band and the conduction band.  In direct-bandgap semiconductor lasers, these radiative inter-band transitions only occur at the same $\mathbf{k}$, shown by the black arrows in \textbf{Supplementary Fig.}~\ref{fig:s3}. It means $\langle B_2,\mathbf{k}_2|\hat{\mathbf{r}}|B_1,\mathbf{k}_1\rangle = \mathbf{0}$ when $\mathbf{k}_1 \ne \mathbf{k}_2$. We also assume spatial symmetry, $\langle B,\mathbf{k}|\hat{\mathbf{r}}|B,\mathbf{k}\rangle = \mathbf{0}$ for $B=v, c$. Upon that, we can derive the (observable) optical polarization $\tilde{\mathbf{P}}$ as
 \begin{CEquation}
\vspace{-2pt}
\tilde{\mathbf{P}} = -eN\text{tr}(\hat{\rho}\hat{\mathbf{r}}) = \sum_{\mathbf{k}} {(\mathbf{P_k} + \mathbf{P_k}^*)}, \vspace{-2pt}
\label{S:Pt}
\end{CEquation} 
where 
\begin{CAlign}
\vspace{-2pt}
\mathbf{P_k} &= N\langle c,\mathbf{k}|\hat{\rho}|v,\mathbf{k}\rangle \mathbf{R_k}^*, \\
\mathbf{R_k} &= -e\langle c,\mathbf{k}|\hat{\mathbf{r}}|v,\mathbf{k}\rangle.
\vspace{-2pt}
\end{CAlign} 

Similar to the derivation of Eqs.~\eqref{Na}--\eqref{P}, we decompose Eq.~\eqref{rho} under $\{|B,\mathbf{k}\rangle\}$ to get the equations for $N_{v\mathbf{k}}$, $N_{c\mathbf{k}}$, and $\mathbf{P_k}$,
 \begin{CAlign}
\vspace{-2pt}
\frac{\partial}{\partial t}N_{v\mathbf{k}} &= \sum_{B',\mathbf{k'}}{\gamma_{(B'\mathbf{k'})(v\mathbf{k})}N_{B'\mathbf{k'}}}-\sum_{B',\mathbf{k'}}{\gamma_{(v\mathbf{k})(B'\mathbf{k'})}N_{v\mathbf{k}}} - \frac{1}{i\hbar}\mathbf{\tilde{E}}\cdot(\mathbf{P_k}-\mathbf{P_k}^*), \label{Nv} 
\vspace{-2pt}
\end{CAlign} 
\begin{CAlign}
\vspace{-2pt}
\frac{\partial}{\partial t}N_{c\mathbf{k}} &= \sum_{B',\mathbf{k'}}{\gamma_{(B'\mathbf{k'})(v\mathbf{k})}N_{B'\mathbf{k'}}}-\sum_{B'\mathbf{k'}}{\gamma_{(v\mathbf{k})(B'\mathbf{k'})}N_{c\mathbf{k}}} + \frac{1}{i\hbar}\mathbf{\tilde{E}}\cdot(\mathbf{P_k}-\mathbf{P_k}^*), \label{Nc} \\
\frac{\partial}{\partial t}\mathbf{P_k} &=- (i\omega_{\mathbf{k}} +\gamma_{\perp \mathbf{k}}) \mathbf{P_k}+ \frac{1}{i\hbar}(\mathbf{E}\cdot\mathbf{R_k})(N_{c\mathbf{k}}-N_{v\mathbf{k}})\mathbf{R}^*_{\mathbf{k}}, \label{Pk}
\vspace{-2pt}
\end{CAlign} 
At each $\mathbf{k}$ point, $\omega_{\mathbf{k}} = \omega_{v}(\mathbf{k})-\omega_{c}(\mathbf{k})$, where $\hbar\omega_{v}(\mathbf{k})$ and $\hbar\omega_{c}(\mathbf{k})$ are the energies of valence and conduction band, respectively. $\gamma_{\perp\mathbf{k}}$ is the dephasing rate. The summation terms in Eq.~\eqref{Nv} and Eq.~\eqref{Nc} describe incoherent transitions, including spontaneous emission, pumping, and intra-band non-radiative scattering caused by atomic collisions. It is neither feasible nor necessary to determine each $\gamma_{(B'\mathbf{k'})(B\mathbf{k})}$. The intra-band scattering happens much faster than inter-band transitions. Therefore, in the time scale of our interest, the electron occupation of each band is at a quasi-equilibrium, allowing us to simplify Eq.~\eqref{Nv} and Eq.~\eqref{Nc} by statistics, 
 \begin{CAlign}
\vspace{-2pt}
\frac{\partial}{\partial t}N_{v\mathbf{k}} &= -\gamma_{v\mathbf{k}}[N_{v\mathbf{k}}-N f_v(\mathbf{k})] - \frac{1}{i\hbar}\mathbf{\tilde{E}}\cdot(\mathbf{P_k}-\mathbf{P_k}^*), \label{FD_Nv} \\
\frac{\partial}{\partial t}N_{c\mathbf{k}} &= -\gamma_{c\mathbf{k}}[N_{c\mathbf{k}}-N f_c(\mathbf{k})] + \frac{1}{i\hbar}\mathbf{\tilde{E}}\cdot(\mathbf{P_k}-\mathbf{P_k}^*), \label{FD_Nc}
\vspace{-2pt}
\end{CAlign} 
where $\gamma_{v\mathbf{k}}$ and $\gamma_{c\mathbf{k}}$ are relaxation rates, and $f_v(\mathbf{k})$ and $f_c(\mathbf{k})$ are Fermi--Dirac distributions,
 \begin{CAlign}
\vspace{-2pt}
f_{v,c}(\mathbf{k}) = \frac{1}{e^{(\hbar\omega_{v,c}(\mathbf{k})-\mu_{v,c})/(k_B T)}+1}.
\vspace{-2pt}
\end{CAlign}
$\mu_{v,c}$ is the chemical potential of valence/conduction band. $k_B$ is the Boltzmann constant, and $T$ is temperature in Kelvin. At thermal equilibrium,  $\mu_{c}=\mu_{v}$, so $f_{v}(\mathbf{k})>f_{c}(\mathbf{k})$. When being pumped, $\mu_{c}>\mu_{v}$. So, the pumping strength can be defined as $\mu_\Delta = \mu_{c}-\mu_{v}$. To turn on the laser, $\mu_\Delta$ must be large enough to achieve population inversion, $f_{v}(\mathbf{k})<f_{c}(\mathbf{k})$. 

Eqs.~\eqref{Pk}--\eqref{FD_Nc} generalize Eqs.~\eqref{P}--\eqref{D} to a direct-bandgap semiconductor, which forms the MB equations together with Maxwell's equations, Eq.~\eqref{MB:E}.

For frequency combs, we consider the trial solution, 
\begin{CAlign}
\vspace{-2pt}
\mathbf{E}(\mathbf{r},t) &= e^{-i\omega_0 t}\sum_{m=-\infty}^{+\infty}\mathbf{E}_m(\mathbf{r})e^{-im\omega_\text{d} t},\label{SCM:E} \\
\mathbf{P_k}(\mathbf{r},t) &= e^{-i\omega_0 t}\sum_{m=-\infty}^{+\infty}\mathbf{R_k}^*P_{\mathbf{k}m}(\mathbf{r})e^{-im\omega_\text{d} t},\label{SCM:P} \\
N_{B\mathbf{k}}(\mathbf{r},t) &= \sum_{m=-\infty}^{+\infty}N_{B\mathbf{k}m}(\mathbf{r})e^{-im\omega_\text{d} t}. \   B=c,v.  \label{SCM:N}
\vspace{-2pt}
\end{CAlign} 
Applying the same mathematical steps that lead to Eqs.~(18-20) in the Method section of the main text, we can show that Eqs.~\eqref{SCM:E}--\eqref{SCM:N} provide an exact solution of the MB equations for a direct-bandgap semiconductor under rotating-wave approximation.
Therefore, we conclude that the same EP comb phenomenon also exists here.

The above derivations lead to
\begin{CAlign}
\vspace{-2pt}
\bar{N}_{v\mathbf{k}}&=N f_v(\mathbf{k})\bar{\delta}-\bar{\bar{\Gamma}}_{v\mathbf{k}}(\bar{\bar{E}}_{\mathbf{k}}^\dag\bar{P}_{\mathbf{k}}-\bar{\bar{E}}_{\mathbf{k}}\bar{P}_{\mathbf{k}-}^*), \label{Svec:Nv} \\
\bar{N}_{c\mathbf{k}}&=N f_c(\mathbf{k})\bar{\delta}+\bar{\bar{\Gamma}}_{c\mathbf{k}}(\bar{\bar{E}}_{\mathbf{k}}^\dag\bar{P}_{\mathbf{k}}-\bar{\bar{E}}_{\mathbf{k}}\bar{P}_{\mathbf{k}-}^*), \label{Svec:Nc} \\
\bar{P}_{\mathbf{k}} &= \bar{\bar{\Gamma}}_{\mathbf{k}+} \bar{\bar{E}}_{\mathbf{k}}\bar{D}_{\mathbf{k}}, \label{Svec:P} \\
\bar{P}_{\mathbf{k}-}^* &= \bar{\bar{\Gamma}}_{\mathbf{k}-}^\dag \bar{\bar{E}}_{\mathbf{k}}^\dag\bar{D}_{\mathbf{k}}. \label{Svec:P*}
\vspace{-2pt}
\end{CAlign}
The notation follows that of the main text: \\
\indent \indent \textbf{Column vectors}: $(\bar{P}_{\mathbf{k}})_m=P_{\mathbf{k}m}$, $(\bar{P}_{\mathbf{k}-}^*)_{m}=P_{\mathbf{k},-m}^*$, $(\bar{N}_{B\mathbf{k}})_m=N_{B\mathbf{k}m}$ for $(B=c, v)$, $\bar{D}_{\mathbf{k}}=\bar{N}_{c\mathbf{k}} - \bar{N}_{v\mathbf{k}}$ and $(\bar{\delta})_m=\delta_{m}$. \\
\indent \indent \textbf{Matrices}: $(\bar{\bar{E}})_{mn}=\mathbf{E}_{m-n}\cdot\mathbf{R}_{\mathbf{k}}$, $(\bar{\bar{\Gamma}}_{B\mathbf{k}})_{mn}=\delta_{m-n}/[\hbar (m\omega_\text{d}+i\gamma_{B\mathbf{k}})]$, $(\bar{\bar{\Gamma}}_{\mathbf{k}\pm})_{mn}=\delta_{m-n}/[\hbar(\pm m\omega_d+\omega_0-\omega_{\mathbf{k}}+i\gamma_{\perp \mathbf{k}})] $. \\
Then we substitute Eqs.~\eqref{Svec:P}--\eqref{Svec:P*} into Eqs.~\eqref{Svec:Nc}--\eqref{Svec:Nv} to solve for $\bar{D}_{\mathbf{k}}$,
\begin{CEquation}
\bar{D}_{\mathbf{k}}=N[\bar{\bar{I}}-(\bar{\bar{\Gamma}}_{v\mathbf{k}}+\bar{\bar{\Gamma}}_{c\mathbf{k}})(\bar{\bar{E}}_{\mathbf{k}}^\dag \bar{\bar{\Gamma}}_{\mathbf{k}+}\bar{\bar{E}}_{\mathbf{k}}-\bar{\bar{E}}_{\mathbf{k}}\bar{\bar{\Gamma}}_{\mathbf{k}-}^\dag\bar{\bar{E}}_{\mathbf{k}}^\dag)]^{-1}[f_c(\mathbf{k})-f_v(\mathbf{k})]\bar{\delta}, \label{SPALT:D}
\end{CEquation} 
which is analogous to Eq.~(12) in the main text.
Substituting the polarization in Eqs.~\eqref{Svec:P}, \eqref{SCM:P}, \eqref{S:Pt} into Maxwell's equations Eq.~\eqref{MB:E}, we get the wave equation
\begin{CEquation}
-\nabla\times\nabla\times \mathbf{E}_{m} + \frac{\omega_{m}^2}{c^2} \left(\varepsilon_c+\frac{i\sigma}{\omega_{m}\varepsilon_0}\right)\mathbf{E}_{m} = -\frac{\omega_{m}^2}{c^2} \sum_{\mathbf{k}}( \bar{\bar{\Gamma}}_{\mathbf{k}+} \bar{\bar{E}}_{\mathbf{k}}\bar{D}_{\mathbf{k}})_m\mathbf{R_k}^* \label{SPALT:E},
\end{CEquation}
which is analogous to Eq.~(11) in the main text.
Note that although Eqs.~\eqref{SPALT:D}--\eqref{SPALT:E} are derived for semiconductor lasers, they also apply to multi-level gain media.    

\begin{figure}[b!]
\centering
\includegraphics[width=1.0\textwidth]{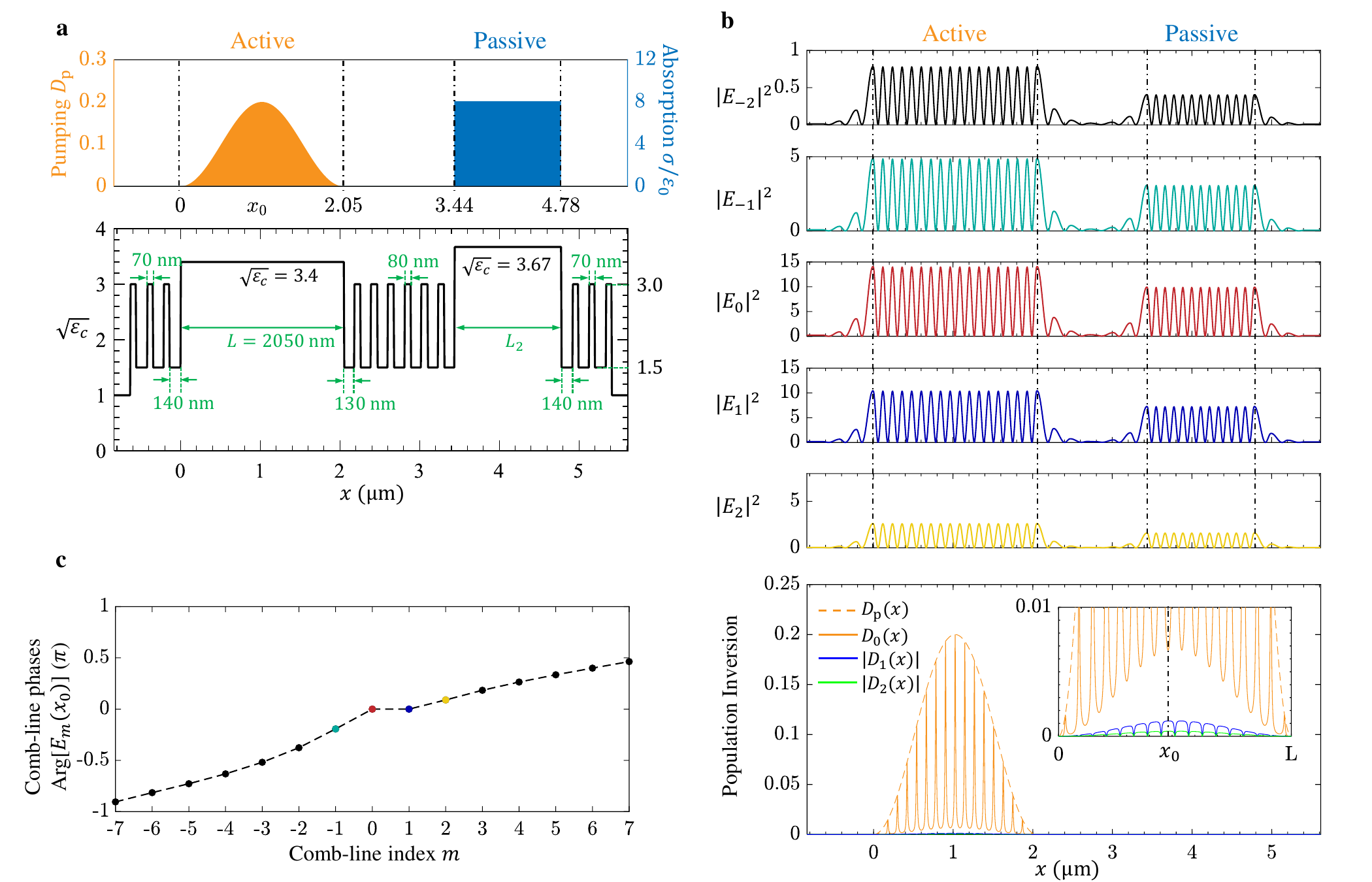}
\vspace{-12pt}
\caption{\textbf{a} Parameters of the EP laser considered in the main text. 
\textbf{b}  Spatial profiles of the intensity and of the inversion at different frequency components, for the EP comb in Fig.~3a of the main text (with $D_{\rm max} = 0.2$, $\sigma/\epsilon_0 = 8.1~{\rm ps}^{-1}$ and $L_2 = 1340 ~{\rm nm}$). \textbf{c} The phases of the electrical fields $E_m(x_0)$ with $-7\le m \le 7$.}
\label{fig:s2}
\end{figure} 

\vspace{-4pt}
\section{System parameters and properties} \label{sec:sys_params}
\vspace{-4pt}
\textbf{Supplementary Fig.}~\ref{fig:s2}\textbf{a} shows the parameters of the one-dimensional (1D) EP laser considered in the main text. The pumping profile is
\begin{CEquation}
D_\text{p} (x)= 
    \begin{cases} 
           0.5 D_\text{max}\left[1-\cos\left(\frac{2\pi x}{L}\right)\right], & 0<x<L \\
           0, & \text{elsewhere}
   \end{cases}
\end{CEquation}
with $L = 2050$ nm being the length of the first cavity. 
The two cavities' lengths, the geometric parameters of the middle distributed Bragg reflector (DBR), and the conductivity $\sigma$ are tuned coarsely (in 10 nm steps) to bring the system close to an EP around $\omega_{ba}$ at the first lasing threshold $D_1^{\text{th}}$. The absorption in the passive cavity is fixed at $\sigma/\epsilon_0 = 4.9~{\rm ps}^{-1}$ when $D_\text{max}$ is below the comb threshold $D^\text{th}_2 = D^\text{th}_{\rm c} = 0.064$ and raised linearly with $D_\text{max}$ above the comb threshold.

\begin{figure}[t]
\centering
\includegraphics[width=0.7\textwidth]{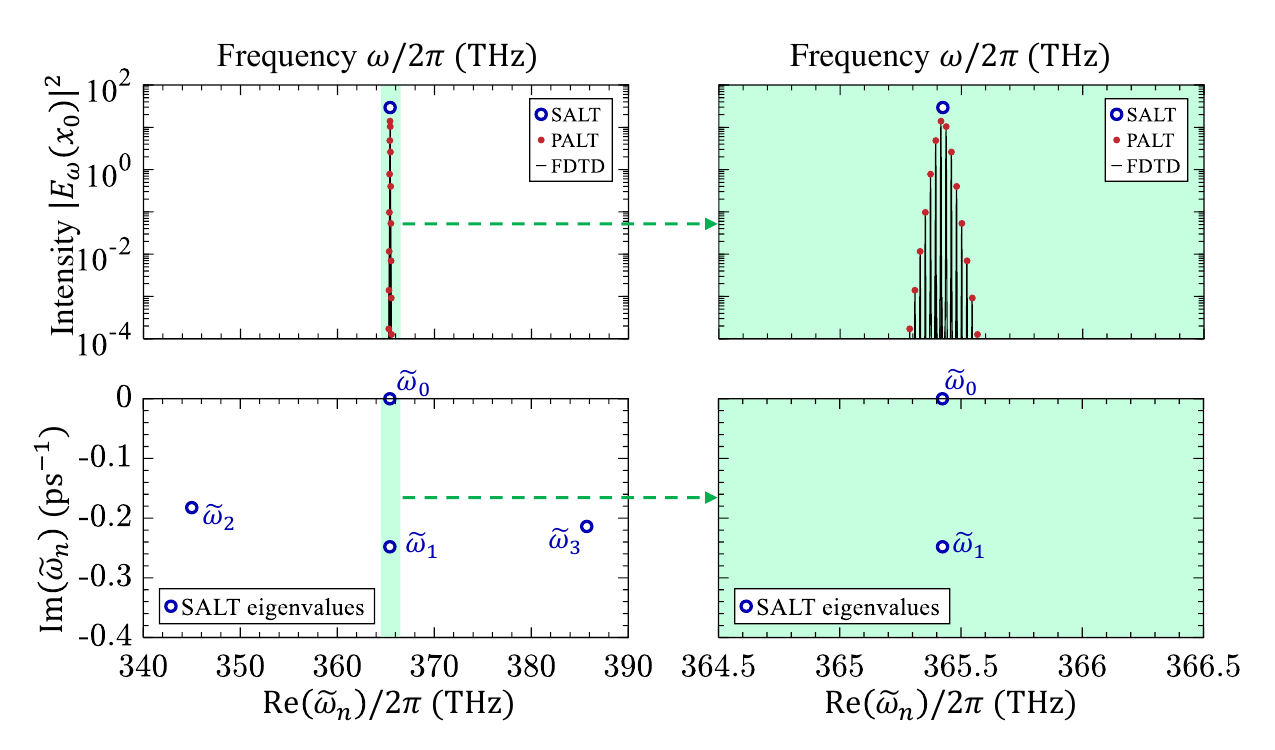}
\vspace{-12pt}
\caption{
{Comparison between the EP comb spectrum (upper panels) and the resonant frequencies of the active cavity from SALT through Eq.~\eqref{psi} (lower panels). 
$\tilde{\omega}_0$ and $\tilde{\omega}_1$ form the near-EP pair. The next nearest resonant modes, $\tilde{\omega}_{2}$ and $\tilde{\omega}_{3}$, are separated from $\tilde{\omega}_{0,1}$ by the free spectral range of the active cavity.
The right panels show zoom-ins near the EP frequency.
}
} 
\label{fig:s4}
\end{figure}

{\textbf{Supplementary Fig.}~\ref{fig:s4} compares (1) the comb spectrum in Fig.~\textbf{3g} of the main text at $\Delta = 0$, $D_\text{max} = 0.2$, $\sigma/\varepsilon_0 = 8.1~\rm ps ^{-1}$ with (2) the nonlinearity-frozen active-cavity eigenvalues $\{\tilde{\omega}_n\}$ of Eq.~\eqref{psi} for the same parameters and using the lasing mode $\mathbf{E}_0$ from SALT (which predicts the system to stay single-mode at $D_\text{max} = 0.2$).
We see that the two center comb lines $\{{\omega}_0, {\omega}_1\}$ lie close to the two near-degenerate active-cavity eigenvalues, $\{{\rm Re}(\tilde{\omega}_0), {\rm Re}(\tilde{\omega}_1)\}$.
The other comb lines do not line up with any additional cavity modes (which are far away in frequency) since they are generated by the nonlinear gain through dynamic four-wave mixing rather than by the other cavity modes.
Note that ${\rm Im}(\tilde{\omega}_1)$ has moved down on the complex-frequency plane ({\it i.e.}, becomes more lossy) compared to its value at the comb threshold $D_\text{max} = 0.064$ (see Fig.~\textbf{2d} of the main text) because of the increased absorption $\sigma$.
}


\vspace{-4pt}
\section{PALT solution through volume integral equation}
\vspace{-4pt}
For the 1D system above, $\mathbf{\uptheta}=\hat{z}, \mathbf{E}_m = E_m\mathbf{\uptheta}^*$, $\partial_y=\partial_z=0$, and Eq.~(15) from the main text can be simplified as
\begin{CEquation}
\frac{d^2}{d x^2}{E}_{m} + \frac{\omega_{m}^2}{c^2} \left(\varepsilon_c+\frac{i\sigma}{\omega_{m}\varepsilon_0}\right){E}_{m} = -\frac{\omega_{m}^2}{c^2} P_m \label{1DPALT:E}. 
\end{CEquation}
Here, the gain-induced polarization $P_m=P_m(E_0,E_{\pm1},...)$ on the right-hand side depends nonlinearly on the fields at all frequencies $\{\omega_m\}$ and is given by Eq.~(14) and Eq.~(12). 
We want to solve this system of nonlinear equations for the complex-valued $\{E_m(x)\}_m$ and the real-valued $\omega_0$ and $\omega_{\rm d}$.

In prior work on SALT, the analogous equations are numerically solved in differential form by expanding $E_m(x)$ in a set of threshold constant flux (TCF) modes~\cite{PhysRevA.82.063824} or with finite-difference frequency-domain (FDFD) discretization~\cite{2014_Esterhazy_PRA}. 
Both approaches would be suitable for a typical system away from EP.
However, the hyper sensitivity near an EP amplifies numerical error and requires an unusually high accuracy, and we found both methods to be inefficient for the present system given the desired accuracy.

Here, we adopt a volume-integral approach instead.
We define the retarded Green's function $G_m(x,x')$ of the cold cavity at frequency $\omega_m$ by
 \begin{CEquation}
\frac{\partial^2}{\partial x^2}G_{m}(x,x') + \frac{\omega_{m}^2}{c^2} \left[\varepsilon_c(x)+\frac{i\sigma(x)}{\omega_{m}\varepsilon_0}\right]G_{m}(x,x') = -\frac{\omega_{m}^2}{c^2} \delta(x-x') \label{Green}, 
\end{CEquation}
with an outgoing boundary condition. Given $G_m(x,x')$, Eq.~\eqref{1DPALT:E} can be written in integral form as
\begin{CEquation}
E_m(x)=\int_{-\infty}^{\infty} G_m(x,x') P_m(x') dx'. \label{E-G}
\end{CEquation}
Since the cold cavity in Eq.~\eqref{Green} consists of a 1D stack of piecewise-constant permittivity values, we can efficiently solve for the Green's function $G_m(x,x')$ semi-analytically with a high precision; such solution captures all linear effects of the cold cavity, including the DBR confinement, the oscillatory nature of the resonances, and their radiation loss.  Then, Eq.~\eqref{E-G} only needs to handle effects of the nonlinear gain, which changes relatively slowly and smoothly. 
Also, since the gain-induced polarization $P_m$ vanishes outside of the pumped region, the integration range in Eq.~\eqref{E-G} can be reduced to the active cavity $0\leq x'\leq L$. 

For numerical calculation, we truncate the frequency summation to $2M+1$ frequencies with $m = -M, -M+1, \cdots, M$ and approximate the integration of Eq.~\eqref{E-G} with the trapezoidal rule
\begin{CEquation}
E_m(x_j)=\sum_{k=0}^{K} G_m(x_j,x_k) P_m(x_k) \Delta x_k,
\label{Dis:E}
\end{CEquation}
where $x_k = kL/K$, $\Delta x_0 = \Delta x_K = L/(2K)$, and $\Delta x_k = L/K$ for $k=1,\cdots, K-1$.
Evaluating Eq.~\eqref{Dis:E} with $j=0, 1, \cdots, K$ yields $(2M+1)(K+1)$ complex-valued equations with $(2M+1)(K+1)$ complex-valued scalar unknowns: $\{E_m(x_j)\}$ with $m\in[-M,M]$ and $j\in[0,K]$.
Additionally, there are two more real-valued unknowns: $\omega_0$ and $\omega_{\rm d}$.
To match the number of equations and the number of unknowns, we recognize that when $\{E_m(x_j)\}$ is a solution, $\{e^{-i\Psi_0-im\Psi_\text{d}}E_m(x_j)$ is also a solution for any real-valued $\Psi_0$ and $\Psi_\text{d}$.
So, we choose $\Psi_0$ and $\Psi_\text{d}$ such that $E_0(x_0)$ and $E_1(x_0)$ are real-valued. 
Then, we have the following $2(2M+1)(K+1)$ real-valued variables
\begin{CEquation}
\text{Variables} =
\begin{bmatrix}
\text{Im}[E_{-M}(x_0)] & \text{Re}[E_{-M}(x_0)] & \dots  &\text{Re}[E_{-M}(x_K)] \\
\vdots                       & \vdots                       & \vdots &\vdots \\
\omega_0   & \text{Re}[E_{0}(x_0)] & ... &\text{Re}[E_{0}(x_K)] \\
\omega_\text{d}   & \text{Re}[E_{1}(x_0)] & ... &\text{Re}[E_{1}(x_K)] \\
\vdots                       & \vdots                       & \vdots &\vdots \\
\text{Im}[E_{M}(x_0)] & \text{Re}[E_{M}(x_0)] & \dots  &\text{Re}[E_{M}(x_K)] \\
\end{bmatrix},
\end{CEquation}
with the same number of real-valued equations.
We solve this system of nonlinear equations with the \verb|fsolve| function in MATLAB.
Given $E_m(x_j)$ at these $K+1$ points, we can evaluate Eq.~\eqref{Dis:E} at other values of $x_j$ to obtain a continuous profile $E_m(x)$ anywhere (including with $x$ outside the pumped cavity or outside the DBR mirrors).
For the example in this paper, we use $M=7$ and $K=200$, which is sufficient to reach four digits of accuracy for $\omega_{\rm d}$.
In contrast, a differential finite-difference approach will require over 100 times more spatial grid points to reach a comparable accuracy.

We adopt the same volume integral approach to solve for $\omega_0$ and $E_0$ in the single-mode lasing regime of Eq.~\eqref{M1:E0} and to solve the stability eigenvalue problem of Eqs.~\eqref{M1:E1}--\eqref{M1:E-1}.

\vspace{-4pt}
\section{FDTD simulations of the Maxwell--Bloch equations}\label{sec:FDTD}
\vspace{-4pt}
To validate the PALT predictions, we additionally carry out direct integration of the MB equations using the finite-difference time-domain (FDTD) method.
Here, we directly work with the real-valued fields, $\tilde{\mathbf{E}}$, $\mathbf{B}$, Re$(\mathbf{P})$, Im$(\mathbf{P})$ and $D$ in Eqs.~\eqref{FDTD:D}--\eqref{FDTD:E} without introducing the rotating-wave approximation. 
Details on the FDTD discretization are given in Ref.~\cite{Cerjan:15}. 
For the 1D simulation here, we implement the outgoing boundary analytically~\cite{mur1981absorbing}. 

Due to the enhanced sensitivity of EPs, a very fine spatial discretization is required for the FDTD simulations to reach enough accuracy to reproduce the PALT results for the EP laser described in Sec.~\ref{sec:sys_params} above.  
Here, we set the spatial grid size to be $\Delta x = 0.25$ nm and the time step size to be $\Delta t = \Delta x/c$. 
This requires over one billion time steps to evolve the system by just one relaxation time $1/\gamma_\parallel = 1$ ns of the gain medium.
Further reducing $\Delta x$ can make FDTD agree even closer to PALT but will incur higher computing costs.

In the absence of noise, the laser spectrum at long times should have zero linewidths. The linewidths in Fig.~2(g,h) of the main text come from a finite temporal window in the FDTD simulations. 
A frequency comb solution features electrical field
\begin{CAlign}
\tilde{\mathbf{E}}(\mathbf{r},t) = {\mathbf{E}}(\mathbf{r},t) + {\mathbf{E}}^*(\mathbf{r},t)= e^{-i\omega_0 t}\sum_m\mathbf{E}_m(\mathbf{r}) e^{-im\omega_d t} + \text{c.c.},\ \  -\infty<t<\infty. \label{FDTD:comb_t}
\end{CAlign}
Fourier transform (FT) of the infinite time series in Eq.~\eqref{FDTD:comb_t} yields
\begin{CAlign}
\text{FT}\Big\{\tilde{\mathbf{E}}(\mathbf{r},t)\Big\} = 2\pi\sum_m\mathbf{E}_m(\mathbf{r}) \delta(\omega-\omega_m) + 2\pi\sum_m\mathbf{E}_m^*(\mathbf{r}) \delta(\omega+\omega_m), \label{FDTD:comb_f}
\end{CAlign}
which has zero linewidths.
In practice, even after the transient behaviors settle away in a FDTD simulation, we will only have access to data within a finite duration: one section of Eq.~\eqref{FDTD:comb_t} in time. Let $T$ be the duration of the long-time data taken from $\tilde{\mathbf{E}}(\mathbf{r},t)$. The truncated data is then expressed as $w(t)\tilde{\mathbf{E}}(\mathbf{r},t)$, where $w(t)$ is a finite-sized window function confined within $-0.5T<t<0.5T$. Let $W(\Delta\omega)$ be the Fourier transform of $w(t)$. Then, the Fourier transform of the practical (truncated) data is 
\begin{CAlign}
\text{FT}\Big\{w(t)\,\tilde{\mathbf{E}}(\mathbf{r},t)\Big\} = \sum_m\mathbf{E}_m(\mathbf{r})W(\omega - \omega_m) + \sum_m\mathbf{E}_m^*(\mathbf{r})W(\omega + \omega_m). \label{FDTD:comb_fw}
\end{CAlign}
The raw FDTD data correspond to a rectangular window for $w(t)$, whose spectrum $W(\Delta\omega)$ is a sinc function with bandwidth $2\pi/T$, which has side peaks that decay slowly as $(\Delta\omega T)^{-1}$.
The side peaks may corrupt the underlying comb spectrum.
A brute-force way to suppress the side peaks is to use a sample duration $T$ that is many orders of magnitude beyond the temporal periodicity $\tau = 2\pi/\omega_{\rm d}$, so that the side peaks are negligible at $\Delta \omega = m \omega_{\rm d}$.
But doing so is wasteful and computationally costly since $\tau \approx 50$ ps is large in the near-EP system studied here.
To more efficiently suppress the side peaks of the window function spectrum, we use a Hann window instead,
\begin{CAlign}
w(t) = 
\begin{cases}
\frac{1}{2}\left[1+\text{cos}\left(\frac{2\pi t}{T}\right)\right], & -0.5T<t<0.5T, \\
0, & \text{otherwise}.
\end{cases}
\end{CAlign}
Its Fourier transform $W(\Delta\omega)$ has a wider center lobe, but the side peaks decay much faster as $(\Delta\omega T)^{-3}$, allowing a much smaller sampling duration $T$.
Here, $W(\Delta\omega=0) = T/2$, so in Fig.~2(g,h) of the main text we plot 
\begin{CAlign}
\mathbf{E}_\omega(\mathbf{r}) \equiv \frac{2}{T}\text{FT}\Big\{w(t)\,\tilde{\mathbf{E}}_{\rm FDTD}(\mathbf{r},t)\Big\},
\end{CAlign}
such that the peaks of $|\mathbf{E}_\omega|^2$ will equal $|\mathbf{E}_m|^2$ in the long-time limit. 

After $\tilde{E}(x,t)$ and $D(x,t)$ settle down, we continue to run the simulation for $T=800$ ps $\approx 16\tau$ while recording $\tilde{E}(x_0,t)$ at $x=x_0$. 
We record one data point per $20\Delta t$, which is over a hundred data points per oscillation, well above the Nyquist sampling rate. 
A fast Fourier transform (FFT)~\cite{2005_Frigo_FFTW3} is then used to compute the spectrum. To ensure that the frequency grid of FFT does not accidentally miss the narrow peak of each comb line, we pad zeros on both sides of the temporal data after the Hann window and before the FFT.

\vspace{-4pt}
\section{Exact-EP laser} \label{sec:exact_EP}
\vspace{-4pt}

As shown in \textbf{Supplementary Fig.}~\ref{fig:s4}, we did not tune to an exact EP for the system considered in the main text.
There, the system is tuned close to an EP at the first threshold ($D_\text{max} = D^{\text{th}}_1$) and stays close to the EP in the single-mode lasing regime when gain saturation is accounted for. The near-EP single-mode lasing state remains stable until when the pumping strength reaches the comb threshold $D_2^{\text{th}} = D_\textrm{c}^{\text{th}}$.

In this section, we consider the less realistic but conceptually interesting question: what happens when we tune the system to an exact EP? An EP below the lasing threshold $D^{\text{th}}_1$ (such as the EP in \textbf{Supplementary Fig.}~\ref{fig:s4}) is not accessible at steady state. So, we are interested in an EP above $D^{\text{th}}_1$.

\begin{figure}[t]
\centering
\includegraphics[width=0.65\textwidth]{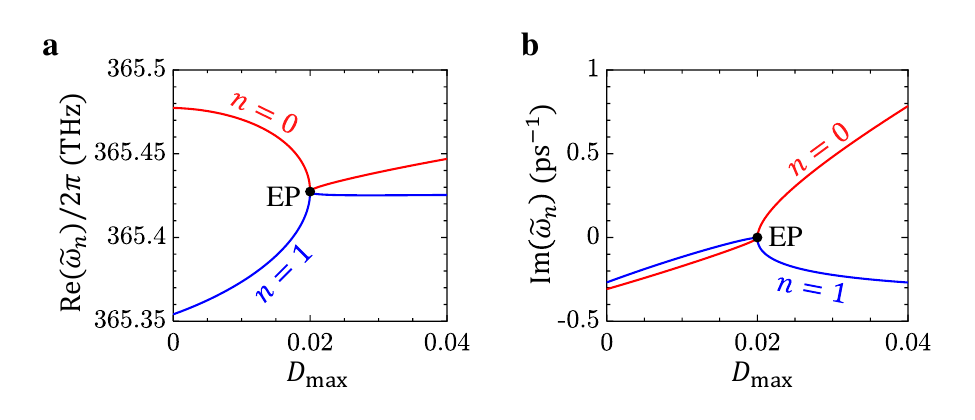}
\vspace{-12pt}
\caption{Evolution of the two eigenvalues $\tilde{\omega}_0$ and $\tilde{\omega}_1$ of Eq.~\eqref{psi} as a function of the pumping strength $D_\text{max}$. \textbf{a} The real part of $\tilde{\omega}_0$ (red) and $\tilde{\omega}_1$ (blue). \textbf{b} The imaginary part of $\tilde{\omega}_0$ (red) and $\tilde{\omega}_1$ (blue). The exceptional point is marked as a black dot in both \textbf{a} and \textbf{b}. The system is the same as Supplementary Fig.~\ref{fig:s2}a except that the passive cavity length is tuned to 1339.985~nm and the material loss is tuned to $\sigma/\varepsilon_0 =4.021~\text{ps}^{-1}$ to bring the system to an exact lasing EP at $D_\text{max}=D_\text{EP}=0.02000$.
The $\mathbf{E}_0(\mathbf{r})$ in Eq.~\eqref{M1:D0} is fixed to the $\mathbf{E}_0(\mathbf{r})$ at $D_\text{max}=D_\text{EP}$ to make the operator $\hat{O}$ linear throughout the evolution.}
\vspace{12pt}
\label{fig:s1}
\end{figure}

To be clear, here we consider an (almost) exact EP of the active-cavity wave operator $\hat{O}$ in Eq.~\eqref{psi}.
For a given pump strength, after solving the nonlinear Eq.~\eqref{M1:E0_O} to obtain the single-mode lasing state $\mathbf{E}_0(\mathbf{r})$, we consider the saturated gain $D_0(\mathbf{r})$ with the $E_0(\mathbf{r})$ in Eq.~\eqref{M1:D0} fixed, which defines the linear operator $\hat{O}$ in Eq.~\eqref{psi}.
We then tune the system parameters with higher precisions to bring this nonlinearity-frozen operator $\hat{O}$ to an EP.
By tuning the length of the passive cavity to $1339.985$~nm and the absorption to $\sigma/\varepsilon_0 =4.021~\text{ps}^{-1}$, we bring the system very close to a single-mode lasing EP at $D_\text{max}=D_\text{EP}=0.02000 >D^\text{th}_1 = 0.01336$.
Here, the eigenvalue difference $|\tilde{\omega}_0 - \tilde{\omega}_1| \approx 0.02~\text{ps}^{-1}$ is around 3000 times smaller than the free spectral range of the overall cavity.
The Petermann factors are $K_1 = 3704$ and 
$K_2 = 369$, much larger than those of the near-EP system in Sec.~\ref{sec:sys_params}.
\textbf{Supplementary Fig.}~\ref{fig:s1} shows the evolution of the two eigenvalues $\tilde{\omega}_0$ and $\tilde{\omega}_1$ of Eq.~\eqref{psi} as a function of $D_\text{max}$, where we make the operator $\hat{O}$ linear throughout the evolution by fixing the ${E}_0(\mathbf{r})$ in Eq.~\eqref{M1:D0} to the ${E}_0(\mathbf{r})$ at the EP with pumping strength $D_\text{max}=D_\text{EP}$.



\begin{figure}[t]
\centering
\includegraphics[width=0.6\textwidth]{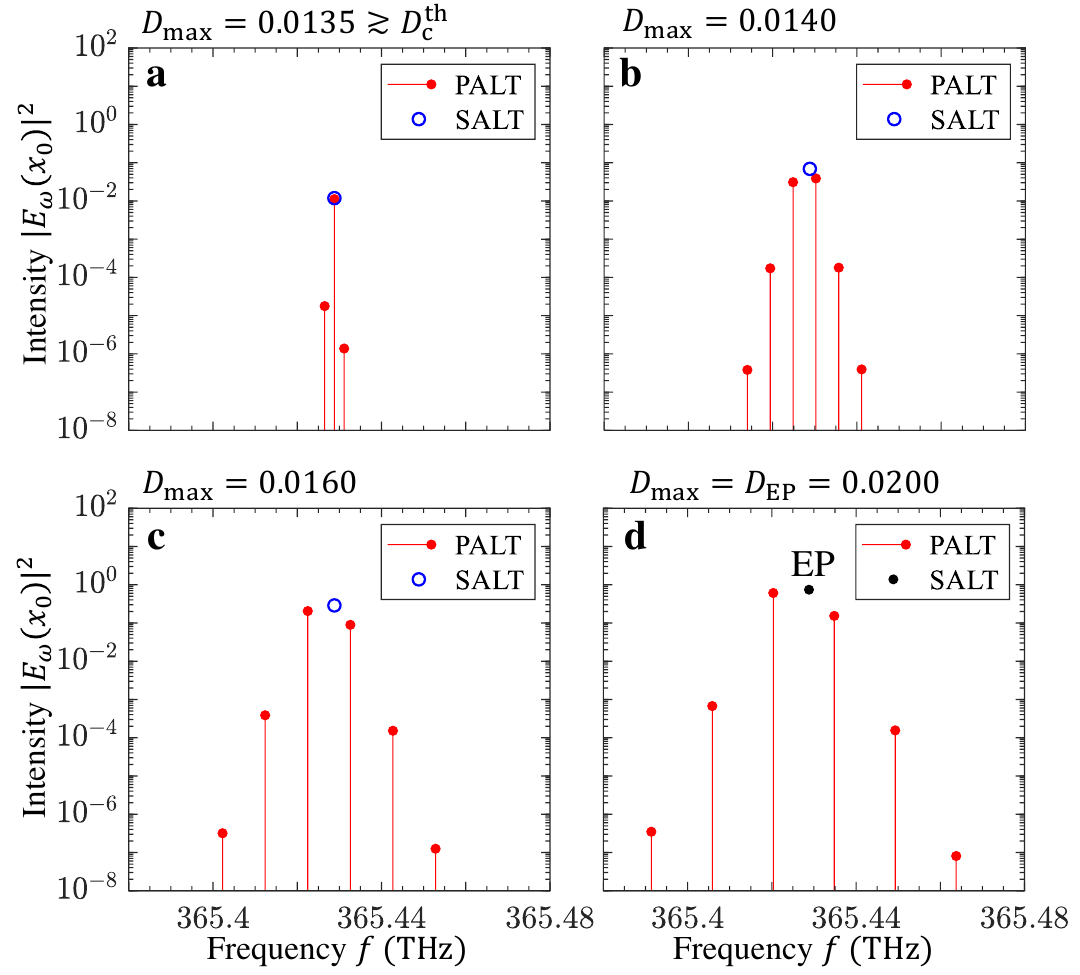}
\vspace{-6pt}
\caption{The spectrum of the exact-EP laser in Supplementary Fig.~\ref{fig:s1} at different pumping strengths, calculated by PALT and SALT. \textbf{a} The spectrum at a $D_\text{max}$ of 0.0135, which is slightly above the comb threshold. \textbf{b}--\textbf{c} The evolution of the spectrum as $D_\text{max}$ increases from the comb threshold towards the exact exceptional point. \textbf{d} The spectrum at the exact exceptional point, $D_\text{max}=D_\text{EP}$, where the laser operates as a stable comb even though SALT incorrectly $D_\text{max}$ increases from predicts a single-mode EP behavior.
$x_0 \approx 1.0$ \textmu m is the location marked in Fig.~2b of the main text.}
\label{fig:s5}
\end{figure} 

Like the near-EP example considered in the main text, if one were to apply SALT to this exact-EP system, one would incorrectly conclude that the system stays in the single-mode lasing regime over a wide range of pumping values with $D_\text{max}>D^\text{th}_1$, including at $D_\text{max}=D_\text{EP}$, for the same reason as described in Sec.~\ref{sec:sys_params}.
However, when applying the PALT single-mode stability analysis by solving Eqs.~\eqref{M1:E1}--\eqref{M1:E-1} on the EP state, we find a stability eigenvalue of $\omega_d = (0.0557 + 0.0254i)~\text{ps}^{-1}$ with a positive imaginary part, indicating that this exact-EP state is in fact unstable.
This result is consistent with Ref.~\citen{2022_Benzaouia_APLph}, which also found a single-mode lasing EP to be unstable in a different coupled cavity system but did not find what state the system would evolve into.

Equipped with PALT, we can now find out what state the laser evolves to when the single-mode lasing EP is unstable.
We fix the other system parameters and gradually increase the pumping strength, starting from zero pump.
With the PALT stability analysis, we find the single-mode solution to be stable only within a very small range of pumping strengths, 
$D^\text{th}_1 = 0.01336 < D_\text{max} < D^\text{th}_2 = D^\text{th}_{\rm c} = 0.01346$.
At the comb threshold $D^\text{th}_{\rm c}$, the laser develops multiple frequencies with a repetition rate of 2.3~GHz.
\textbf{Supplementary Fig.}~\ref{fig:s5} shows the spectra at different $D_\text{max}$, comparing the exact PALT and the SALT predictions.
Because the absorption is not raised together with the pumping strength, the comb spacing increases above the comb threshold.
At $D_\text{max}=D_\text{EP} = 0.02000$, the laser operates as a stable comb with 14.5 GHz repetition rate.  
In other words, the exact lasing EP is unreachable. The system evolves into an EP comb before reaching the unstable exact EP.


\clearpage

\bibliographystyle{naturemag}
\bibliography{supplementary}